\documentclass[nofootinbib,aps,prd,reprint,superscriptaddress]{revtex4-2}

\usepackage[utf8]{inputenc} 
\usepackage{graphicx,overpic,mathtools}
\usepackage{amsthm,amsmath,amssymb,hyperref}
\usepackage{braket,bm,bbm,setspace}
\usepackage[normalem]{ulem} 
\usepackage{physics}
\usepackage{float}
\usepackage[makeroom]{cancel}
\usepackage[english]{babel}
\usepackage{xcolor}
\usepackage{tensor}
\graphicspath{ {images/} }
\addto\captionsspanish{}
\hypersetup{
	colorlinks=true,
	pdfborder={0 0 0},
	citecolor=blue,
	linkcolor=blue,
	filecolor=blue,
	urlcolor=black,
}
\usepackage{subfigure}

\makeatletter
\def\l@subsection#1#2{}
\def\l@subsubsection#1#2{}
\makeatother

\begin{document}
\title{Generalized Quantum PageRank Algorithm with Arbitrary Phase Rotations}
\author{Sergio A. Ortega}
\email{sergioan@ucm.es}
\affiliation{Departamento de Física Teórica, Universidad Complutense de Madrid, 28040 Madrid, Spain}
\author{Miguel A. Martin-Delgado}
\email{mardel@ucm.es}
\affiliation{Departamento de Física Teórica, Universidad Complutense de Madrid, 28040 Madrid, Spain}
\affiliation{CCS-Center for Computational Simulation, Universidad Politécnica de Madrid, 28660 Boadilla del Monte, Madrid, Spain.}

\begin{abstract}
	{The quantization of the PageRank algorithm is a promising tool for a future quantum internet. Here we present a modification of the quantum PageRank introducing arbitrary phase rotations (APR) in the underlying Szegedy's quantum walk. We define three different APR schemes with only one phase as a degree of freedom. We have analyzed the behavior of these algorithms in a small generic graph, observing that a decrease of the phase reduces the standard deviation of the instantaneous PageRank, so the nodes of the network can be distinguished better. However, the algorithm takes more time to converge, so the phase can not be decreased arbitrarily. With these results we choose a concrete value for the phase to later apply the algorithm to complex scale-free graphs. In these networks, the original quantum PageRank is able to break the degeneracy of the residual nodes and detect secondary hubs that the classical algorithm suppresses. Nevertheless, not all of the detected secondary hubs are real according to the PageRank's definition. Some APR schemes can overcome this problem, restoring the degeneration of the residual nodes and highlighting the truly secondary hubs of the networks. Finally, we have studied the stability of the new algorithms. The original quantum algorithm was known to be more stable than the classical. We have found that one of our algorithms, whose PageRank distribution resembles the classical one, has a stability similar to the original quantum algorithm.}
\end{abstract}

\maketitle

\section{Introduction}
\label{sec-introduction}

The revolution of search engines to surf the internet originates from novel algorithms inspired by the PageRank algorithm \cite{Brin1,Brin2,Brin3,Google_book}. Contrary to its competitors, whose ranking of pages was quite subjective, the PageRank algorithm classifies in an objective manner, taking into account the structure of links between the pages. Beyond its importance in the World Wide Web (WWW), it is worth mentioning that the PageRank algorithm has a lot of applications, such as in bibliometrics \cite{PR-Biblio-1,PR-Biblio-2}, finances \cite{PR-Finances}, metabolic networks \cite{PR-Metabolism}, drug discovery \cite{PR-Drug}, protein interaction networks \cite{PR-BQ}, and social networks \cite{PR-Twitter}.

In the early era of quantum computing there has been great interest in the development of large-scale quantum networks with the perspective of a future quantum internet \cite{QN-QInternet,QN-QKD,QN-Teleportation}. Quantum networks require less resources than a full quantum computer, and thus are expected to become available before having a fault-tolerant quantum computer. As happens with classical information, the quantum information of the quantum internet will need to be classified. In this sense, the classical PageRank could be used in a classical computer to classify the quantum information. However, it is sensible to think that a quantized version of the PageRank algorithm will classify the quantum information better, taking into account the effects of superposition and interference of quantum mechanics. With that purpose, in 2012 a quantization of the PageRank algorithm was proposed \cite{Paparo1}. This quantization was based on a quantum walk introduced by Szegedy \cite{Szegedy} as a generalization of the Grover algorithm \cite{Grover}, since the classical PageRank algorithm can been understood as a random walk in the network formed by the pages (nodes) and their links (edges). The classical simulation of the quantum algorithm belongs to the computational complexity class $P$, so it can be simulated in a classical computer to classify the quantum information of near-term quantum networks, even lacking a fault-tolerant quantum computer.

The quantum PageRank was first implemented in small networks, showing intriguing properties such as a violation in the nodes ranking \cite{Paparo1}. Later, it was scaled to complex networks, showing further properties such as a better resolving of the network structure \cite{Paparo2}. Since then, there has been a recent interest in the quantum PageRank. For example, other quantizations such as coined discrete-time quantum walks \cite{DTQW-PR}, and continuous-time quantum walks \cite{CT-QPR,Comparing-CQ,TF_QPR} have been proposed, and it has even been realized experimentally in the continuous-time version using photons \cite{Exp_QPR}. It has also been coupled to quantum search as a further step towards a quantum search engine \cite{Searchrank}. On one hand, the results obtained with the quantum algorithm are expected to be more sensible for quantum networks. However, until larger quantum networks become available, this is still an open problem. On the other hand, if we consider that the information is classical, then the quantum PageRank shows features that can even enhance the classical algorithm in this context. Thus, the quantum PageRank is interesting not only for the future quantum networks, but also for the current classical ones.

Szegedy's quantum walk has a wide range of applications, such as optimization and artificial intelligence, for example. With the purpose of going beyond this quantum walk, in this paper we propose a modification of the algorithm, introducing arbitrary phase rotations (APR). This technique was introduced in the context of Grover quantum search \cite{Grover,Grover2,PM-Search,MA_Grover,PM-Search-2}, and has been used to improve its performance \cite{PM-Landa,PM-Multi,PM-FP} and even make it deterministic \cite{PM-Zero,DGrover}. In order to show how the arbitrary phase rotations can enhance this quantum walk, we apply this modification to the quantum PageRank. We will compare our results with those of the original quantum PageRank using first the same small graph studied in \cite{Paparo1}, and later using scale-free networks, which are complex networks that model the WWW \cite{SF-WWW}. We will be comparing the performance of the new quantum algorithms with classical information, keeping in mind that the results can be of interest for future quantum networks as well.

This paper is structured as follows. In Section \ref{PageRank} we introduce the PageRank algorithm and our new approach. In Section \ref{SGG} we study the effect of the new quantum algorithm in a small generic graph. In Section \ref{SF} we apply this algorithm to complex scale-free networks. In Section \ref{Stability} we study the stability of the new algorithms. Finally, we summarize and conclude in Section \ref{Conclusions}. In several appendices we elaborate more on the details when necessary and apply our generalized quantum PageRank to Erdős-Rényi networks.

\section{New Quantum PageRank Algorithms}\label{PageRank}

The aim of this work is to study the effect of introducing APR on Szegedy's quantum walk applied to quantum PageRank. Before describing this modification, let us review the classical algorithm and its quantization, whose complete details can be found in \cite{Paparo1}.

In the classical algorithm $I_c$ is defined as the vector whose entries are the classical importance or PageRanks of every page $P_i$. The naive definition of the PageRank is the following:
\begin{equation}\label{PR}
	I_c(P_i) := \sum_{P_j \in B_i} \frac{I_c(P_j)}{\text{outdeg}(P_j)},
\end{equation}
where $B_i$ is the set of nodes linking to the node $P_i$ and outdeg($P_j$) is the outdegree of the node $P_j$. This formula means that the importance of a node depends of the nodes that link to it. The more important a linking node is, the greater is its contribution to the PageRank. However, its contribution is equally distributed between all the nodes it links to.

In order to compute the PageRank we use a random walk in a directed graph whose nodes represent the pages $P_i$, and the associated connectivity $N\times N$ matrix $H$, defined as
\begin{equation}
	H_{i,j} := 
	\left\lbrace\begin{array}{c}
		1/\text{outdeg}(P_j) \ \ \ \text{if} \ P_j \in B_i,\\
		0 \ \ \ \ \ \ \ \ \ \ \ \ \ \ \ \ \ \ \ \text{otherwise},
	\end{array}
	\right.
\end{equation}
where $N$ is the number of nodes in the network. With this matrix equation \eqref{PR} can be expressed as $I_c = HI_c$, so we can apply a power method to obtain the eigenvector $I_c$. However, for the algorithm to work this matrix must be patched. First, it needs all columns where all the elements are zeros (which correspond to nodes whose outdegree is zero) to be substituted with columns with all entries equal to $1/N$. This results in a (column-) stochastic matrix $E$, where all columns sum up to one. Secondly, before performing the random walk, this matrix $E$ is mixed with another matrix \textbf{1} where all entries are equal to $1$, obtaining a primitive and irreducible matrix called the Google matrix $G$:
\begin{equation}\label{G}
	G := \alpha E + \frac{(1-\alpha)}{N} \text{\textbf{1}}.
\end{equation}
The parameter $\alpha$ is called the damping parameter corresponding to the previous mixing, and its value lies in $[0,1]$. It was found by Brin and Page that the optimal value is $\alpha = 0.85$. In our paper this value of the damping parameter is considered unless otherwise stated. This mixing can be interpreted as that the random walk is performed in the network of interest driven by $E$ with probability $\alpha$, and with probability $1-\alpha$ the walker makes a random hopping driven by the matrix \textbf{1}. We perform the random walk with the patched matrix $G$, so we redefine the vector of PageRanks satisfying $I_c = G I_c$. Thus, it is the eigenvector with eigenvalue $1$ of the matrix $G$. Thanks to the mixing with the random hopping matrix, a random walk performed with the matrix $G$ over any probability distribution will converge to this eigenvector. Then, now we can use a power method to obtain this eigenvector. We only have to take a probability distribution and repeatedly apply the matrix G until it converges. In this work, we choose the equally probability distribution.

The quantization of this algorithm that we are going to deal with is based on the quantum walk of Szegedy \cite{Szegedy}, using as transition matrix $G$. The Hilbert space is the span of all the vectors representing the $N \times N$ directed edges of the duplicated graph, i.e., $\mathcal{H} = \text{span}\lbrace\left|i\right>_1\left|j\right>_2,\ i,j \in N \times N\rbrace = \mathbb{C}^N \otimes \mathbb{C}^N$, where the states with indexes $1$ and $2$ refer to the nodes on two copies of the original graph.

We define the vectors:
\begin{equation}
	\left|\psi_i\right> := \left|i\right>_1 \otimes \sum_{k=1}^N \sqrt{G_{ki}}\left|k\right>_2,
\end{equation}
which are a superposition of the vectors representing the edges outgoing
from the $i^{th}$ vertex, whose coefficients are given by the square root of the $i^{th}$ column of the matrix $G$. From these vectors we define the projector operator onto the subspace generated by them:
\begin{equation}
	\Pi := \sum_{k=1}^N \left|\psi_k\right>\left<\psi_k\right|.
\end{equation}
The quantum walk operator $U$ is defined as
\begin{equation}
	U := S(2\Pi - \mathbbm{1}),
\end{equation}
where $S$ is the swap operator between the two quantum registers, i.e., $S = \sum_{i,j=1} \left|i,j\right>\left<j,k\right|$. Since the swap operator changes the directedness of the graph, the operator $U$ must be applied an even number of times, so the actual time evolution operator is chosen as $W := U^2$.

The initial state of the system is chosen to be
\begin{equation}
	\left|\psi_0\right> = \frac{1}{\sqrt{N}}\sum_{i=1}^N\left|\psi_i\right>,
\end{equation}
and the final state is $\left|\psi_f(t)\right> = W^t\left|\psi_0\right>$. The position of the walker after the quantum evolution is described by the register $2$ of the quantum state, so the projection onto the computational basis of the second register will give us the quantum PageRanks for each node:
\begin{equation}\label{InstPR}
	I_q(P_i,t) := ||\tensor[_2]{\left<i\right|\left.\psi_f(t)\right>}{}||^2.
\end{equation}
This quantum PageRank depending on time is called the instantaneous PageRank \cite{Paparo1}, and it fluctuates in time instead of converging. For that reason, the time-averaged quantum PageRank is defined as
\begin{equation}\label{AvPR}
	I_q(P_i) := \frac{1}{T}\sum_{t=0}^T I_q(P_i,t).
\end{equation}
This quantity converges for a sufficiently large value of $T$ \cite{Paparo2}. In the following, when we refer to the quantum PageRank we mean the time-averaged quantum PageRank unless we mention explicitly the instantaneous PageRank.

The introduction of arbitrary phase rotations is done in Grover's algorithm by introducing complex phases in the reflection operators. Then, they apply an arbitrary complex phase to the perpendicular component of a vector, instead of changing its sign. The part $2\Pi-\mathbbm{1}$ of the quantum walk operator corresponds to a reflection over the subspace generated by the states $\left|\psi_i\right>$, so a natural modification of Szegedy's quantum walk to introduce APR is to define a new unitary operator as:
\begin{equation}
	U(\theta) := S\left(\left[1-e^{i\theta}\right]\Pi - \mathbbm{1}\right),
\end{equation}
where $\theta \in (-\pi,\pi]$.

As we have mentioned, the quantum PageRank needs to preserve the directedness of the graph, so we need to apply this operator an even number of times. Let us define the actual unitary time evolution operator of the quantum PageRank algorithm as
\begin{equation}
	W(\theta_1,\theta_2) := U(\theta_2)U(\theta_1),
\end{equation}
where in general $\theta_1 \neq \theta_2$. Thus, now the quantum PageRanks will be obtained in the same manner as before, but the final state will be $\left|\psi_f(t)\right> = W^t(\theta_1,\theta_2)\left|\psi_0\right>$. This gives rise to a new family of quantum PageRank algorithms with $\theta_1$ and $\theta_2$ as two degrees of freedom.

If we chose $\theta_1 = \theta_2 = \pi$, the original quantum PageRank algorithm is recovered, so we are going to name that particular case the standard case. To study the effect of choosing different phases, we are going to define three APR schemes with only one degree of freedom $\theta$:

\begin{itemize}
	\item Equal-Phases scheme: both phases are equal, i.e., $\theta_1 = \theta_2 = \theta$.
	\item Opposite-Phases scheme: phases have the opposite signs, i.e., $\theta_1 = -\theta_2 = \theta$.
	\item Alternate-Phases scheme: the first phase is fixed to $\pi$, while the second phase is free, i.e., $\theta_1 = \pi, \theta_2 = \theta$. We have found that the results for $\theta_1 = \theta, \theta_2 = \pi$ are similar, so we do not take into account that case in this work.
\end{itemize}

Since the operator $\Pi$ is real, inverting the sign of the phase $\theta$ would result in complex conjugating the operator $W(\theta_1,\theta_2$). The initial statevector is also real, so the final result would just be the complex conjugated. However, the quantum PageRanks are real probabilities, so there would not be any effect. For that reason, w.l.o.g. $\theta \in [0,\pi]$.

The matrix representing the unitary operator $W(\theta_1,\theta_2)$ is of size $N^2\times N^2$, so to simulate it classically we need memory resources that scale as $\mathcal{O}(N^4)$, as opposed to the classical algorithm whose memory requirements scale as $\mathcal{O}(N^2)$. Thus, the naive classical simulation of the quantum PageRank is very difficult when dealing with complex networks. This problem can be overcome using the spectral decomposition of the operator, avoiding the use of such a big matrix in order to save memory resources. Such a method for simulating the standard quantum PageRank is explained in \cite{Paparo1}, and in Appendix \ref{Ap_spectral} we show a generalized version that takes into account the APR. The optimized method only needs to store $2N$ eigenvectors from the dynamical subspace of the unitary operator, and since these vectors are $N^2$-dimensional, the total memory requirements scale as $\mathcal{O}(N^3)$.

Regarding the application of this algorithm in a quantum computer, there have been advances in constructing efficient circuits for Szegedy's quantum walk for certain kinds of graphs \cite{Q_circuits}. The proposed circuits consist of the construction of a unitary operator that diagonalizes the reflection operator, so that the reflection can be implemented with a multicontrolled-$\pi$ gate. Then, to take into account the APR in the circuit, it would only be necessary to change the multicontrolled-$\pi$ gate to a multicontrolled-$P(\theta)$ gate, where $P(\theta)$ is a gate that adds a phase $e^{i\theta}$ to the controlled qubit. This means that given a circuit for implementing the standard quantum PageRank algorithm, it can be easily modified to implement the new algorithms with APR.

In the following section we will study the effect of the phase $\theta$ in a simple small generic graph, and we will choose a concrete value of $\theta$ for studying the algorithm on complex graphs.

\section{Small Generic Graph}\label{SGG}

\begin{figure}[hbpt]
	\centering
	\includegraphics[scale=0.5]{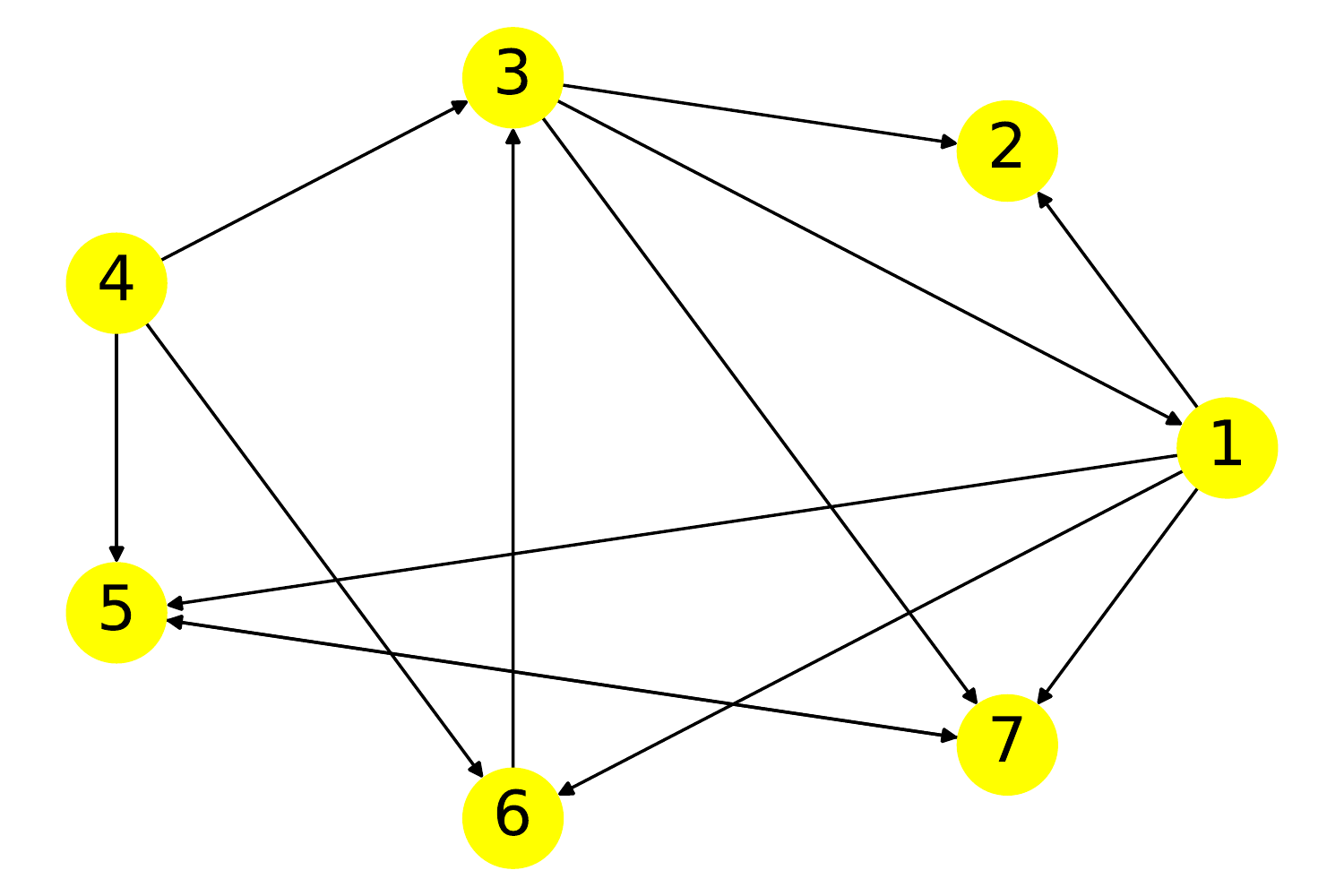}
	\caption{Small generic graph with seven nodes, where each node represent a web page and the directed edges represent links between web pages.}
	\label{F:General}
\end{figure}

Once we have defined the new family of quantum PageRanks that we are going to study, we want to see the effect of the free parameter $\theta$ in the three APR schemes described above. For this task we are going to use a small generic graph with seven nodes (see Figure \ref{F:General}) that was previously introduced in \cite{Paparo1} to study the standard quantum PageRank.

\begin{figure}[htpb]
	\centering
	\includegraphics[scale=0.5]{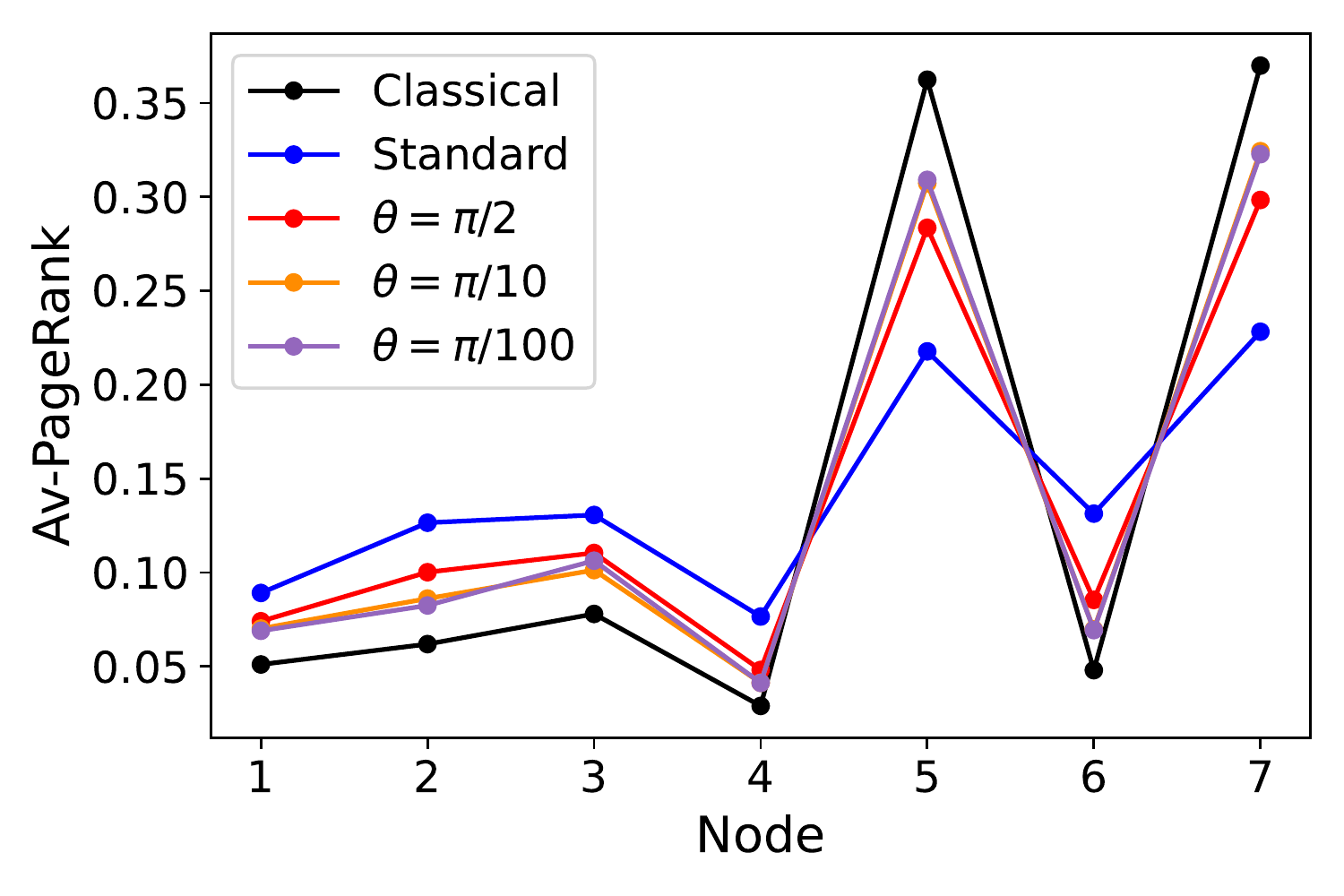}
	\caption{Time-averaged quantum PageRanks for the Alternate-Phases scheme with $\theta = \pi/2$, $\pi/10$, and $\pi/100$ for the small generic graph with seven nodes. They are compared with the classical PageRanks and the standard quantum PageRanks.}
	\label{F:PR_Alt}
\end{figure}

\begin{figure*}
	\subfigure[]{\includegraphics[scale=0.5]{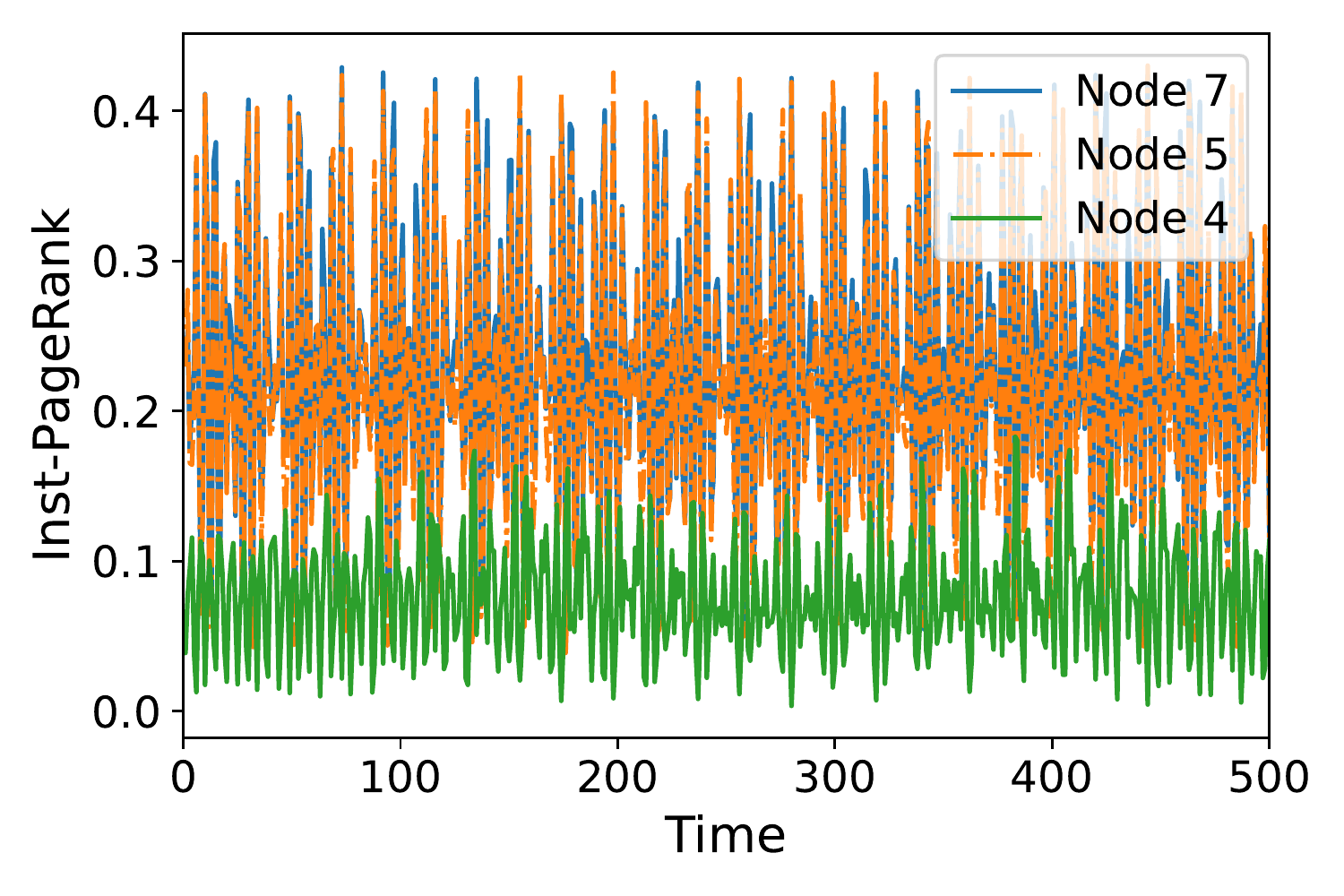}\label{F:Inst_s}}
	\subfigure[]{\includegraphics[scale=0.5]{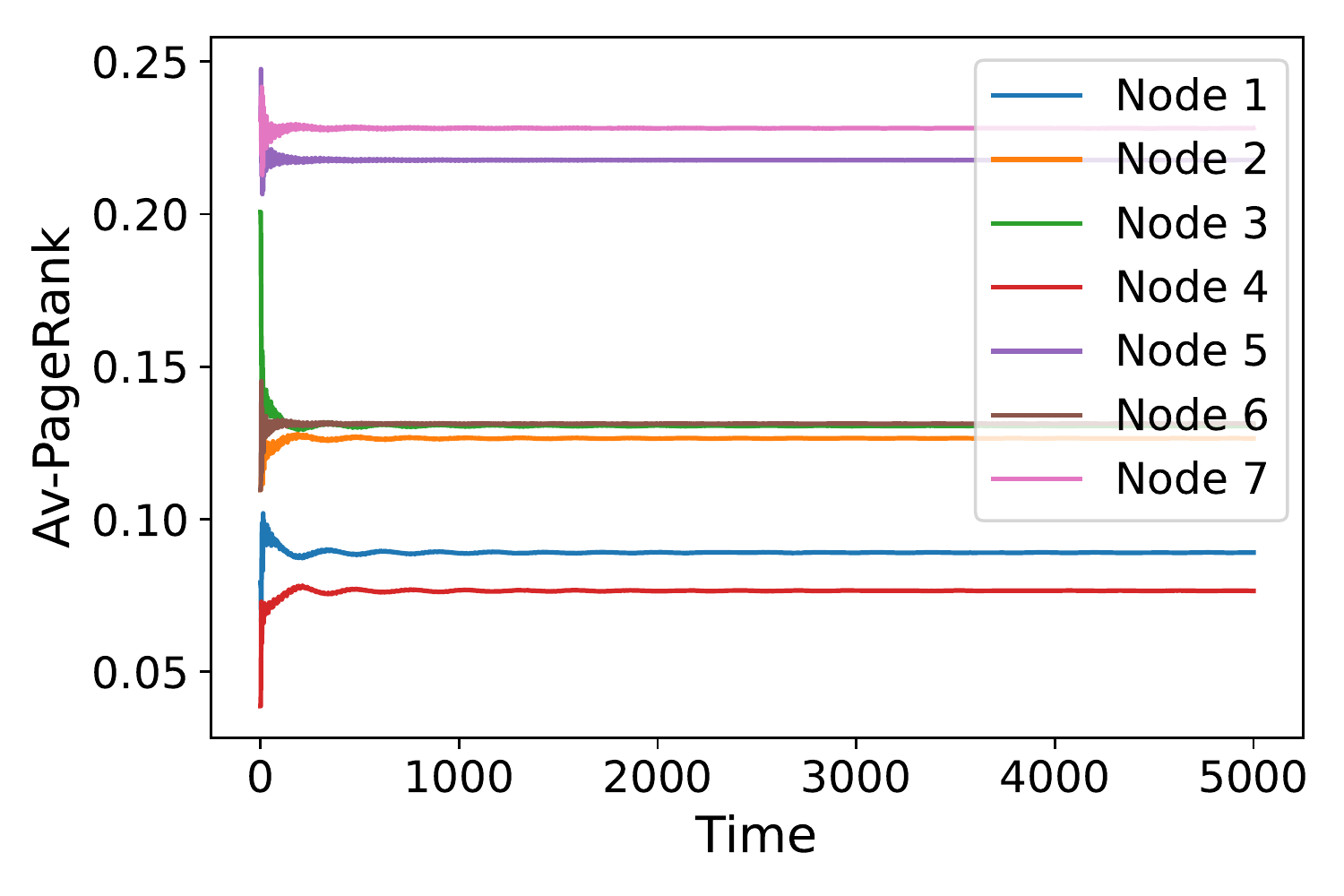}\label{F:Conv_s}}
	\subfigure[]{\includegraphics[scale=0.375]{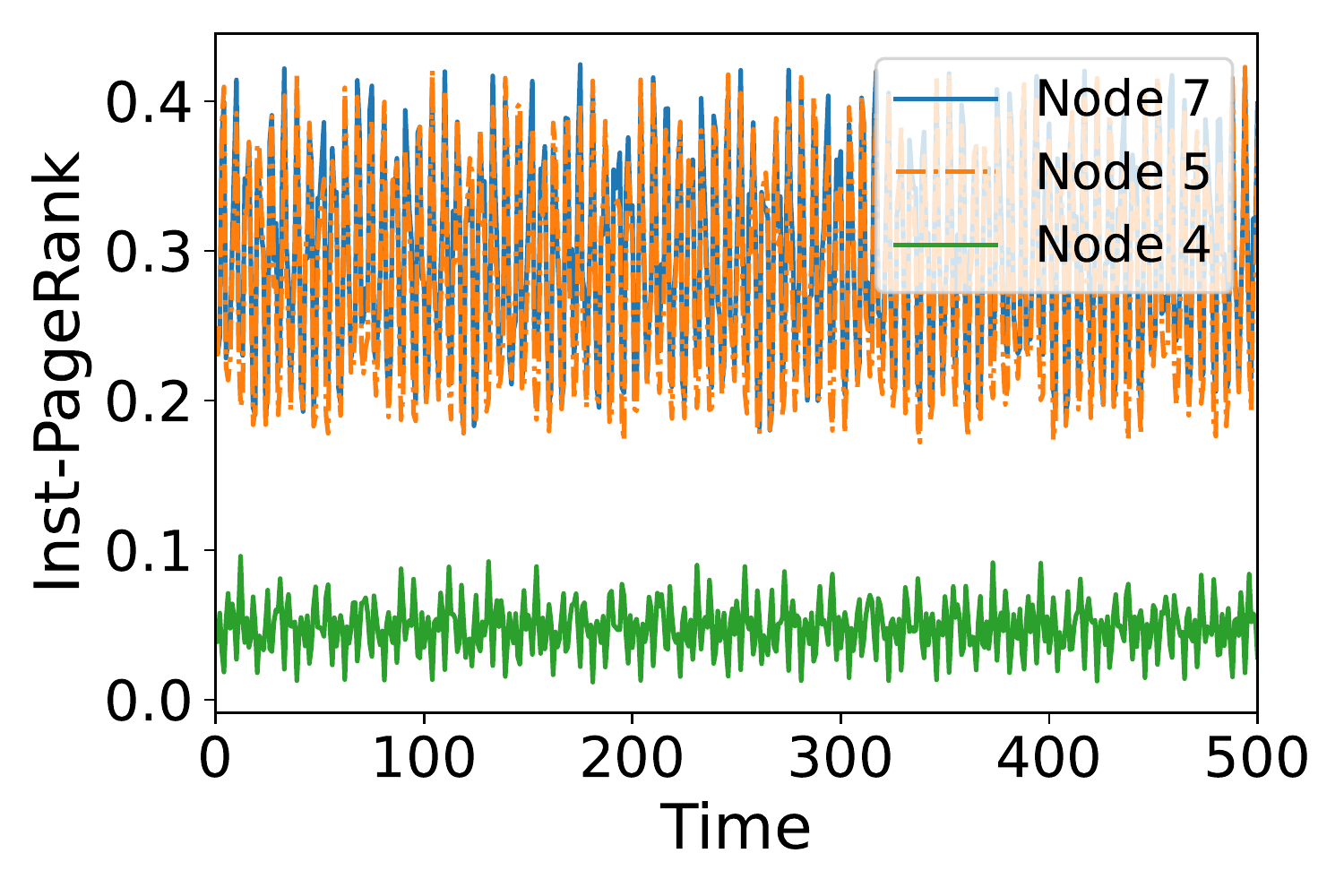}\label{F:Inst_Alt_pi2}}
	\subfigure[]{\includegraphics[scale=0.375]{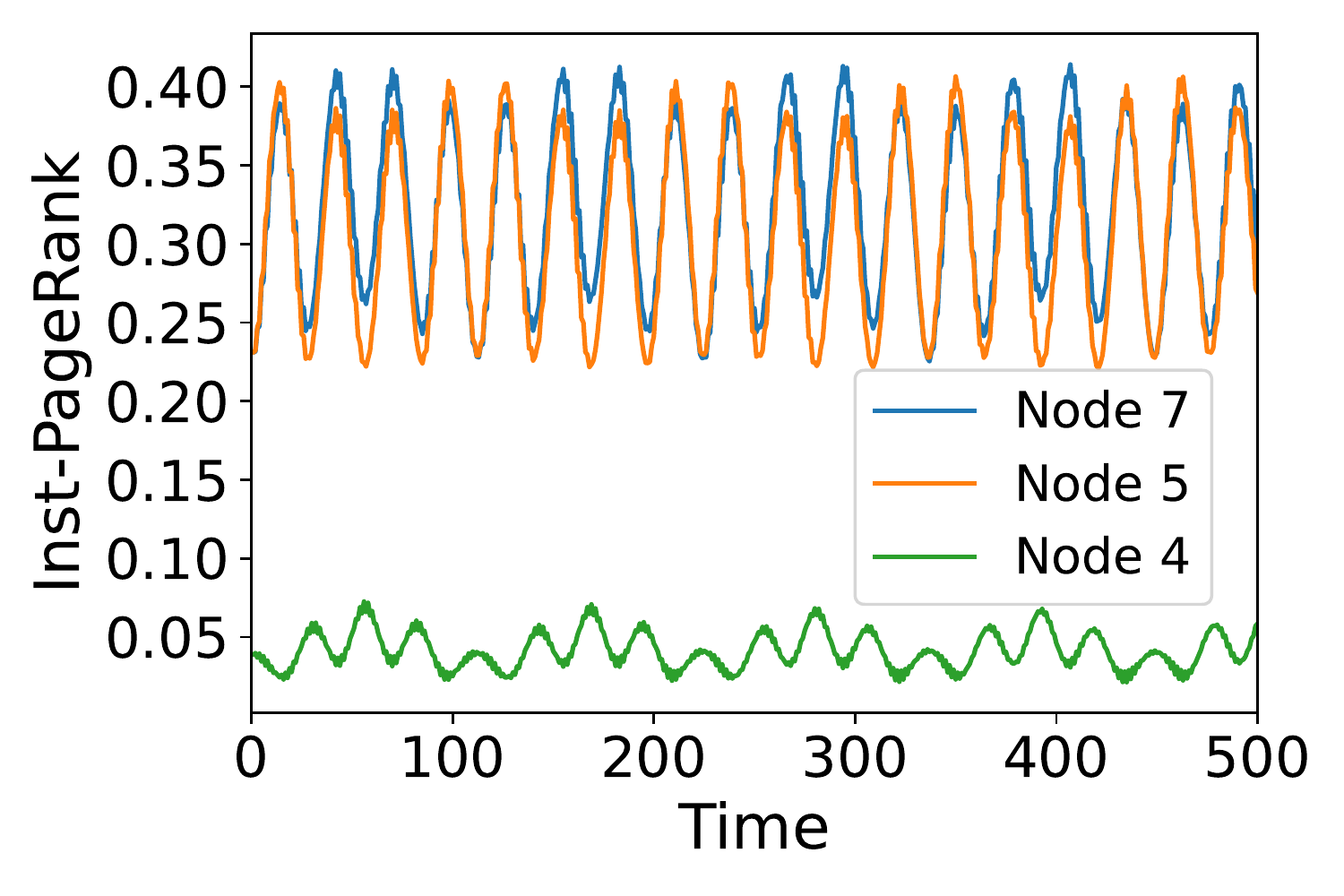}\label{F:Inst_Alt_pi10}}
	\subfigure[]{\includegraphics[scale=0.375]{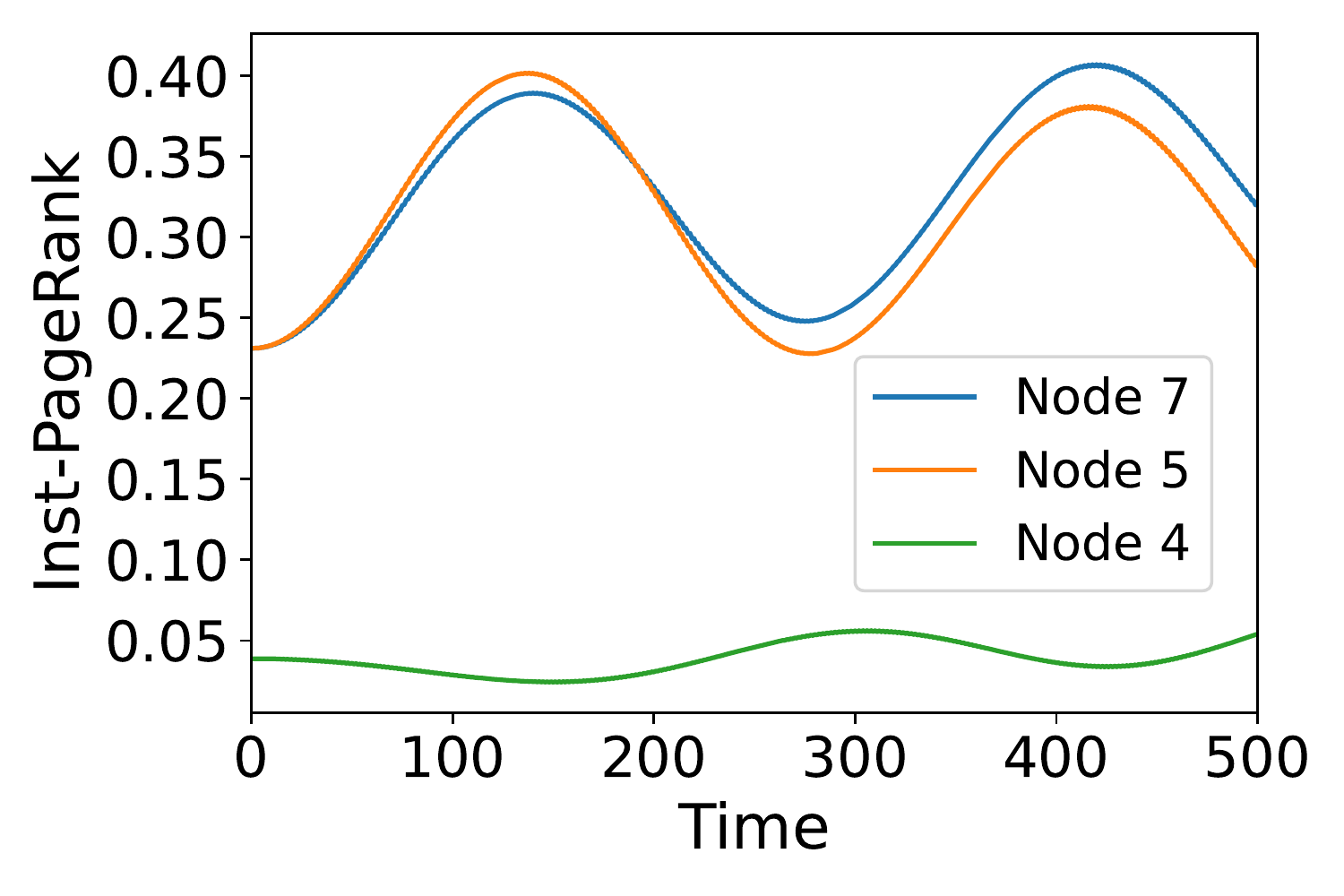}\label{F:Inst_Alt_pi100}}
	\subfigure[]{\includegraphics[scale=0.375]{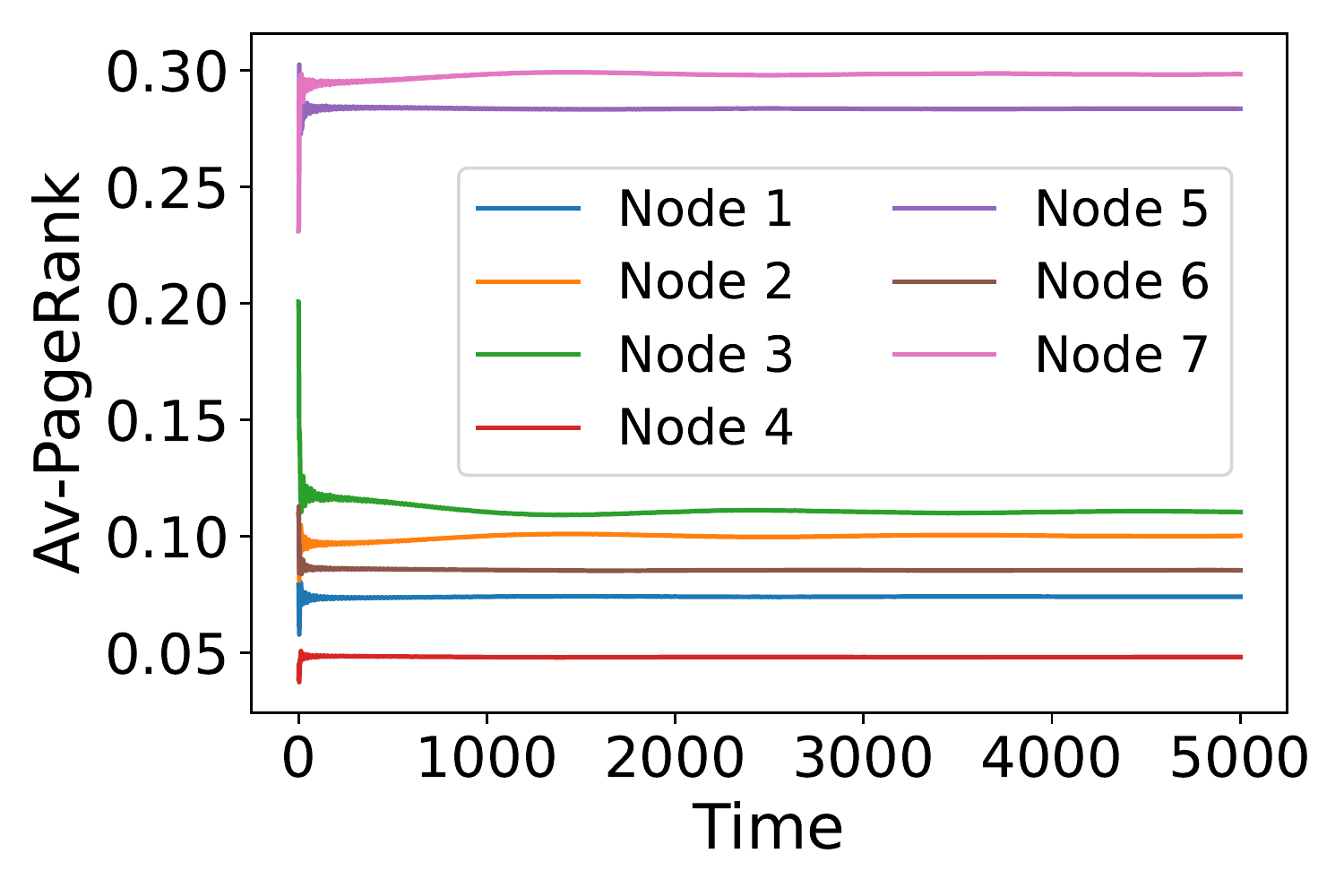}\label{F:Conv_Alt_pi2}}
	\subfigure[]{\includegraphics[scale=0.375]{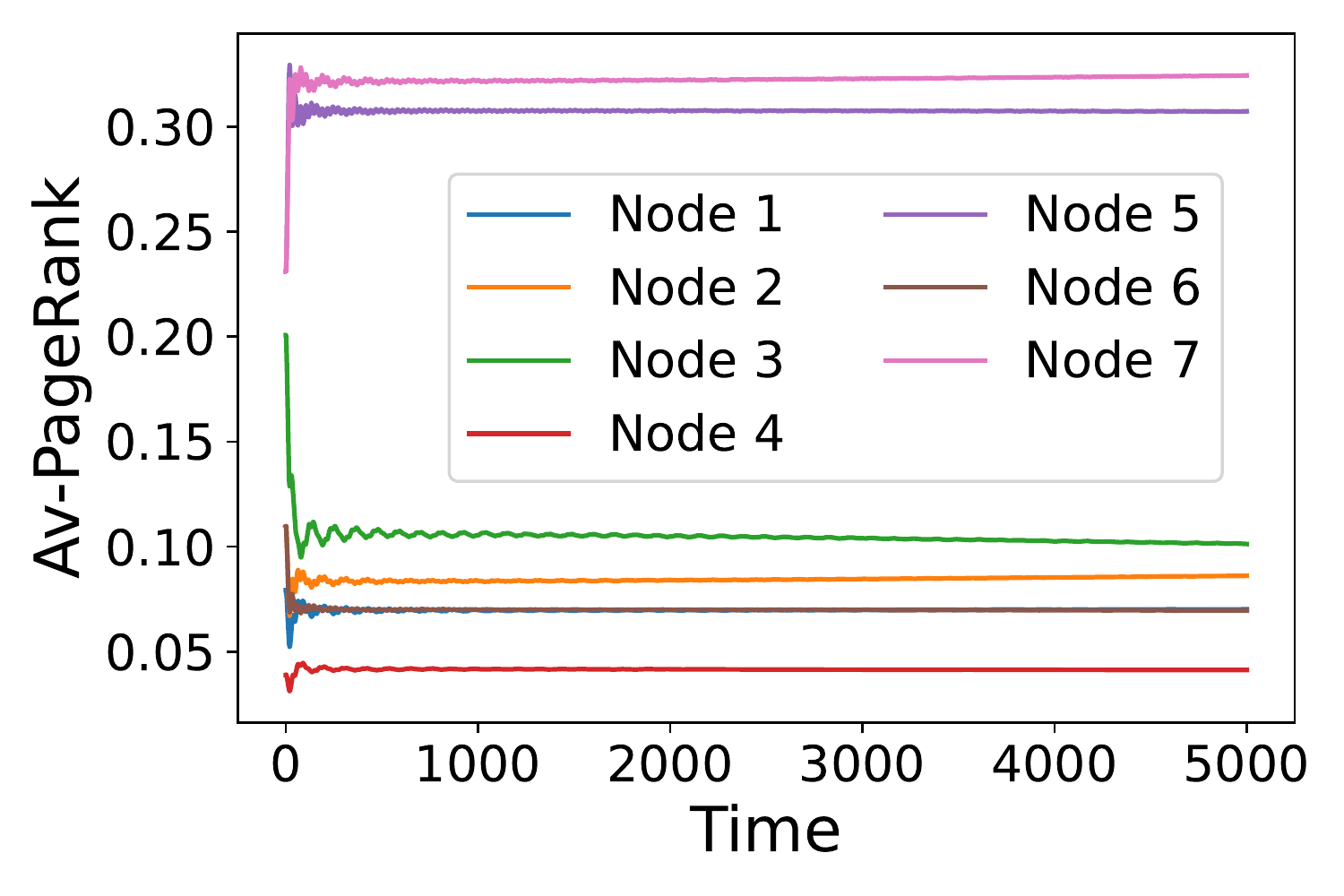}\label{F:Conv_Alt_pi10}}
	\subfigure[]{\includegraphics[scale=0.375]{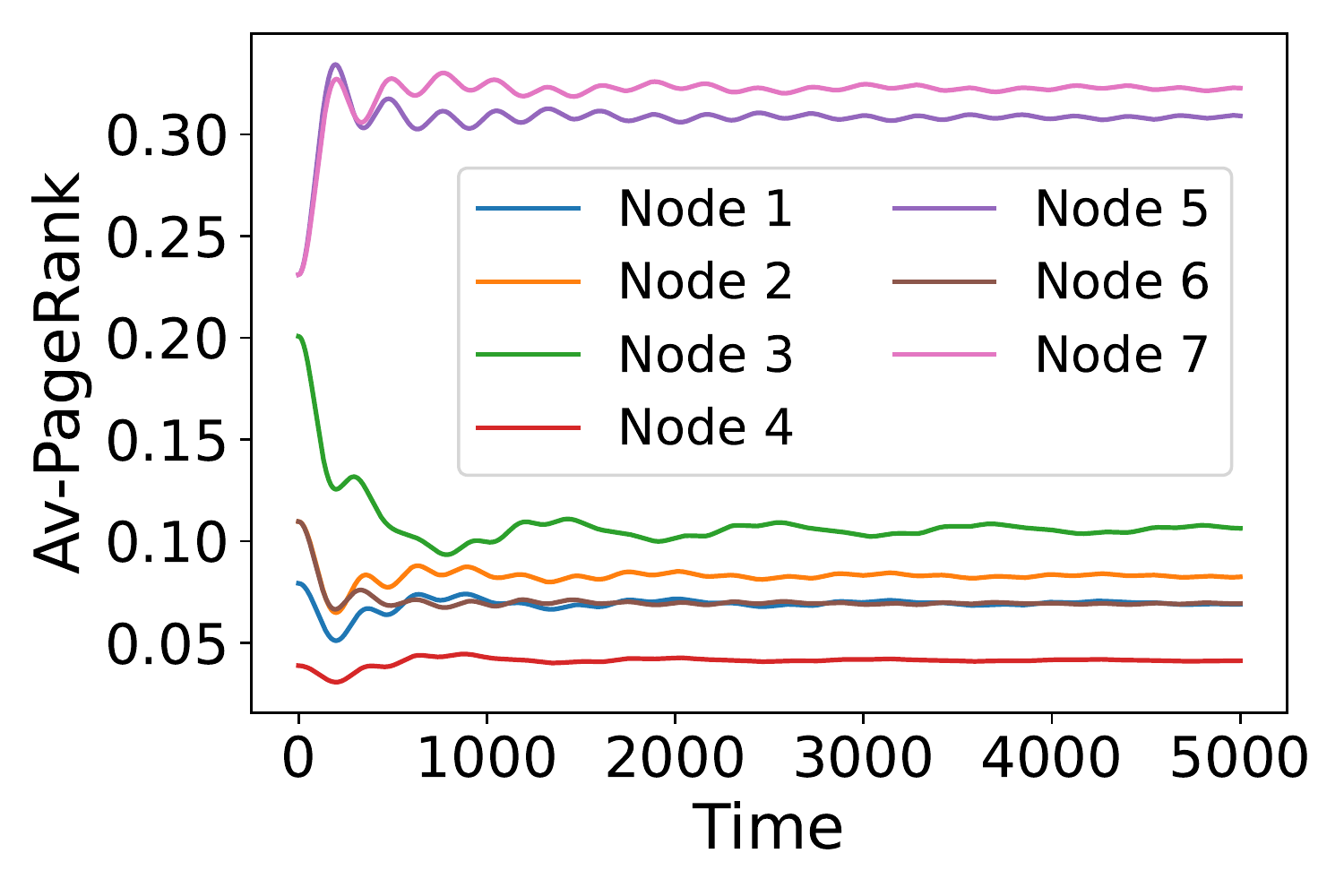}\label{F:Conv_Alt_pi100}}
	\caption{Instantaneous PageRanks of nodes $7$, $5$ and $4$ of the small generic graph for a) the standard quantum algorithm, c) the Alternate-Phases algorithm with $\theta = \pi/2$, d) the Alternate-Phases algorithm with $\theta = \pi/10$, and e) the Alternate-Phases algorithm with $\theta = \pi/100$. Time-averaged quantum PageRanks for all nodes vs time for b) the standard quantum algorithm, f) the Alternate-Phases algorithm with $\theta = \pi/2$, g) the Alternate-Phases algorithm with $\theta = \pi/10$, and h) the Alternate-Phases algorithm with $\theta = \pi/100$. It is observed that as $\theta$ decreases, the quantum fluctuations get slower and the algorithm takes more time to converge.}
	\label{...}
\end{figure*}

In \cite{Paparo1} it was shown that the classically most important node was node $7$, followed by node $5$, whereas the least important node was node $4$, as can be seen in Figure \ref{F:PR_Alt}. Their instantaneous PageRanks defined by equation \eqref{InstPR} for the standard quantum PageRank are reviewed in Figure \ref{F:Inst_s}, where it can be seen that the fluctuation changes the relative importance not only between nodes $5$ and $7$, but also between these nodes and the least important node $4$. For that reason, the decision was made to take the time-averaged quantum PageRank \eqref{AvPR} to rank these nodes. This quantity converges in time, as can be seen in Figure \ref{F:Conv_s}.

To see the effect of the complex phase $\theta$ on the instantaneous quantum PageRanks, we show the results for the Alternate-Phases scheme. However, similar results are obtained for the other APR schemes, which are shown in Appendix \ref{Ap_general}. As three interesting cases we choose a decreasing $\theta$ like $\theta = \pi/2$, $\pi/10$, and $\pi/100$. For $\theta = \pi/2$ the instantaneous quantum PageRanks in Figure \ref{F:Inst_Alt_pi2} have a smaller amplitude, allowing to properly distinguish the most important nodes from the least important node, and it is also shown that the fluctuation is slower. This oscillation gets slower and slower as we reduce the phase $\theta$, as can be seen for $\theta = \pi/10$ and $\theta = \pi/100$ in Figures \ref{F:Inst_Alt_pi10} and \ref{F:Inst_Alt_pi100}, respectively.

The time-averaged quantum PageRanks for the Alternate-Phases scheme are shown in Figure \ref{F:PR_Alt}, together with the standard quantum and the classical PageRanks. We observe that as the phase decreases, the distribution gets closer to the classical one, up to a limit. Indeed, from $\theta = \pi/10$ the ranking of the nodes is the same as that in the classical case, restoring the ranking violation showed by the standard quantum PageRank \cite{Paparo1}. However, the smaller $\theta$ is, the slower is the oscillation of the instantaneous PageRank, so we need more time to average the dynamics properly. This results in a slower convergence of the algorithm. In Figures \ref{F:Conv_Alt_pi2}, \ref{F:Conv_Alt_pi10}, and \ref{F:Conv_Alt_pi100} we show how the averaged quantum PageRank converges more slowly as the complex phase decreases. For this reason, we cannot decrease the angle $\theta$ arbitrarily.

\begin{table}[htbp]
	\centering
	\caption{Fidelity with the classical PageRank distribution for the standard quantum algorithm and the three APR schemes with $\theta = \pi/2$, $\pi/10$ and $\pi/100$, for the small generic graph.}
	\begin{tabular}{c|c|c|c}
		\hline
		Quantum case & $\theta = \pi/2$ & $\theta = \pi/10$ & $\theta = \pi/100$ \\
		\hline
		Standard & \multicolumn{3}{c}{$0.9546$} \\
		\cline{2-4}    Equal-Phases & $0.9874$ & $0.9886$ & $0.9887$ \\
		Opposite-Phases & $0.9638$ & $0.9622$ & $0.9621$ \\
		Alternate-Phases & $0.9870$ & $0.9940$ & $0.9941$ \\
		\hline
	\end{tabular}%
	\label{T:G_fidelities}%
\end{table}%

We can use the classical fidelity defined as

\begin{equation}\label{fidelity}
	f(I_1,I_2) := \sum_{i=1}^N\sqrt{I_1(P_i)I_2(P_i)},
\end{equation}
to measure the similarity between the quantum distributions and the classical one. The results for the three values of $\theta$ analyzed with the three APR schemes are summarized in Table \ref{T:G_fidelities}. From this table and Figures \ref{F:PR_Alt}, \ref{F:PR_Eq}, and \ref{F:PR_Op} (see Appendix \ref{Ap_general}) we can see that the major effect of the APR is achieved with $\theta = \pi/2$. Then, we can use this value, allowing the algorithm to converge relatively quickly, in the same manner as the standard quantum algorithm.

\begin{figure}[htpb]
	\centering
	\includegraphics[scale=0.5]{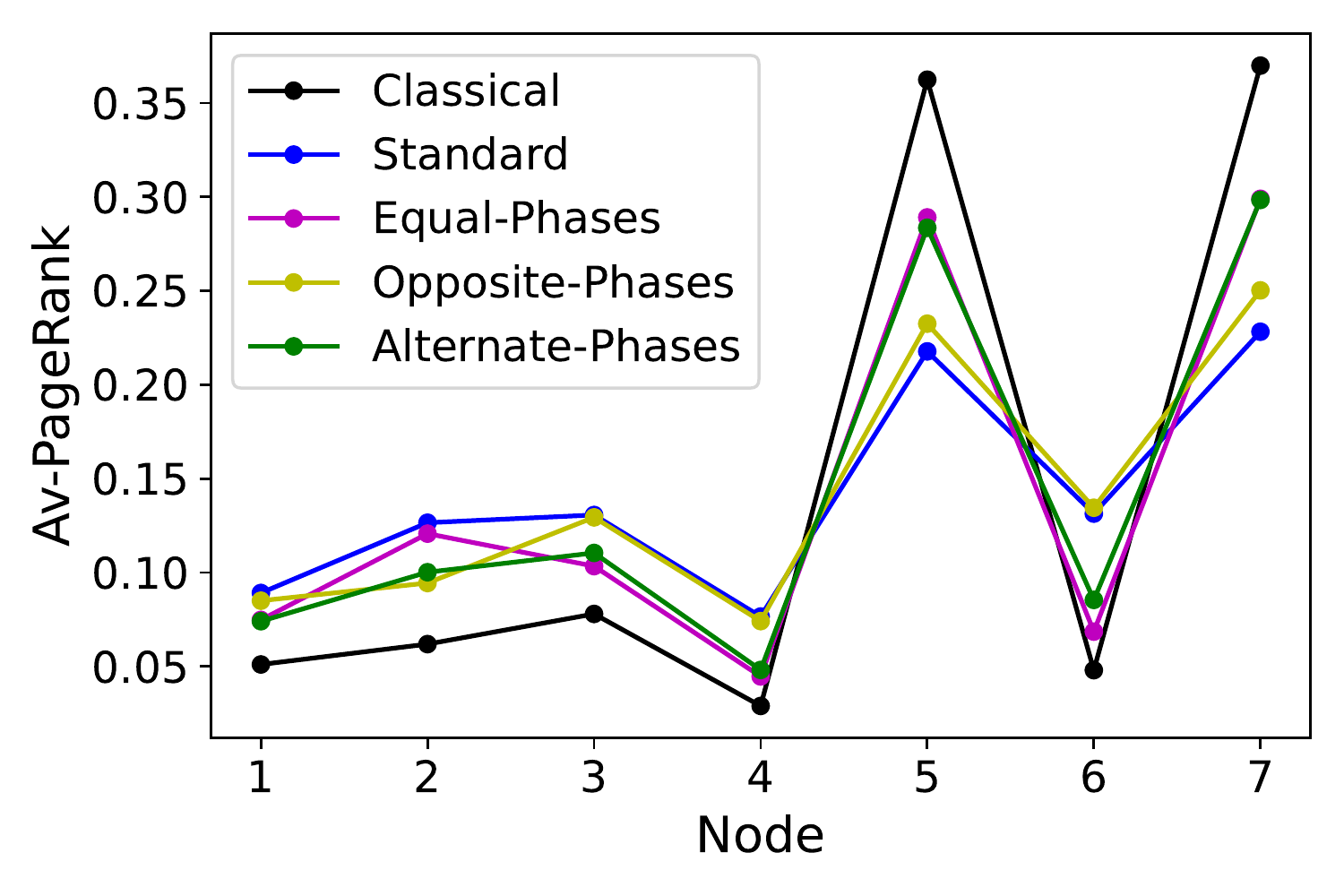}
	\caption{Time-averaged quantum PageRanks for the three APR schemes with $\theta = \pi/2$ for the small generic graph with seven nodes. They are compared with the classical PageRanks and the standard quantum PageRanks.}
	\label{F:PR_pi2}
\end{figure}

Once we have chosen a concrete value of $\theta$, we can compare the results for the three APR schemes with $\theta = \pi/2$. The averaged quantum PageRanks are shown in Figure \ref{F:PR_pi2}. We can see that for these graphs the Equal-Phases and Alternate-Phases cases gets qualitatively closer to the classical distribution, whereas the Opposite-Phases case resembles more the standard quantum distribution. As we will see later, this behavior depends on the type of graph, and will be different for complex networks.

If we consider the rankings of nodes, which are summarized in Table \ref{T:order}, we observe that the classical ranking is violated in all the quantum cases. The node that is shifted in the ranking varies between APR schemes. However, it is worth mentioning that all algorithms detect the two most important nodes as well as the less important node properly.

\begin{table}[H]
	\centering
	\caption{Ranking of the nodes of the small generic graph for the classical algorithm, the standard quantum algorithm, and the three APR schemes with $\theta = \pi/2$. The shifted node with respect to the classical ranking is marked in red.}
	\begin{tabular}{c|c|c|c|c}
		\hline
		Classical & Standard & Eq.-phases & Opp.-phases & Alt.-phases \\
		\hline
		7     & 7     & 7     & 7     & 7 \\
		5     & 5     & 5     & 5     & 5 \\
		3     & \textcolor[rgb]{ 1,  0,  0}{6} & \textcolor[rgb]{ 1,  0,  0}{2} & \textcolor[rgb]{ 1,  0,  0}{6} & 3 \\
		2     & 3     & 3     & 3     & 2 \\
		1     & 2     & 1     & 2     & \textcolor[rgb]{ 1,  0,  0}{6} \\
		6     & 1     & 6     & 1     & 1 \\
		4     & 4     & 4     & 4     & 4 \\
		\hline
	\end{tabular}%
	\label{T:order}%
\end{table}%

Another important result of the introduction of the APR is that the standard deviation of the fluctuation of the instantaneous quantum PageRank is reduced. In Table \ref{T:standards} we summarize the standard deviations for all nodes in the three APR schemes with $\theta = \pi/2$. The most significant reduction occurs in the Alternate-Phases case, although in general the three schemes show a tendency to decrease the standard deviation. This allows to better distinguish between different nodes, improving the performance of the quantum algorithm.

\begin{table}[htpb]
	\centering
	\caption{Standard deviations for the PageRanks of the nodes of the small generic graph using the standard quantum algorithm, and the three APR schemes with $\theta = \pi/2$.}
	\begin{tabular}{c|c|c|c|c}
		\hline
		Node  & Standard & Eq.-phases & Opp.-phases & Alt.-phases\\
		\hline
		1     & $0.046$ & $0.044$ & $0.030$ & $0.029$\\
		2     & $0.071$ & $0.071$ & $0.033$ & $0.044$ \\
		3     & $0.063$ & $0.053$ & $0.055$ & $0.046$ \\
		4     & $0.039$ & $0.026$ & $0.029$ & $0.016$ \\
		5     & $0.105$ & $0.081$ & $0.088$ & $0.072$ \\
		6     & $0.070$ & $0.034$ & $0.064$ & $0.034$ \\
		7     & $0.102$ & $0.078$ & $0.084$ & $0.068$\\
		\hline
	\end{tabular}%
	\label{T:standards}%
\end{table}%

\section{Application to Complex Scale-free Graphs Networks}\label{SF}

\subsection{PageRank distributions}

Now that we have seen the effect of the APR in a small network, we want to study the behavior of the new algorithms in complex networks, where the standard quantum algorithm has shown a good performance \cite{Paparo2}. In this context, we are going to use scale-free networks, which not only are good models of the World Wide Web \cite{SF-WWW}, but also have a wide range of applications such as in neuronal connections \cite{SF-Brain}, metabolomics \cite{SF-Metabolism-1,SF-Metabolism-2}, and finances \cite{SF-Finances}. These kinds of graphs are characterized by a power law distribution in the connectivity of nodes \cite{SF}, and it has been observed that the classical and quantum PageRanks also show a power law behavior \cite{Classical-PL,Paparo2}.

Scale-free networks are formed by continuously adding nodes, which are connected to the existing nodes with a probability that is proportional to the in- and out-degree of the existing nodes. Thus, it is expected that the first nodes added to the model have the largest number of nodes linking to them, which turns out in a higher ranking. These nodes will constitute the main hubs of the network, where the term hub refers to a node with a relatively large number of links \cite{Paparo2}. In our work we implement the model described in \cite{Directed-SF} to create random directed scale-free graphs using the python library NetworkX \cite{NetworkX} with the default parameters. This model considers networks with multiple edges and loops. In order to be in concordance with the PageRank's definition in \eqref{PR}, we have eliminated duplicated edges, but there would not be a major difference if we considered it. At the same time, no major difference in the results has been found whether we eliminate the loops or not, and we have decided to keep them.

\begin{figure}
	\subfigure[]{\includegraphics[scale=0.5]{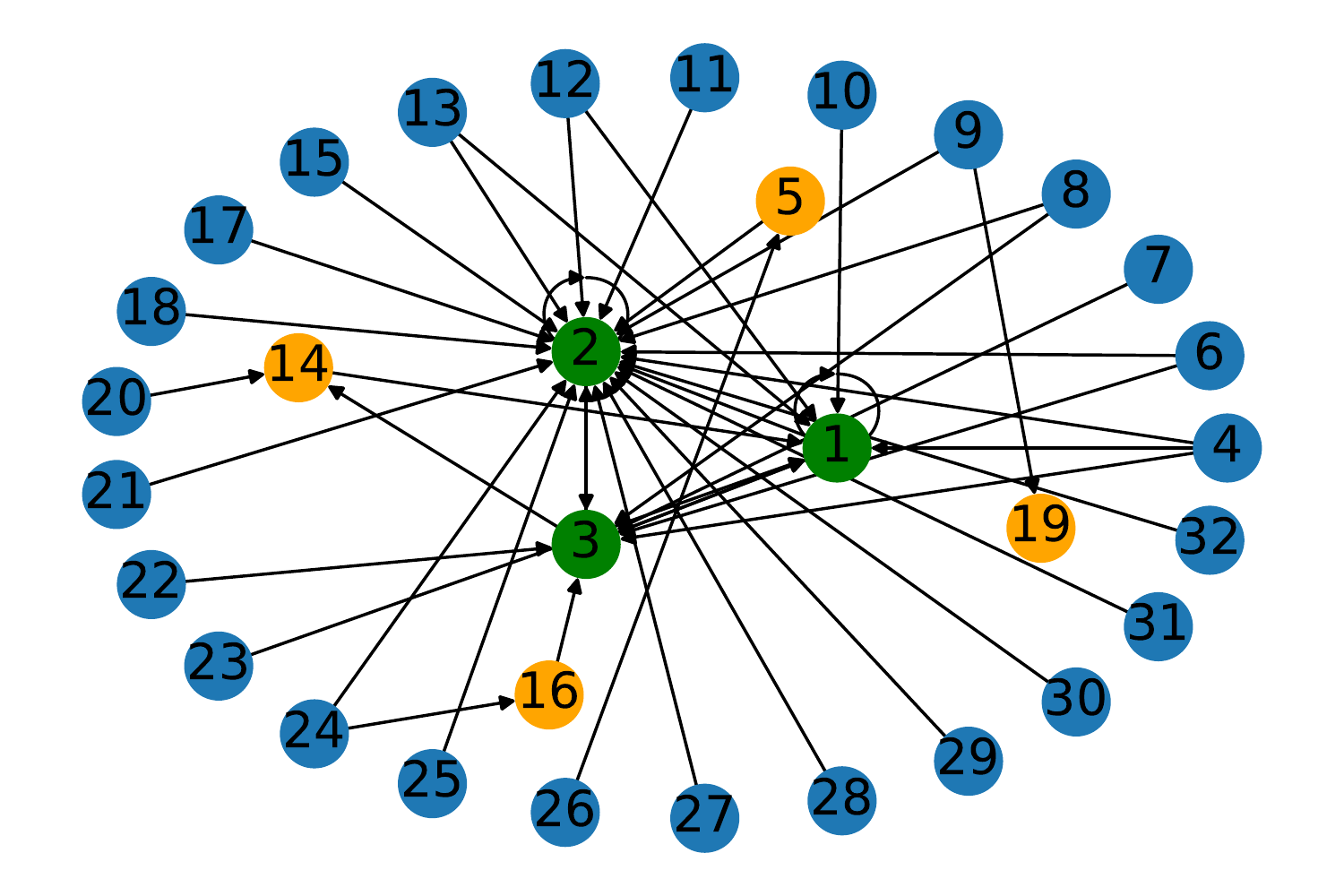}\label{F:PR_SF_32_graph}}
	\subfigure[]{\includegraphics[scale=0.25]{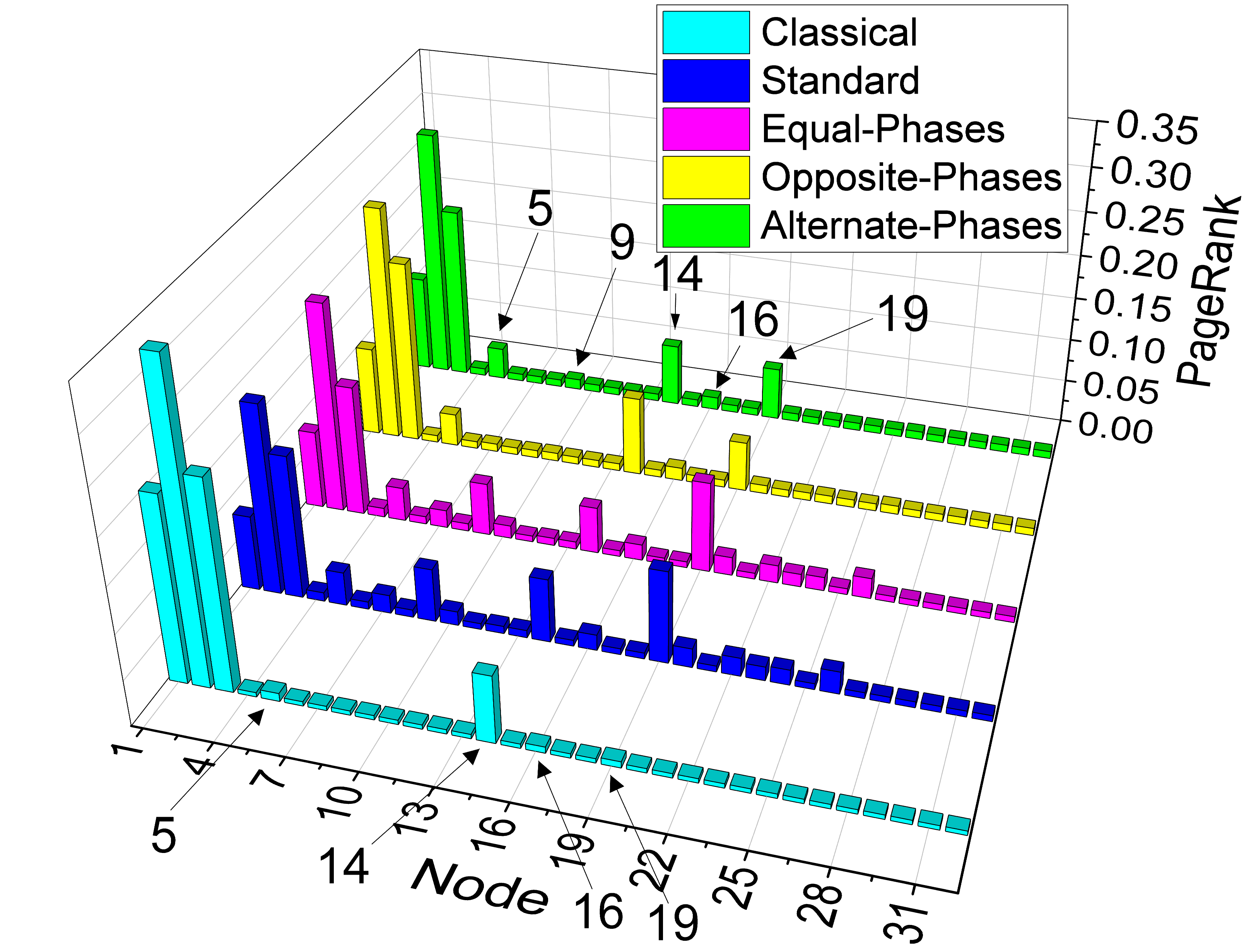}\label{F:PR_SF_32_hist}}
	\caption{a) Scale-free network with 32 nodes. The inner (green) nodes correspond to the main hubs. The middle (orange) nodes correspond to secondary hubs. The outer (blue) nodes correspond to residual nodes without links pointing to them. b) PageRank distributions of the scale-free network. The classical distribution is compared with all the quantum distributions, using $\theta = \pi/2$ in the three APR schemes. We see a partial restoration of the degeneracy of the less important nodes for the Opposite-Phases and Alternate-Phases schemes.}
	\label{F:PR_SF_32}
\end{figure}

In \cite{Paparo2} it was observed that with the classical PageRank the less important nodes are quite degenerate. However, the standard quantum PageRank could break this degeneracy. This meant that the quantum algorithm could unveil the structure of the graph in more detail. To study the effect of the APR with this kind of graphs, we have constructed a random scale-free graph with 32 nodes. The resulting network is shown in Figure \ref{F:PR_SF_32_graph}, whereas the classical PageRank and all the quantum PageRanks are shown in the histogram of Figure \ref{F:PR_SF_32_hist}. We can see effectively that the classical distribution has a degeneracy of the less important nodes, broken in the quantum distribution. When we look at the new algorithms with APR we find intriguing properties. Whereas the Equal-Phases case shows a pattern similar to the standard quantum algorithm, the Opposite-Phases and Alternate-Phases cases show a partial restoring of the degeneration of the less important nodes, resembling the classical distribution. This can be seen explicitly, for example, in nodes $20-26$, where the standard and Equal-Phases algorithms find differences in importance not present in the other distributions. Regarding the main hubs, as expected, they correspond to the three first nodes, and are detected properly by all the algorithms. Moreover, the classical relative importance of the three main hubs is kept by all the quantum algorithms, although this may not be the case for other graphs of the same class.

Let us look deeper at the structure of the graph. According to the classical definition of PageRank in \eqref{PR}, those nodes without links pointing to them would have null PageRank. Due to the patches introduced to built the Google matrix, these nodes have a small non-null PageRank, which is the same for all of them. These nodes are the outer (blue) nodes of Figure \ref{F:PR_SF_32_graph}. In the histogram we can effectively see that all of them are degenerate in the classical distribution. Node 14 is a secondary hub with two nodes linking to it, and since one of the linking nodes is a main hub (node 3), its PageRank is high enough to distinguish it. However, nodes 5, 16, and 19, which have a node linking to them, have a small classical PageRank that is very similar to the less important nodes. This means that the classical PageRank is not able to identify all secondary hubs properly. In the case of the standard quantum algorithm, it lifts the importance of these secondary hubs. Nevertheless, it breaks the degeneracy of the less important nodes in a manner that some of these residual nodes overshadow the secondary hubs. See, for example, how node 9, which should be residual, has a greater importance than nodes 5 and 16. The fact that nodes which are equal from the point of view of \eqref{PR} are different in the quantum PageRank can make us think that the quantum algorithm is sensitive not only to the nodes linking to a concrete node, but also to the nodes it points to.

\begin{figure}[htpb]
	\centering
	\includegraphics[scale=0.5]{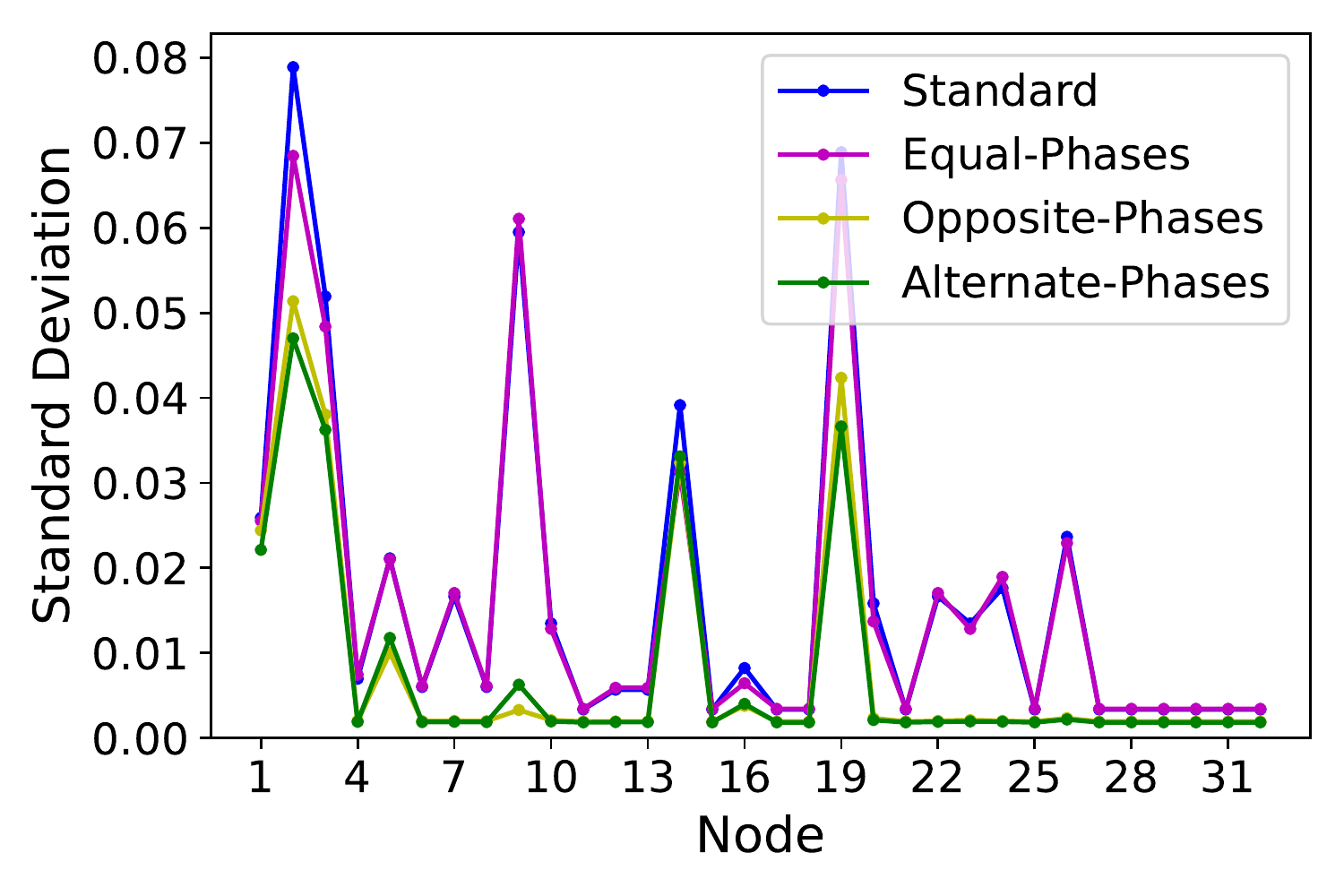}
	\caption{Standard deviations for the quantum PageRanks of a random scale-free graph with 32 nodes. $\theta = \pi/2$ has been used for the three APR schemes. The standard deviations decreases for the Opposite-Phases and Alternate-Phases schemes.}
	\label{F:Std_SF_32}
\end{figure}

\begin{figure*}
	\subfigure[]{\includegraphics[scale=0.5]{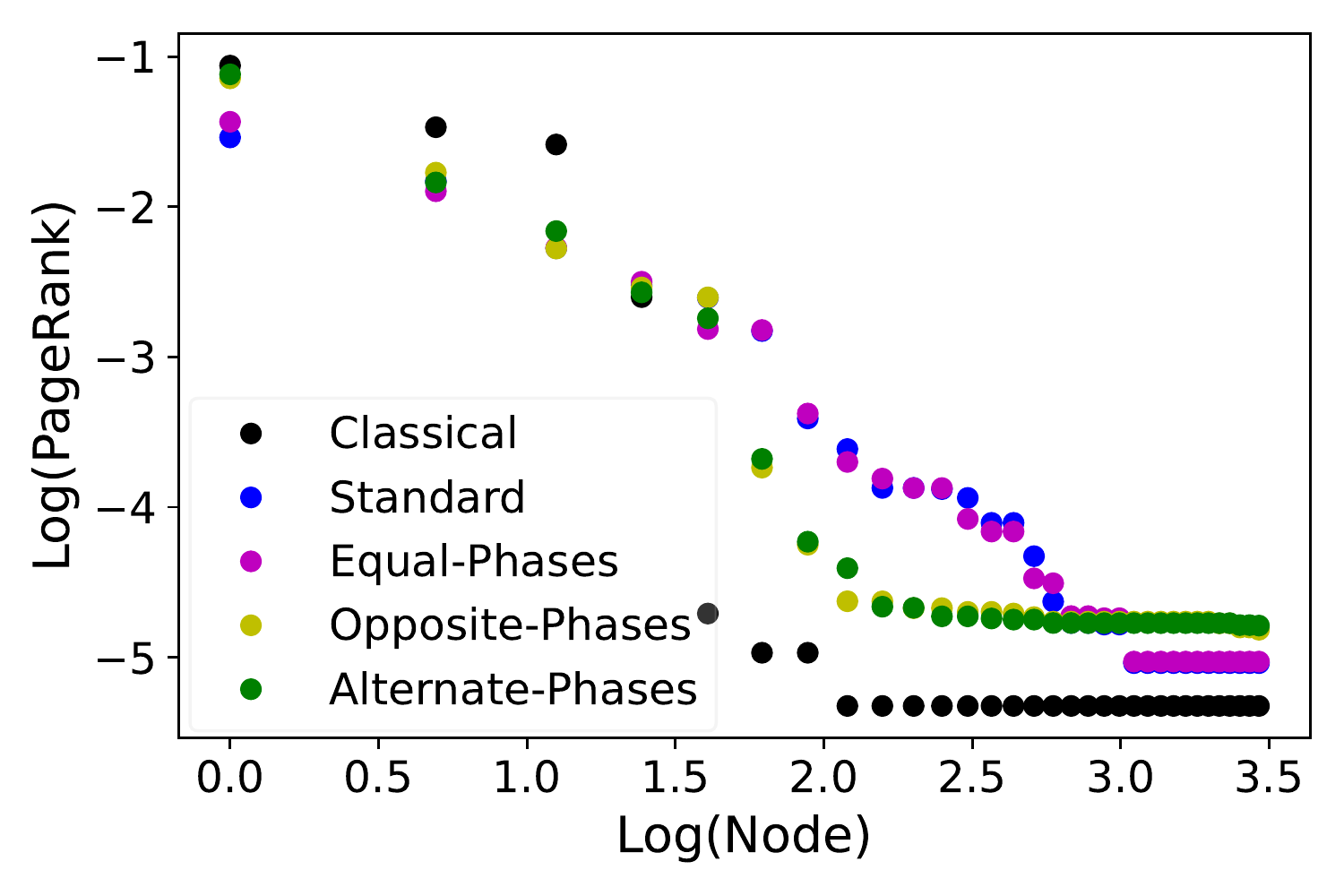}\label{F:PL_SF_32}}
	\subfigure[]{\includegraphics[scale=0.5]{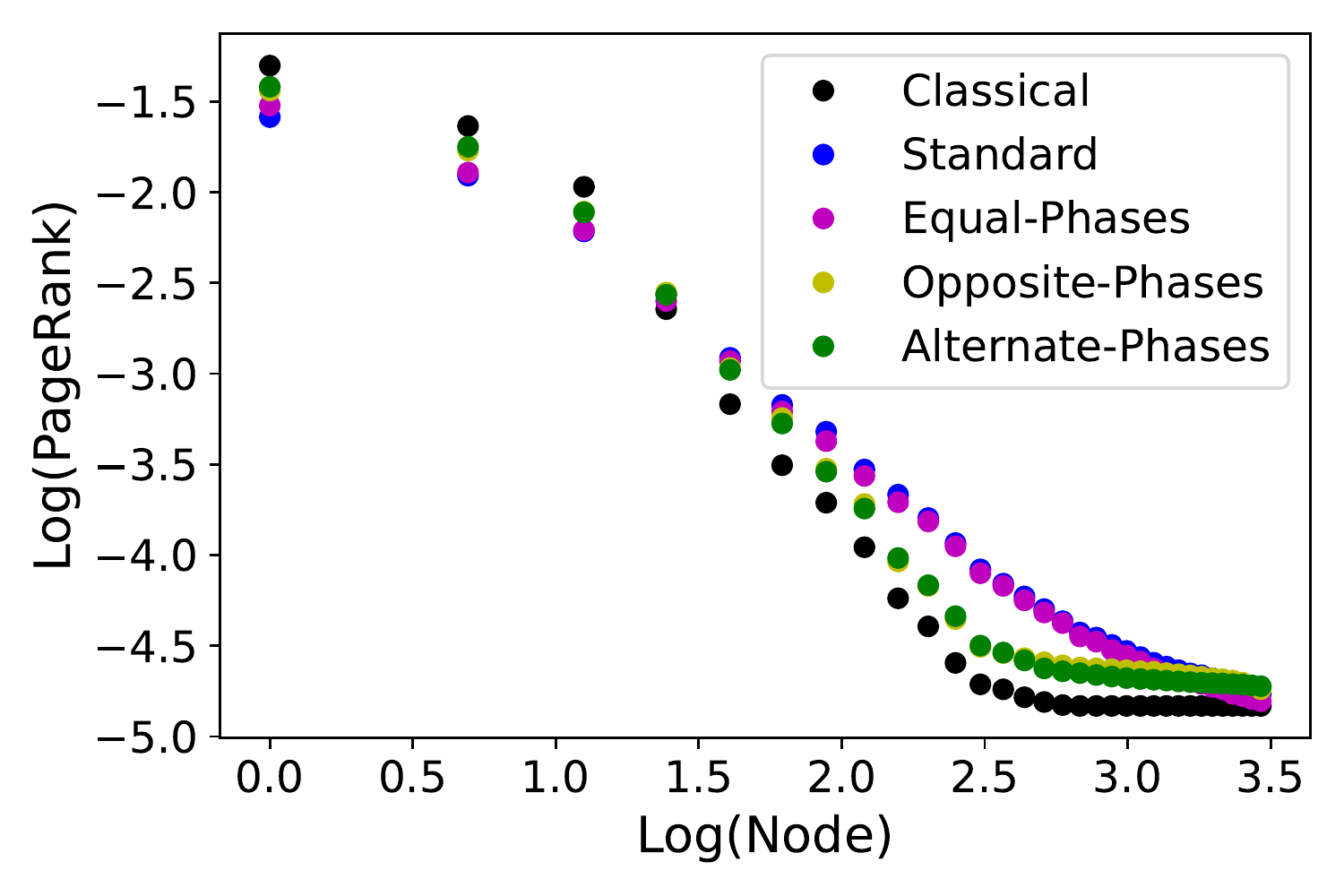}\label{F:PL_SF_32_av}}
	\caption{a) Logarithmic plot of the PageRanks vs the node index (after sorting) for a random scale-free graph with 32 nodes. b) Averaged  logarithmic plot of the PageRanks vs the node index (after sorting) for an ensemble of 50 random scale-free graphs with 32 nodes. The classical distribution is compared with all the quantum distributions, using $\theta = \pi/2$ in the three APR schemes. We see a partial restoration of the degeneracy of the less important nodes for the Opposite-Phases and Alternate-Phases schemes.}
	\label{...}
\end{figure*}

\begin{figure*}
	\subfigure[]{\includegraphics[scale=0.5]{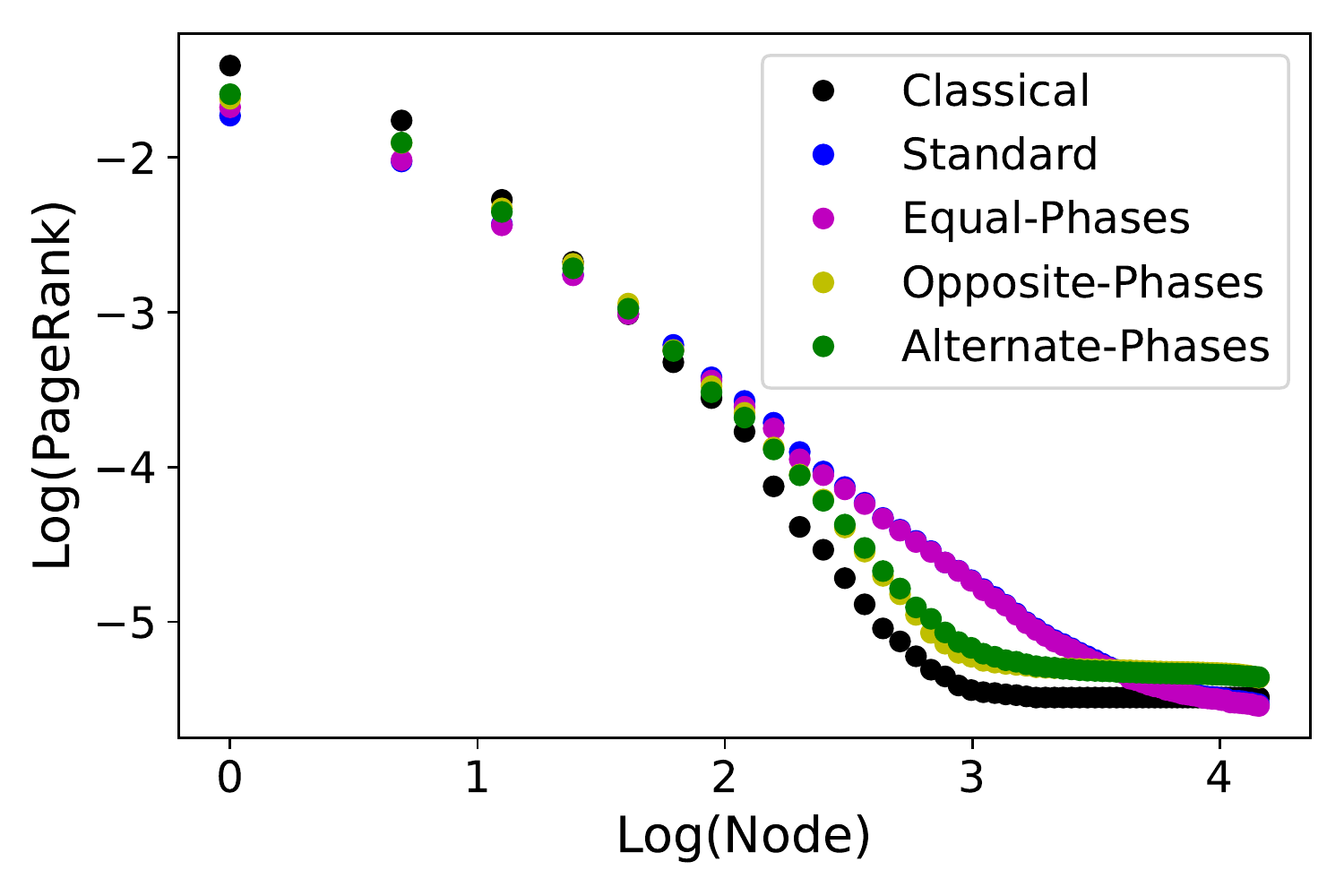}\label{F:PL_SF_64_av}}
	\subfigure[]{\includegraphics[scale=0.5]{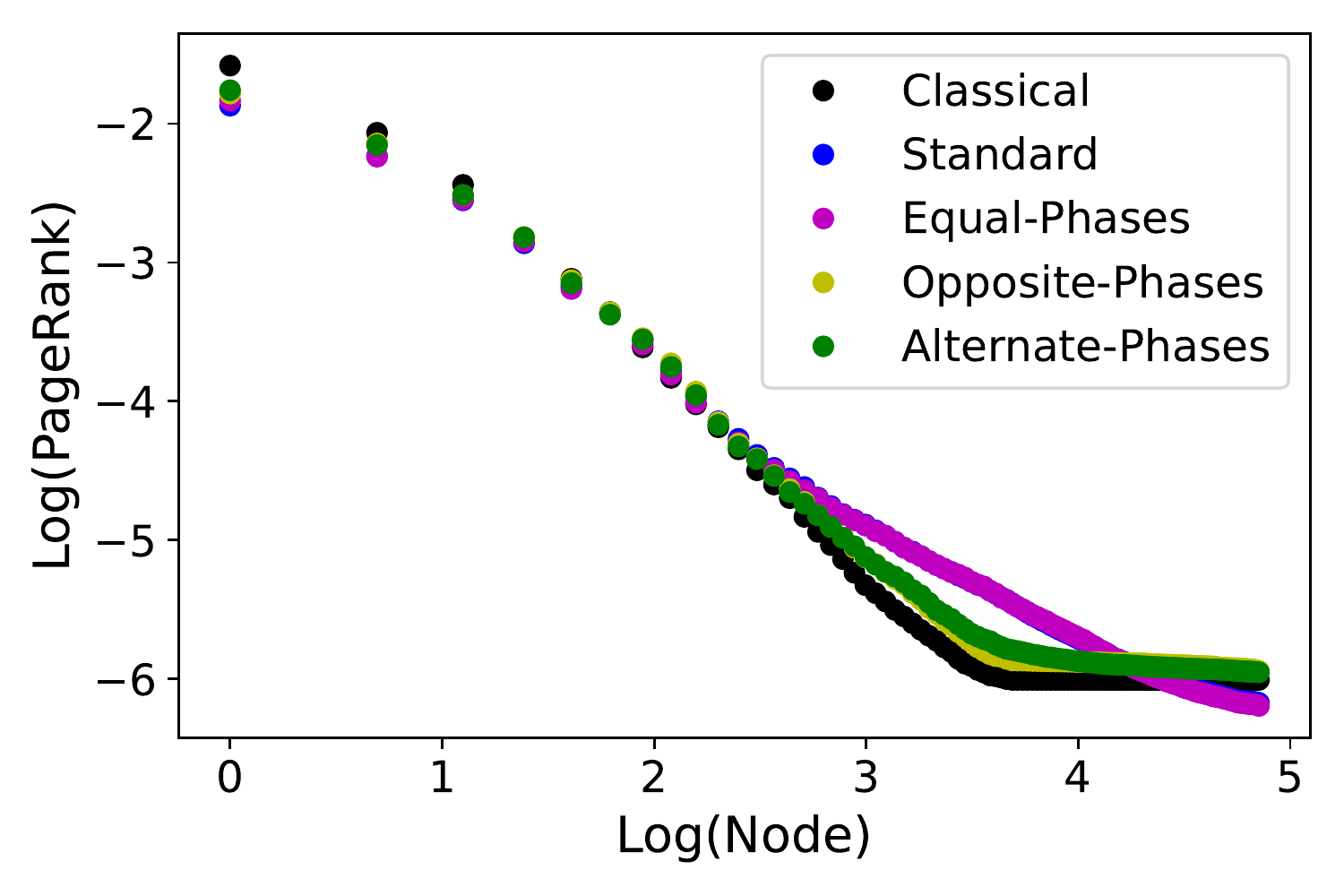}\label{F:PL_SF_128_av}}
	\caption{Averaged logarithmic plot of the PageRanks vs the node index (after sorting) for an ensemble of 50 random scale-free graphs with a) 64 nodes and b) 128 nodes. The classical distribution is compared with all the quantum distributions, using $\theta = \pi/2$ in the three APR schemes. We see that the behavior is independent on the size of the network.}
	\label{F:PL_64_128}
\end{figure*}

When we add the APR to the quantum algorithm, we do not find a significant difference in the Equal-Phases case. However, in the Opposite-Phases and Alternate-Phases algorithms the residuals nodes are again degenerate in majority. Note that node 9 has a slightly greater importance in the Alternate-Phases algorithm than the other residual nodes, but it is still less important than the truly secondary hubs. This means that in these APR schemes the quantum algorithms are practically only sensitive to the in-degree distribution of the nodes, as the classical one. Moreover, these two schemes maintain the quantum property of highlighting secondary hubs with respect to the classical algorithm, as can be seen in nodes 5, 16, and 19. This makes these algorithms a valuable tool for ranking nodes in a scale-free network, because they improve the classical deficiencies while solving the problematic quantum sensitivity to the out-degree distribution.

Note that in the small generic graph the APR schemes that more resembled the classical distribution were the Equal-Phases and Alternate-Phases cases. However, here the Equal-Phases scheme is more similar to the standard quantum case, and the Opposite-Phases scheme is more similar to the classical distribution. This suggests that the behavior of the APR schemes depends on the kind of graph.

As happened with the small graph, we have found that the standard deviation of the quantum PageRank can be decreased using certain APR schemes. In Figure \ref{F:Std_SF_32} the standard deviations for all the nodes are shown for the four quantum algorithms. While the Equal-Phases scheme seems to have standard deviations similar to the standard case, the Opposite-Phases and Alternate-Phases schemes show a clear improvement, decreasing the standard deviations. Recall that these last two schemes are those that have a partial restoration of the degeneration. This highlights the valuable importance of the Opposite-Phases and Alternate-Phases schemes as APR alternatives to the standard quantum algorithm.

\subsection{Power law distribution of the PageRanks}

\begin{figure*}
	\subfigure[]{\includegraphics[scale=0.375]{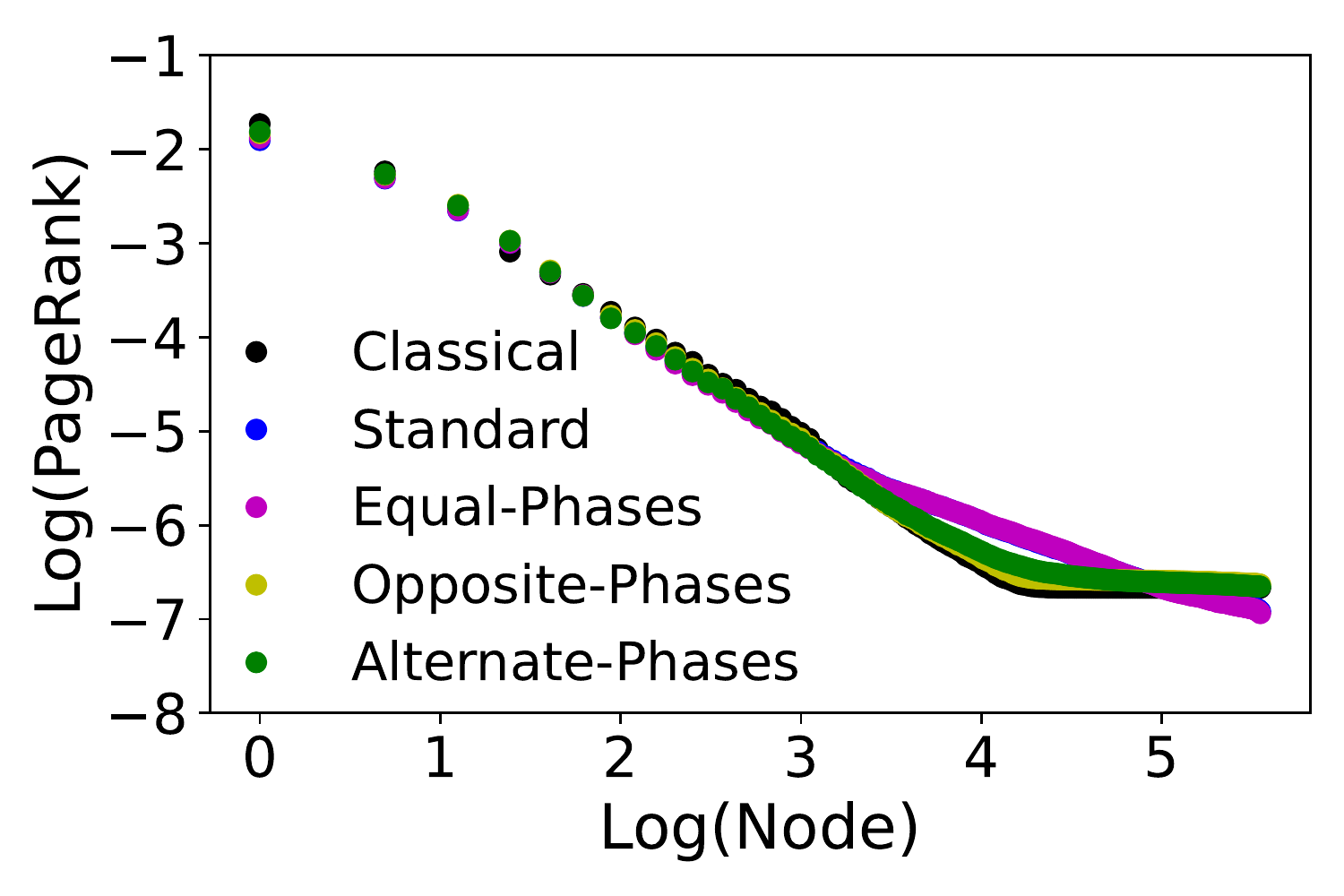}}
	\subfigure[]{\includegraphics[scale=0.375]{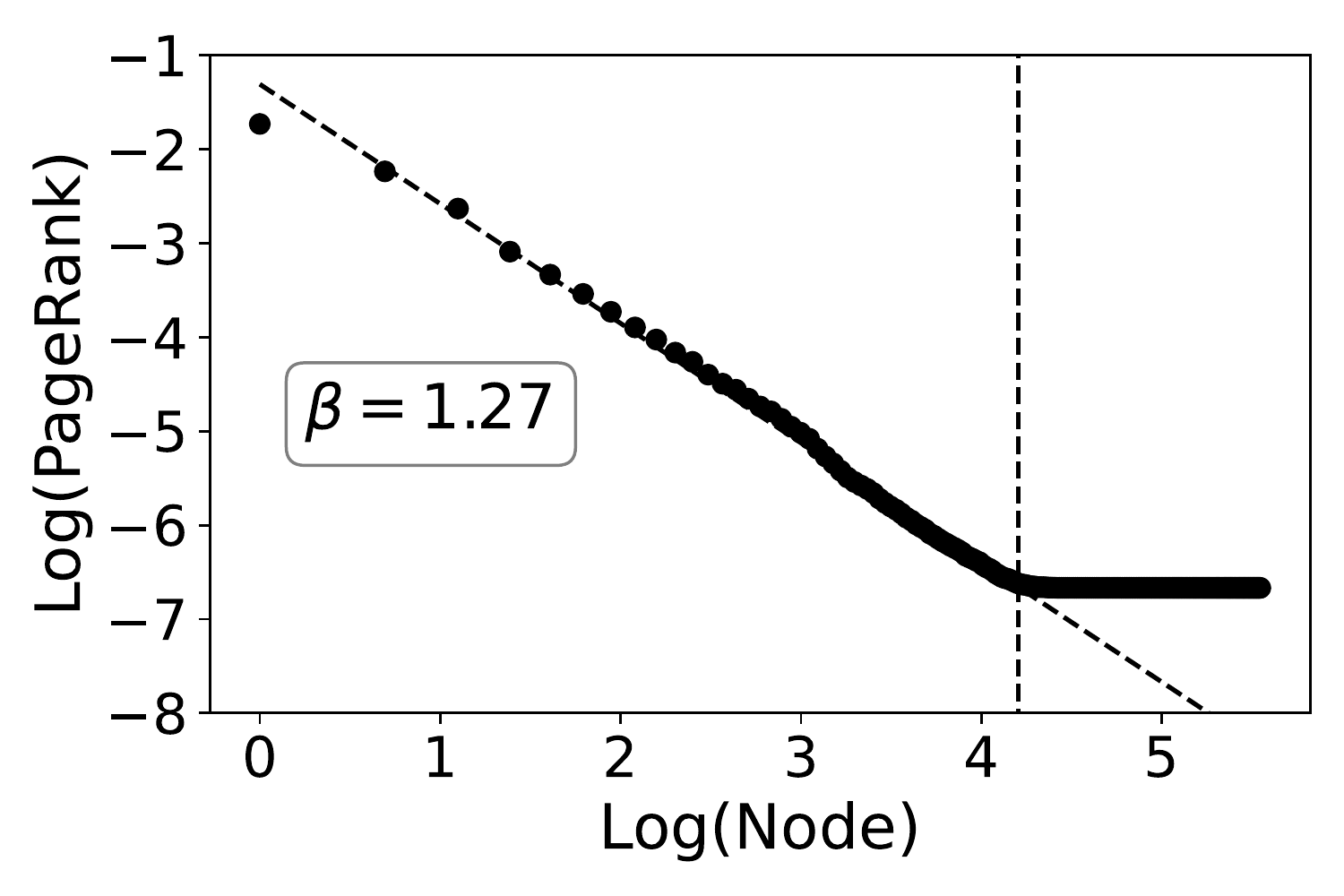}}
	\subfigure[]{\includegraphics[scale=0.375]{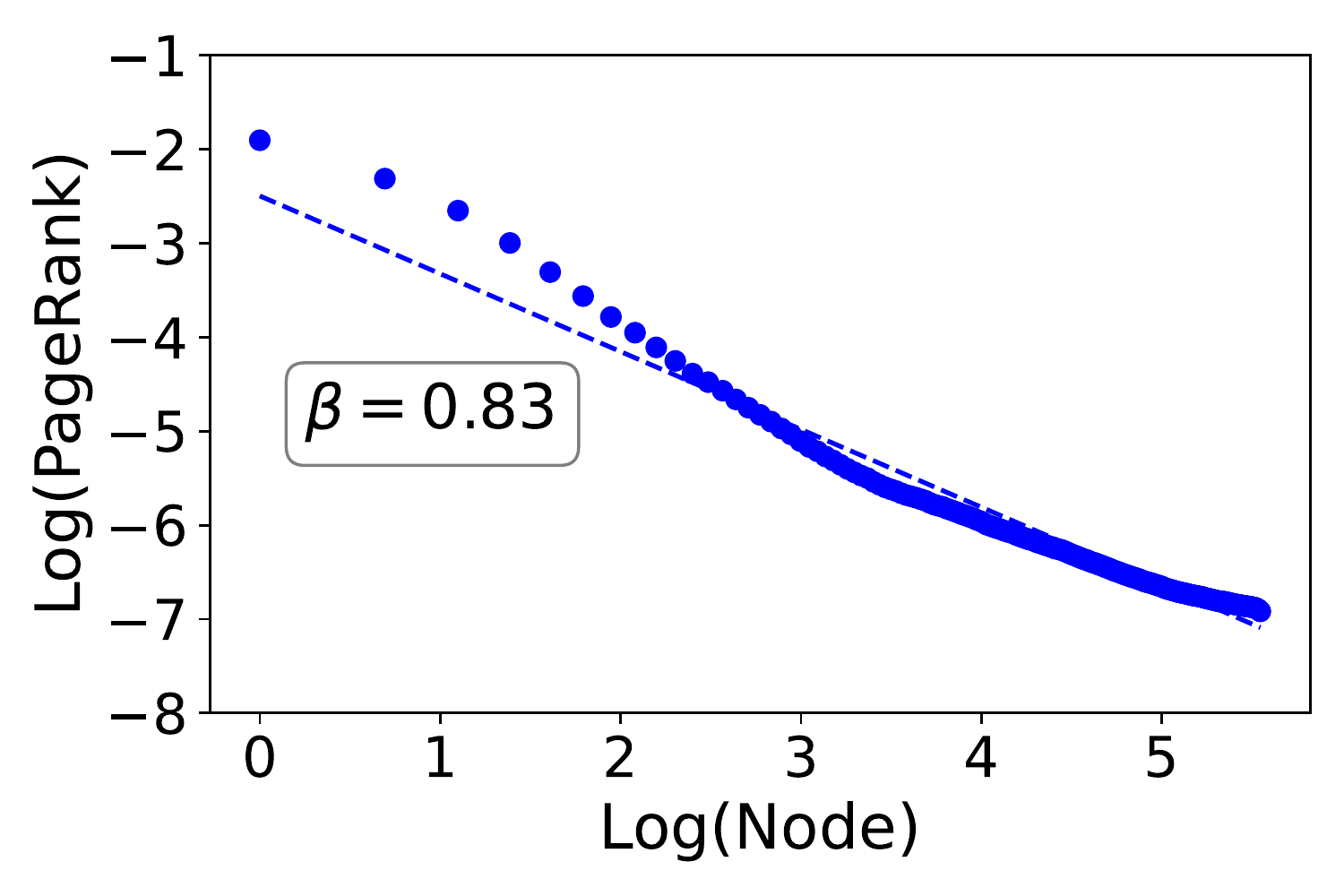}}
	\subfigure[]{\includegraphics[scale=0.375]{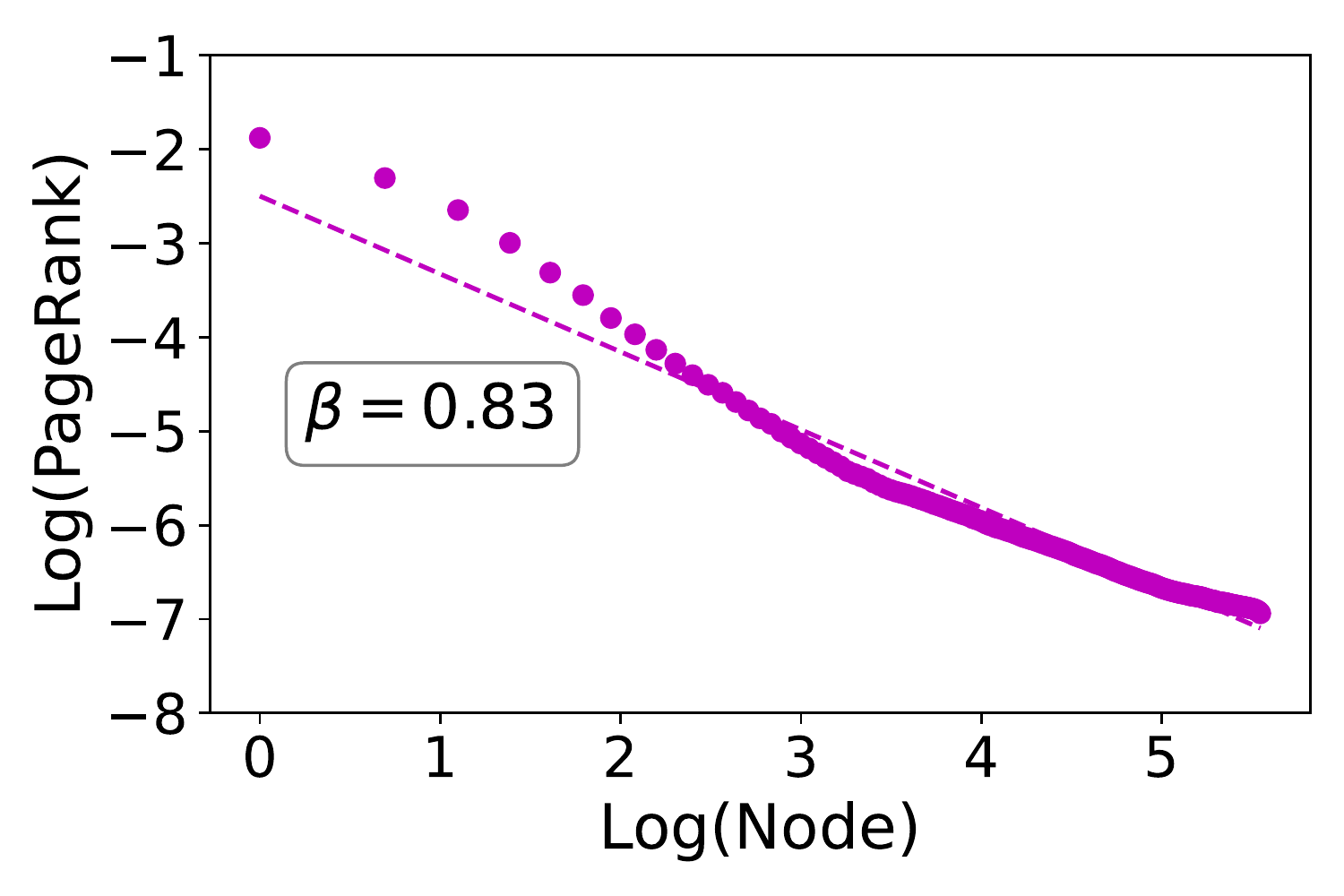}}
	\subfigure[]{\includegraphics[scale=0.375]{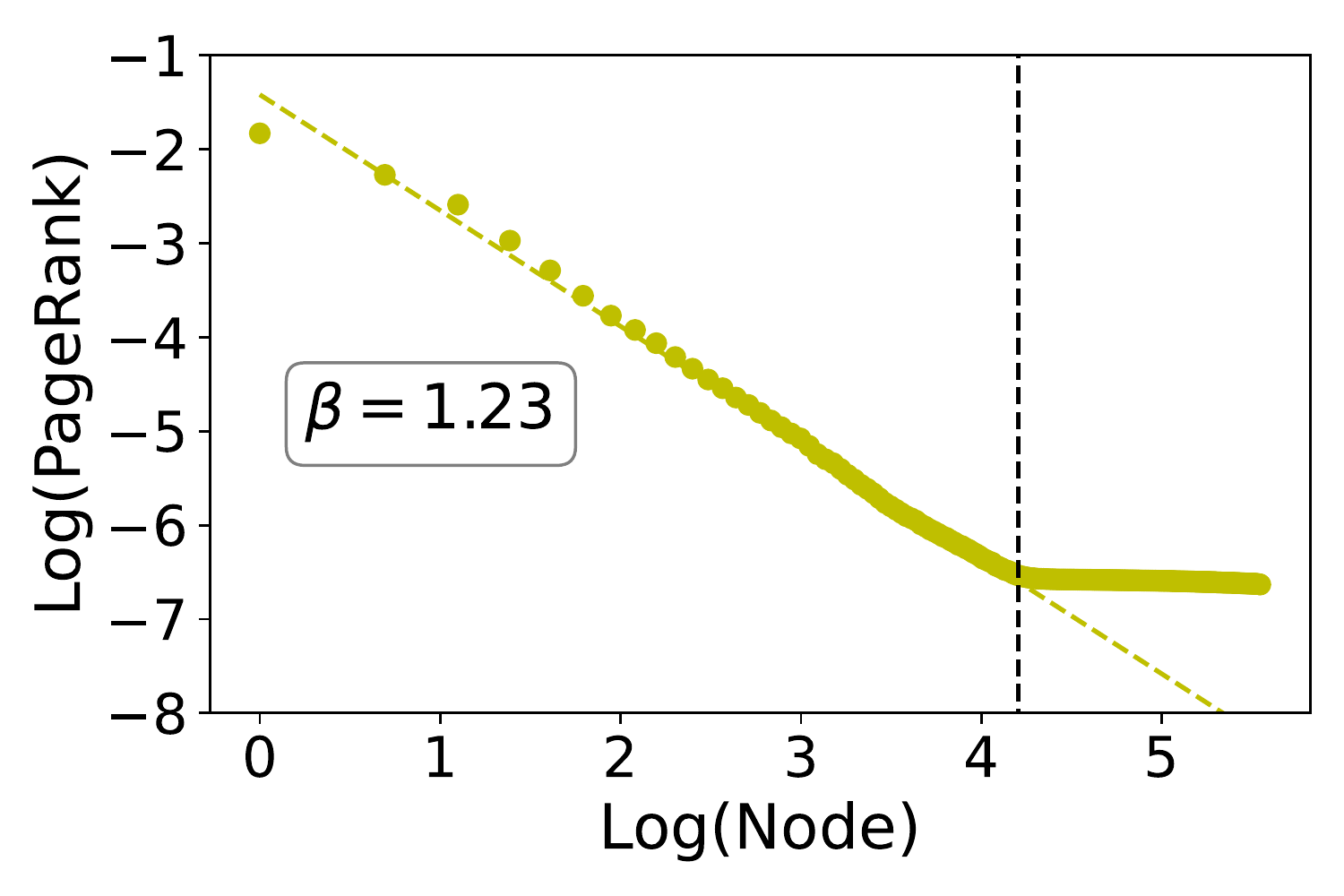}}
	\subfigure[]{\includegraphics[scale=0.375]{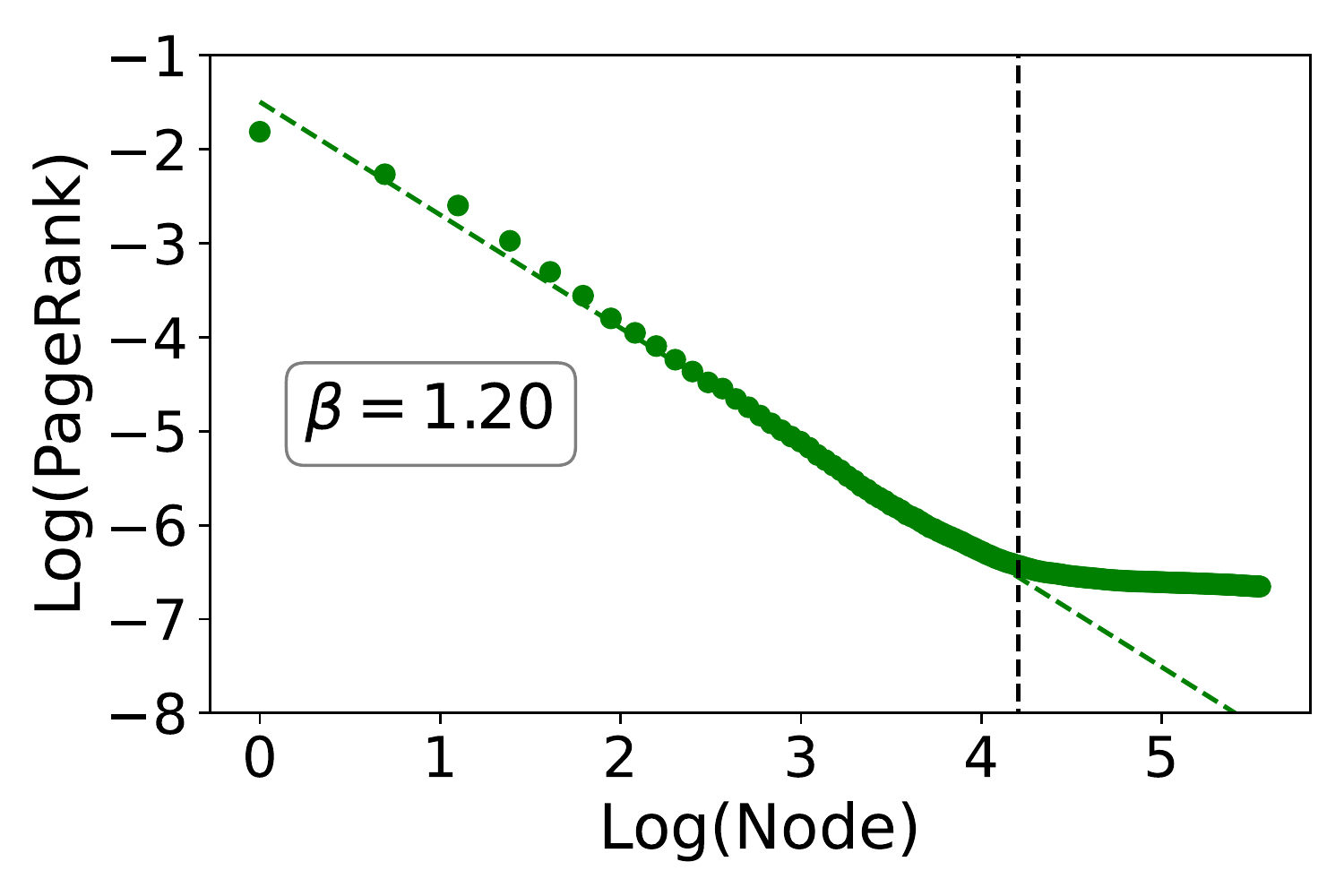}}
	\caption{a) Averaged logarithmic plot of the PageRanks vs the node index (after sorting) for an ensemble of 50 random scale-free graphs with 256 nodes. b)-f) Linear fitting to equation \eqref{LPL} for b) the classical algorithm, c) the standard quantum algorithm, d) the Equal-Phases algorithm, e) the Opposite-Phases algorithm, and f) the Alternate-Phases algorithm. In b), e), and f), only the nondegenerate region has been take into account. $\theta = \pi/2$ has been used in the three APR schemes. We see that the standard quantum algorithm and the Equal-Phases algorithm follow a smoother power law.}
	\label{F:LF}
\end{figure*}

Since scale-free networks follow a power law distribution in the connectivity of the nodes, they also have a similar behavior in the PageRanks distribution \cite{Classical-PL,Paparo2}. Then, the PageRank can be expressed as:
\begin{equation}\label{PL}
	I \sim i^{-\beta},
\end{equation}
where $i$ is the index of the node after sorting them by importance, and $\beta$ is a constant coefficient. Taking logarithms to both sides of \eqref{PL}, we obtain
\begin{equation}\label{LPL}
	\log{I} \sim -\beta\log{i}.
\end{equation}
Then, plotting the sorted nodes in a logarithmic way, we expect to see a linear behavior. This plot is shown in Figure \ref{F:PL_SF_32} for the graph with 32 nodes used previously. We can see that the standard and Equal-Phases quantum algorithms show a smoother behavior due to the degeneracy breaking of the less important nodes. The classical, Opposite-Phases and Alternate-Phases algorithms have a big degeneration in the less important nodes, so the decay in PageRank before the degenerate region is more abrupt. Moreover, these last two APR schemes were able to highlight truly secondary hubs, and the less important nodes have a higher PageRank than in the classical distribution, so the distribution is also smoother with respect to the classical one.

To ensure that this behavior is not particular for this concrete graph, but for the majority of the scale-free graphs with 32 nodes, we have averaged the sorted PageRanks from an ensemble of 50 random scale-free graphs. The averaged results are shown in Figure \ref{F:PL_SF_32_av}. This confirms that the discussion above is valid for the scale-free graphs class with 32 nodes, rather than for a concrete graph. Furthermore, we claim that the above discussion holds for scale-free graphs with a higher number of nodes. With this purpose, we show the averaged logarithmic plot of the sorted nodes for scale-free graphs with 64 and 128 nodes in Figures \ref{F:PL_SF_64_av} and \ref{F:PL_SF_128_av}, respectively, obtaining similar results as with 32 nodes.

\begin{figure*}
	\subfigure[]{\includegraphics[scale=0.5]{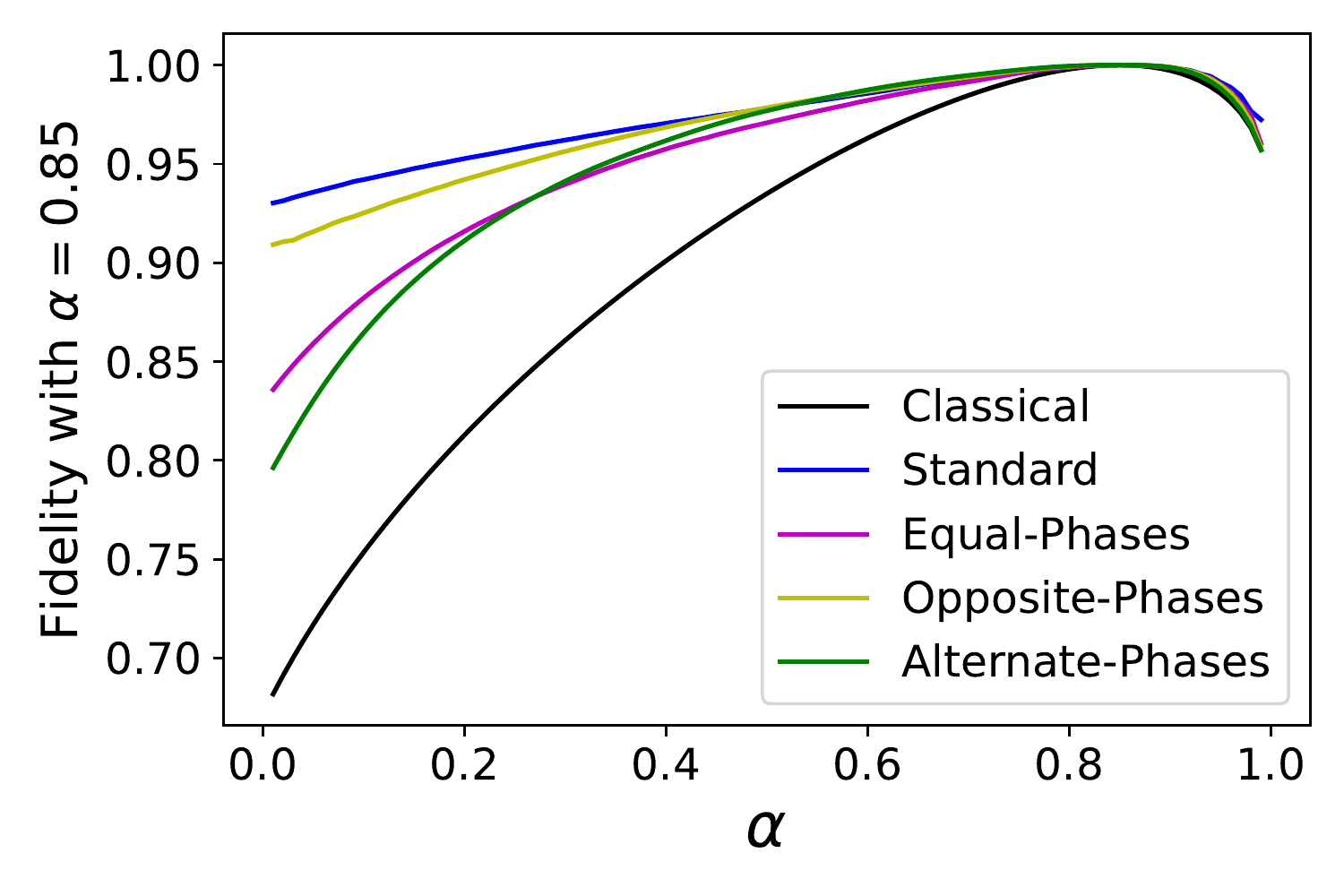}\label{F:Stability_32}}
	\subfigure[]{\includegraphics[scale=0.5]{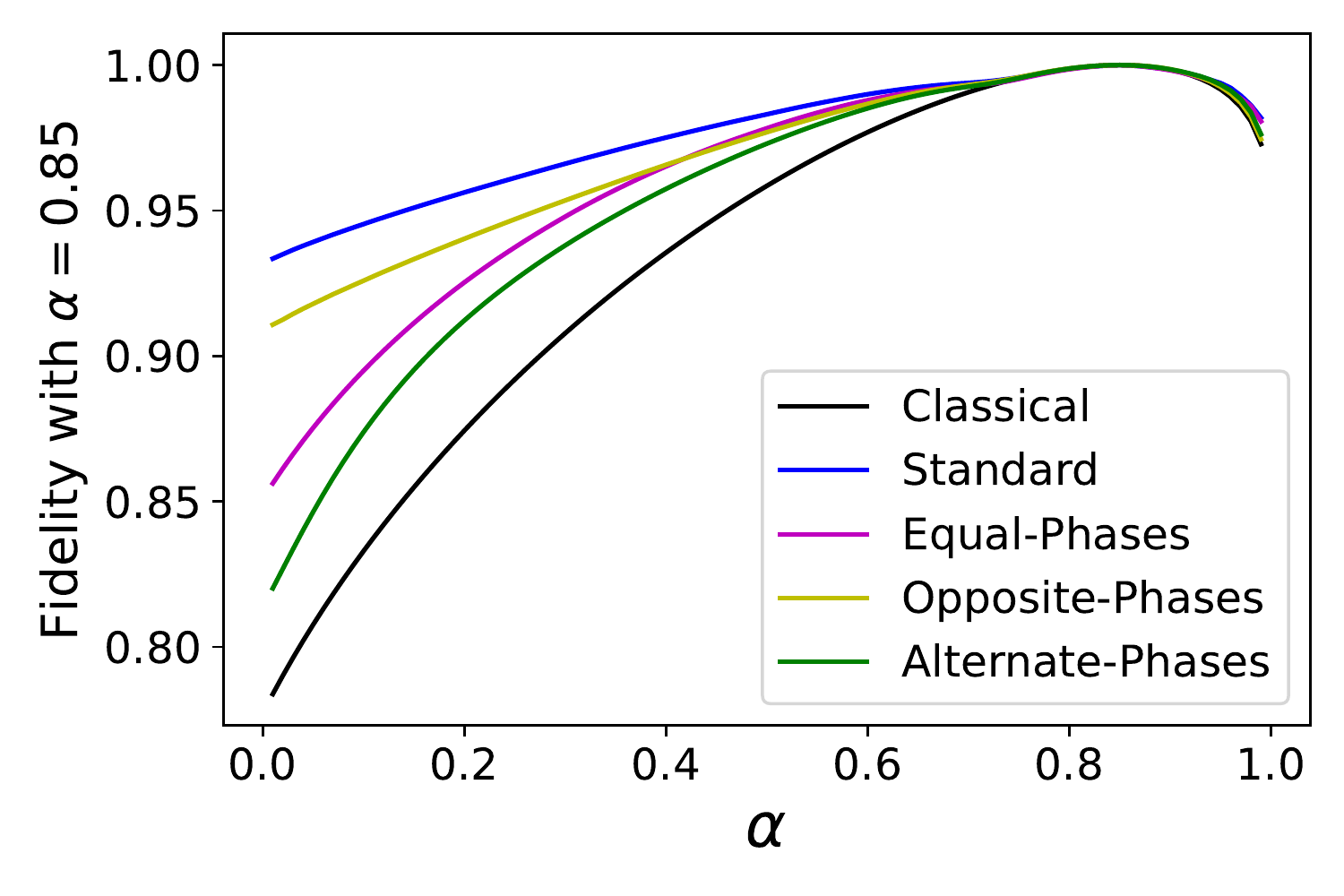}\label{F:Stability_32_av}}
	\caption{a) Fidelity of the PageRank distributions vs the damping parameter $\alpha$, with respect to the distribution with $\alpha = 0.85$, for a random scale-free graph with 32 nodes. b) Averaged fidelity of the PageRank distributions vs the damping parameter $\alpha$, with respect to the distribution with $\alpha = 0.85$, for an ensemble of 50 random scale-free graphs with 32 nodes. The classical distribution is compared with all the quantum distributions, using $\theta = \pi/2$ in the three APR schemes. We see that all the quantum algorithms are more stable than the classical one, with the standard and the Opposite-Phases algorithms being the most stable.}
	\label{...}
\end{figure*}

The coefficient $\beta$ in \eqref{LPL} is a measurement of the smoothness of the power law distribution. We can obtain the coefficient $\beta$ for each algorithm after a linear fitting of the logarithmic plots. In Figure \ref{F:LF} we show the power law distribution for an ensemble of 50 graphs with 256 nodes, as well as the linear fitting for each algorithm. In the classical algorithm and in the Opposite-Phases and Alternate-Phases schemes the fitting has been made only with the nondegenerate nodes, since the degenerate region has a constant distribution. The separation between both regions is shown with a vertical line. We can see that the standard quantum algorithm as well as the Equal-Phases case have a smoother behavior with respect to the other algorithms, which is characterized by a smaller value of $\beta$. Moreover, the power law distribution extends to the less important nodes since they are not degenerate. This can be related with the previous discussion that these algorithms are more sensitive to both the in- and out-degree distributions of the nodes, thus better capturing the power law of the vertex connectivity. On other hand, as expected, the Opposite-Phases and Alternate-Phases distributions are smoother than the classical one, having a slightly smaller value of $\beta$. Finally, we have also done the linear fitting for ensembles of 50 graphs with 32, 64, and 128 nodes. The values of $\beta$ for each case are summarized in Table \ref{T:betas}. In the four cases we see a similar qualitative behavior of $\beta$ between the different algorithms, although the absolute values can change slightly with the number of nodes.

\begin{table}[htbp]
	\centering
	\caption{Values of the coefficient $\beta$ in the power-law distribution of the PageRank for all the algorithms using ensembles of scale-free graphs with 32, 64, 128 and 256 nodes. $\theta = \pi/2$ has been used in the three APR schemes.}
	\begin{tabular}{c|c|c|c|c}
		\hline
		Algorithm & $N = 32$    & $N = 64$    & $N = 128$   & $N = 256$\\
		\hline
		Classical & \ $1.53$ \ & \ $1.55$ \  & \ $1.38$ \  & \ $1.27$ \ \\
		Standard & $1.04 $ & $1.02 $ & $0.90 $ & $0.83$ \\
		Equal-Phases & $1.06 $ & $1.02 $ & $0.90 $ & $0.83$ \\
		Opposite-Phases & $1.34 $ & $1.38 $ & $1.28 $ & $1.23$ \\
		Alternate-Phases & $1.35 $ & $1.35 $ & $1.25 $ & $1.20$\\
		\hline
	\end{tabular}%
	\label{T:betas}%
\end{table}%

\section{Stability of Generalized Quantum PageRanks}\label{Stability}

We have fixed the damping parameter $\alpha$ in \eqref{G} to $\alpha = 0.85$ since that is the value that showed an optimal performance in the classical algorithm. It is known that the classical algorithm is very sensitive to the value of this parameter \cite{Georgeot}, and the authors of \cite{Paparo2} found that the standard quantum PageRank algorithm is more stable for scale-free networks. The aim of this section is to study how the introduction of the APR schemes affects the stability of the quantum PageRank.

For this purpose, we start with the random scale-free graph with 32 nodes of Figure \ref{F:PR_SF_32_graph}. We can measure the similarity between two distributions obtained with different damping parameter $\alpha$ using the fidelity defined in \eqref{fidelity}. We analyze the behavior of this fidelity between the distribution obtained for a value of $\alpha \in [0.1,0.99]$ and the usual distribution with $\alpha = 0.85$. These fidelities are represented in Figure \ref{F:Stability_32} for the classical and all the quantum algorithms, with $\theta = \pi/2$ in the APR schemes. We observe that the fidelity for the classical algorithm decreases very quickly with the parameter $\alpha$, reaching a value under $0.70$. However, as it was also shown in \cite{Paparo2}, the standard quantum algorithm is by far more stable, with a minimum fidelity of approximately $0.93$.

Regarding the quantum algorithms with APR, we find that they are more stable than the classical too. On one hand, both the Equal-Phases and Alternate-Phases cases show a similar behavior (the former slightly better), which is intermediate between the standard quantum and the classical algorithms. On the other hand, the Opposite-Phases algorithm seems to be approximately as stable as the standard quantum algorithm. We find this result surprising, since the Equal-Phases algorithm is the one that shows a distribution of PageRanks similar to the standard quantum case, whereas the Opposite-Phases algorithm restores the degeneracy of the nodes in a pattern similar to the Alternate-Phases case. Then, the Opposite-Phases algorithm seems to be very promising for scale-free graphs, since it resembles the classical distribution of PageRanks highlighting truly secondary hubs, and maintains the quantum stability. To ensure that this intriguing behavior is not particular for this concrete network, but is a property of the scale-free networks class, we have averaged the stability results for 50 random scale-free graphs with 32 nodes in Figure \ref{F:Stability_32_av}, finding results that fit in the discussion above.

\begin{figure*}
	\subfigure[]{\includegraphics[scale=0.5]{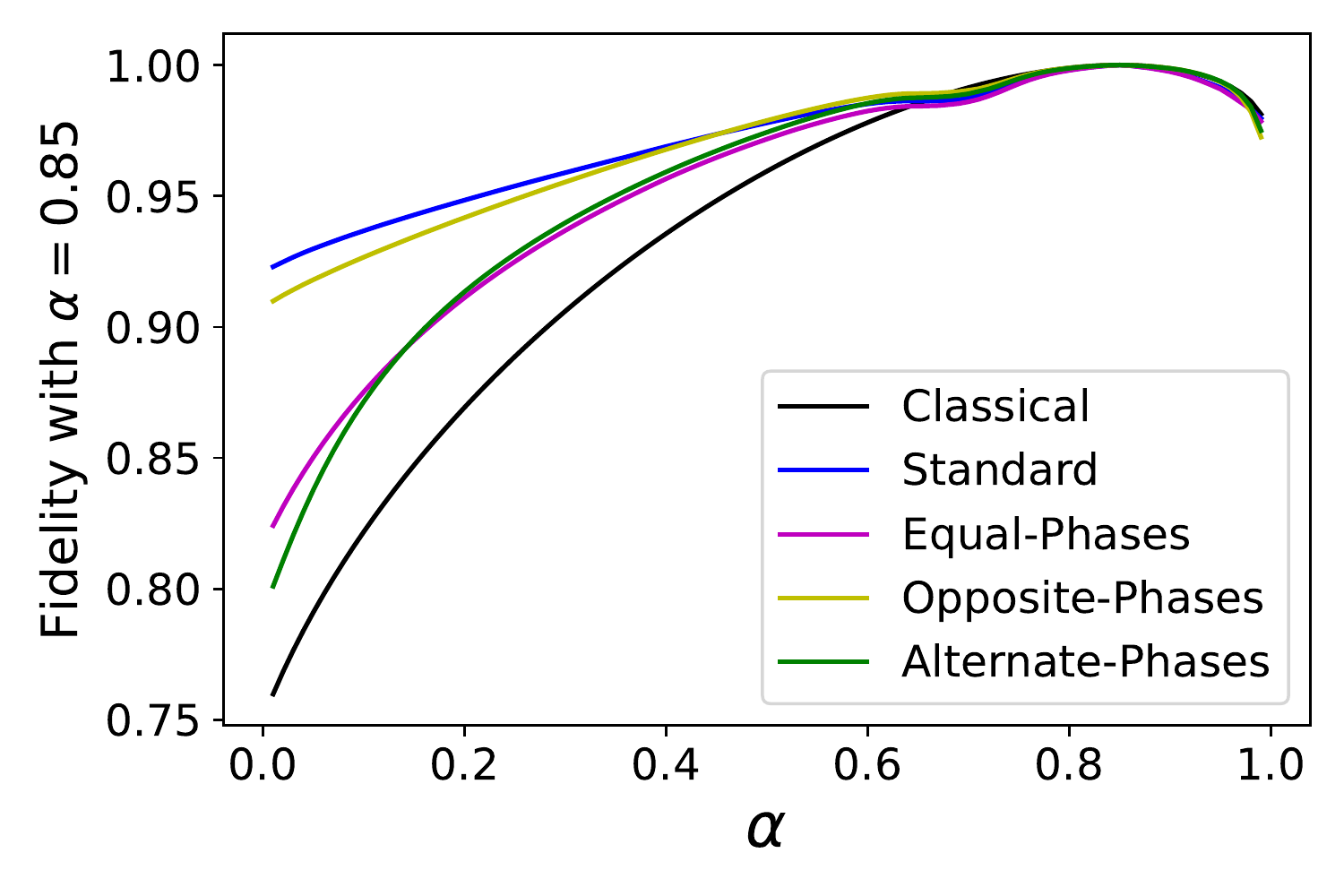}\label{F:Stability_64_av}}
	\subfigure[]{\includegraphics[scale=0.5]{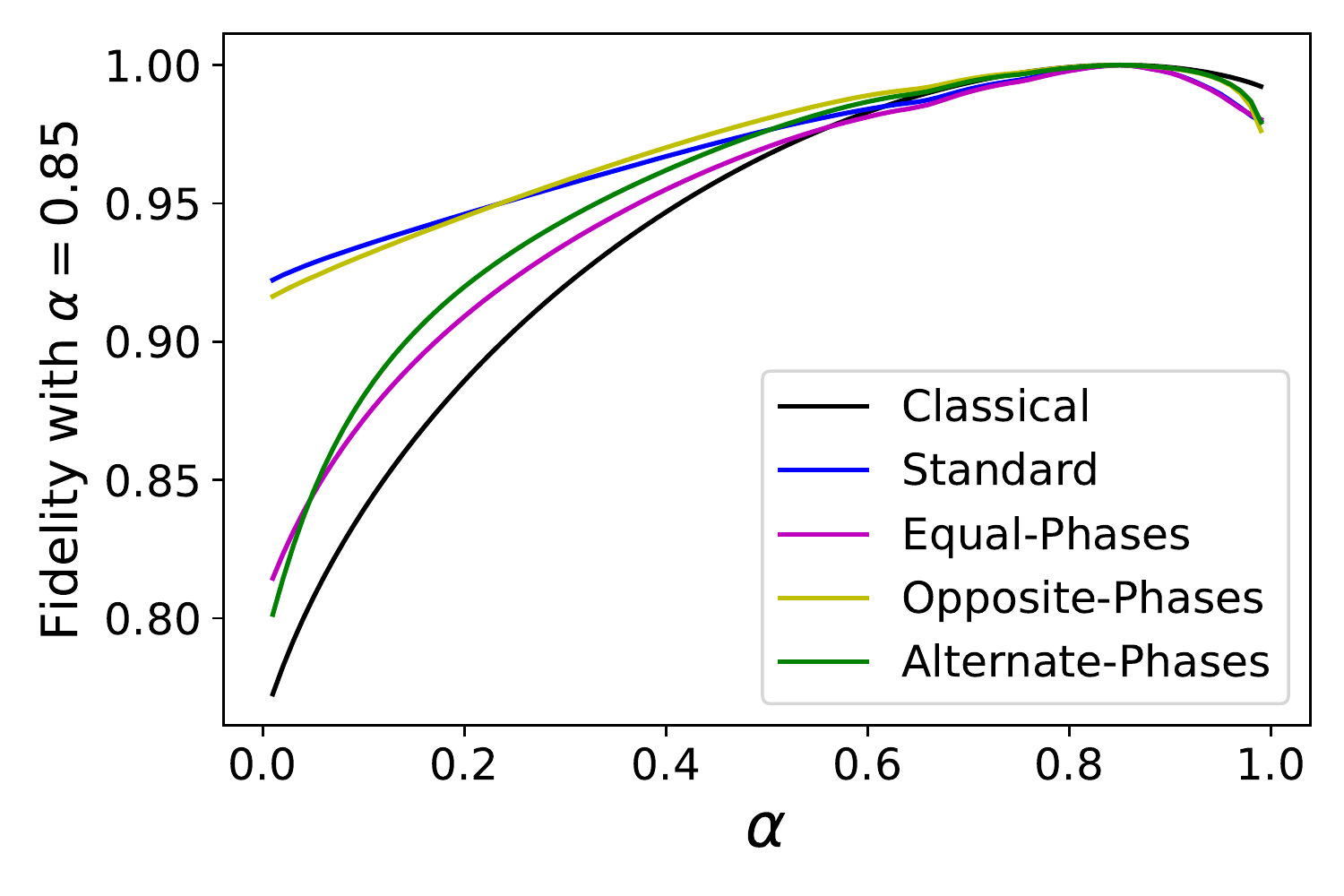}\label{Stability_128_av}}
	\caption{Averaged fidelity of the PageRank distributions vs the damping parameter $\alpha$, with respect to the distribution with $\alpha = 0.85$, for an ensemble of 50 random scale-free graphs with a) 64 nodes and b) 128 nodes. The classical distribution is compared with all the quantum distributions, using $\theta = \pi/2$ in the three APR schemes. We see that the main behavior is independent of the network size, albeit the Opposite-Phases scheme can outperform the standard quantum algorithm for 128 nodes in a certain range of $\alpha$.}
	\label{...}
\end{figure*}

To ensure that this behavior holds for bigger scale-free networks, we have averaged the results for 50 scale-free graphs with $64$ nodes in Figure \ref{F:Stability_64_av} and with $128$ nodes in Figure \ref{Stability_128_av}. We effectively find a similar behavior than before. It is worth noting that for the networks with $128$ nodes there is a region where the Opposite-Phases algorithm outperforms the standard quantum algorithm. Indeed, we have found for a lot of graphs in this class that the curve with the Opposite-Phases scheme is slightly above the curve of the standard quantum case for all values of $\alpha$.

\begin{figure*}
	\subfigure[]{\includegraphics[scale=0.375]{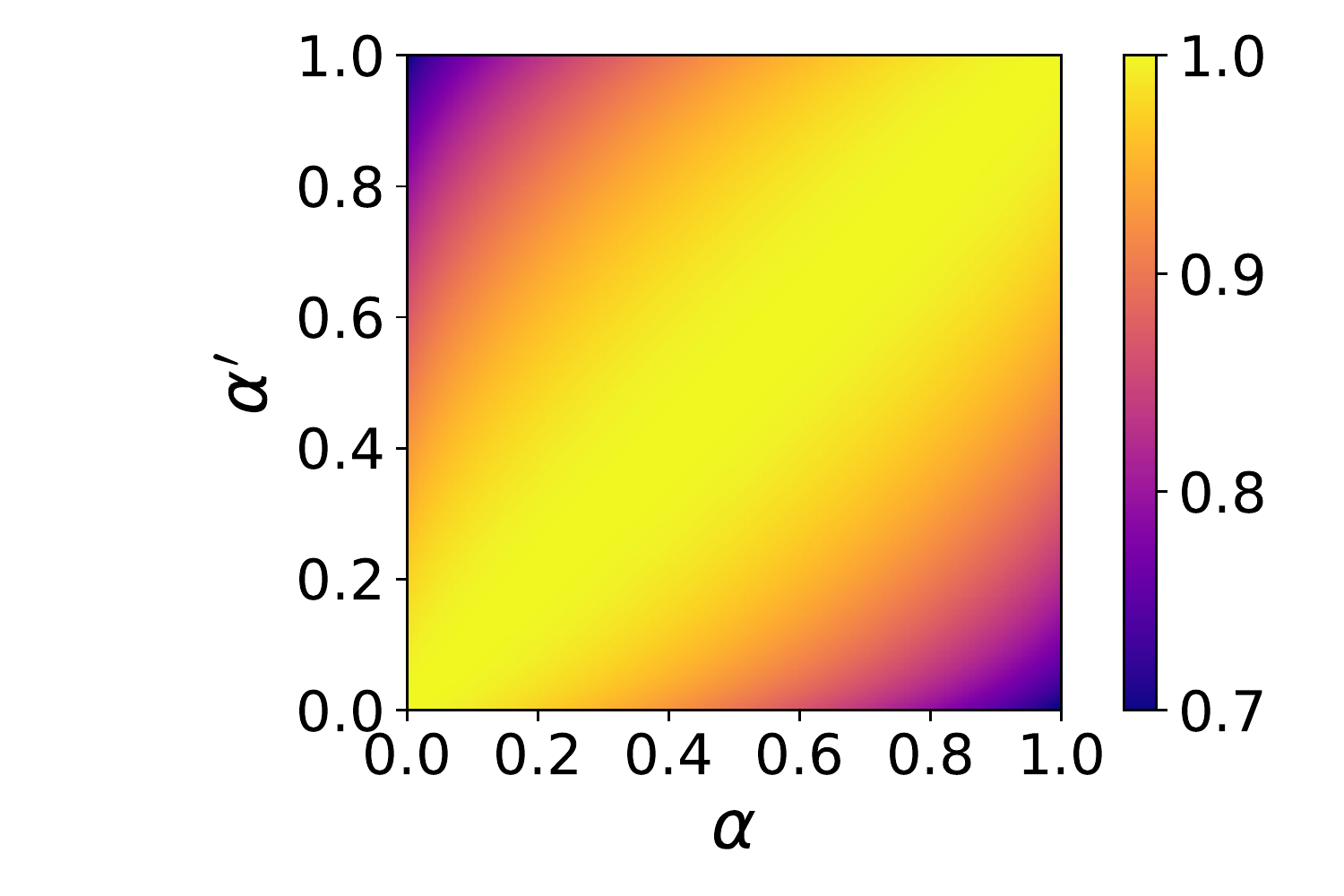}}
	\subfigure[]{\includegraphics[scale=0.375]{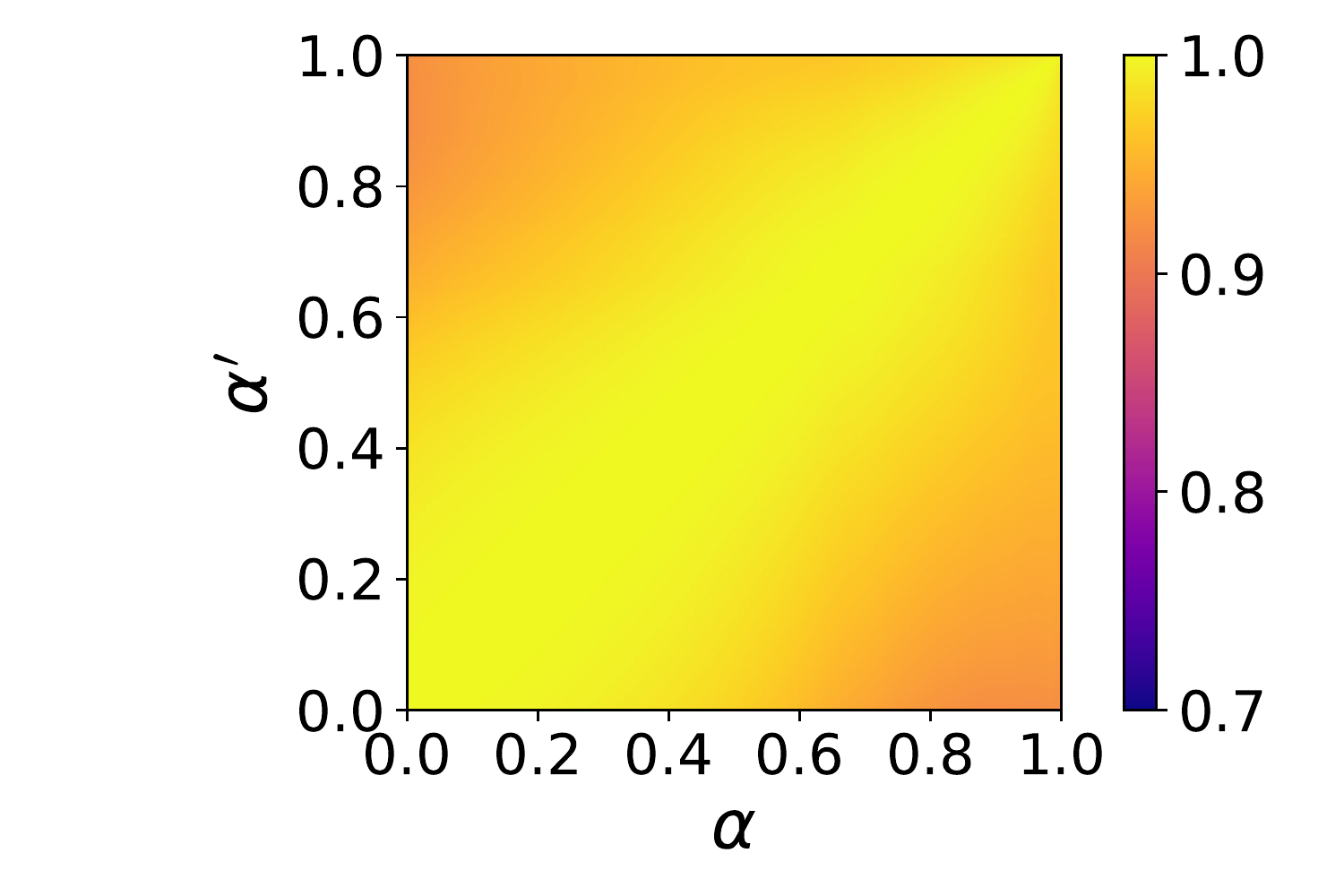}}
	\subfigure[]{\includegraphics[scale=0.375]{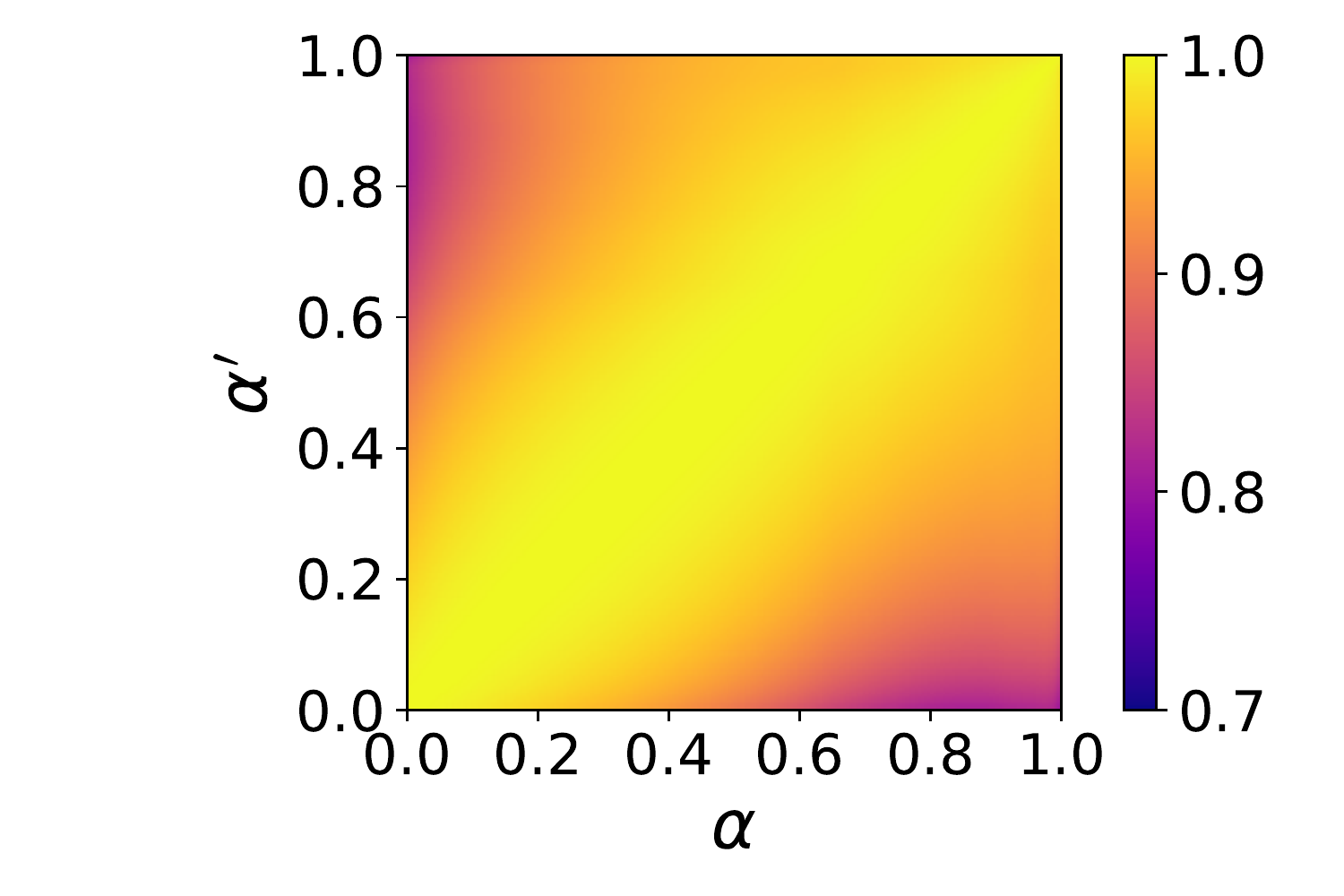}}
	\subfigure[]{\includegraphics[scale=0.375]{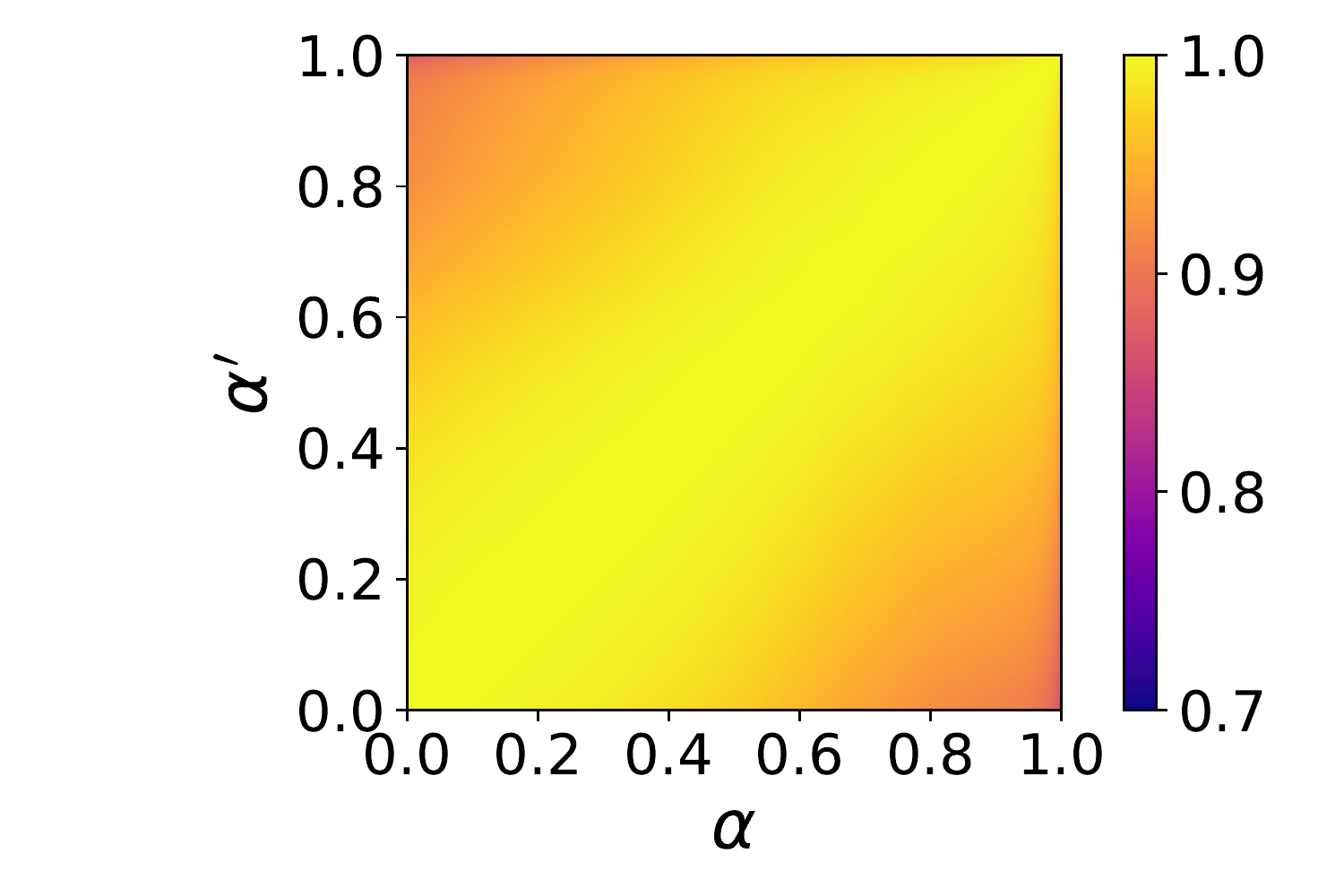}}
	\subfigure[]{\includegraphics[scale=0.375]{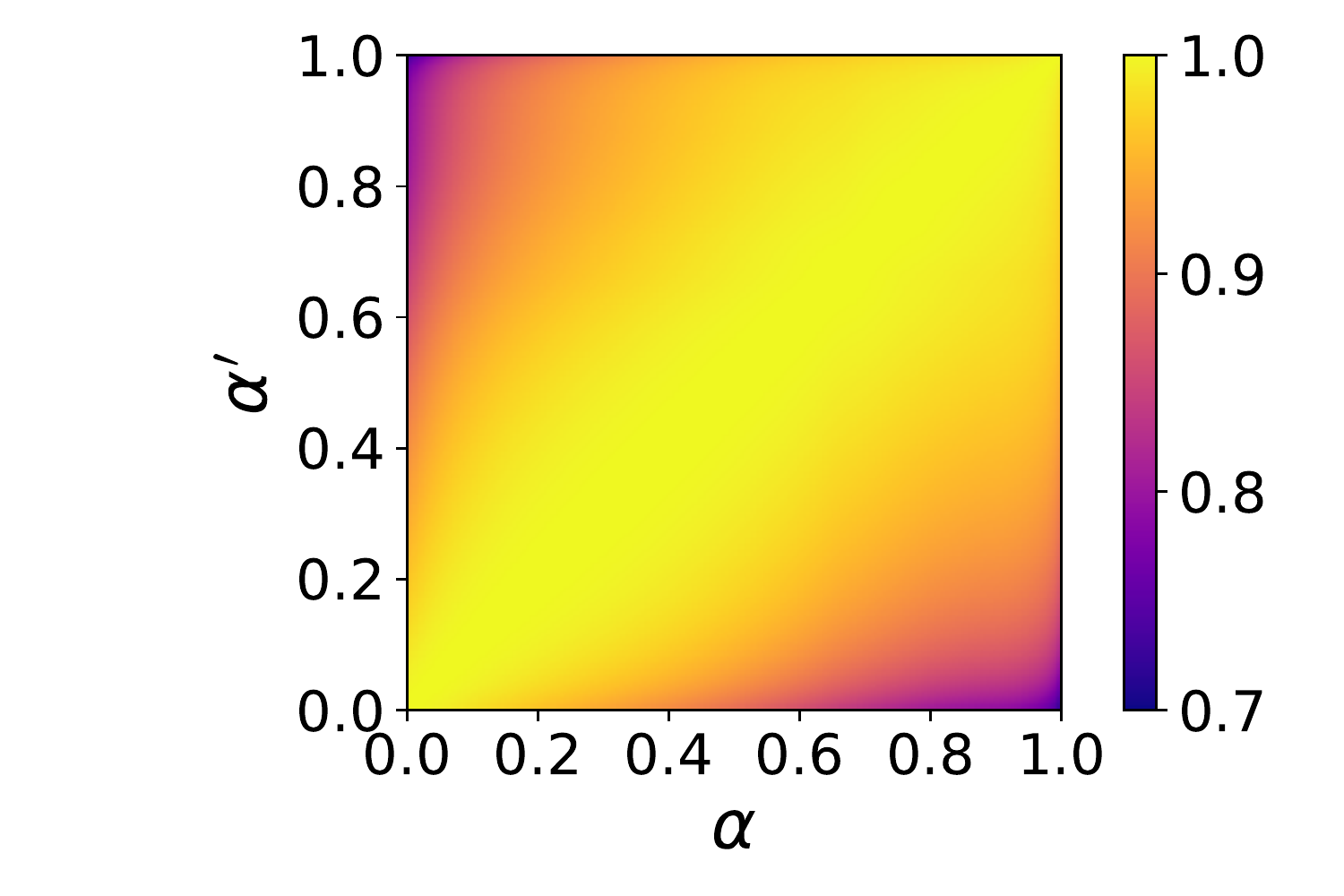}}
	\caption{Averaged fidelity of the PageRank distributions for all pairs $(\alpha,\alpha')$, for an ensemble of 50 random scale-free graphs with 128 nodes, using a) the classical algorithm, b) the standard quantum algorithm, c) the Equal-Phases algorithm, d) the Opposite-Phases algorithm, and e) the Alternate-Phases algorithm. $\theta = \pi/2$ has been used in the three APR schemes. We see that all the quantum algorithms are more stable than the classical one, with the standard and the Opposite-Phases algorithms being the most stable.}
	\label{F:cmap}
\end{figure*}

Finally, we shall see what happens with the fidelity not only for $\alpha = 0.85$, but for any pair $(\alpha,\alpha')$, with $\alpha, \alpha' \in [0.1,0.99]$. We show the averaged results for the ensemble of graphs with 128 nodes in Figure \ref{F:cmap} using heatmaps. The standard quantum algorithm shows a good stability region that extends to all values of $\alpha$, having a minimal fidelity of $0.92$ in the extreme pairs. As expected, the Opposite-Phases algorithm has a similar pattern, but the fidelity drops slightly at the extremal pairs of $\alpha$, reaching a minimal fidelity of $0.87$. The classical algorithm is the least stable, with the fidelity falling quickly when we move out of the central region. The minimal value of fidelity reached is $0.70$. Last, the Equal-Phases and Alternate-Phases algorithms show an intermediate behavior between the classical and the standard quantum algorithms. The minimal values of fidelity are $0.80$ and $0.73$, respectively. These results seem to reinforce the previous discussion about the stability of the PageRank algorithm.

\section{Conclusions}\label{Conclusions}
\label{sec-conlusions}

We have reviewed the quantization of Google's PageRank algorithm in order to modify it, introducing arbitrary phases rotations (APR) in the underlying Szegedy's quantum walk. This modification introduces two degrees of freedom in the algorithm. However, we have defined three simple schemes, called Equal-Phases, Opposite-Phases, and Alternate-Phases schemes, with only one parameter $\theta$. Furthermore, we have shown a method to simulate these new algorithms in a classical computer.

We have applied the new quantum algorithms with APR to a small generic graph with seven nodes, comparing the results with those described in the literature. We have found that the decrease in the value of $\theta$ reduces the standard deviation of the instantaneous quantum PageRank, allowing to better distinguish between nodes. However, the oscillation of the instantaneous PageRank gets slower, so that the time-averaged quantum PageRanks needs more time to converge. This means that we cannot reduce the phase $\theta$ arbitrarily. We have chosen a value of $\theta = \pi/2$ as a value where the APR scheme has a great effect on the time-averaged PageRank while the convergence time is short. With this value we have seen that the Equal-Phases and Alternate-Phases distributions more resemble the classical one, whereas the Opposite-Phases distribution resembles the standard quantum one.

We have studied the time-averaged quantum PageRank with APR in scale-free complex networks since they are good models of the World Wide Web. It was known that the standard quantum algorithm was able to highlight secondary hubs of the networks, whose PageRank was suppressed in the classical distribution. Moreover, the quantum algorithm breaks the degeneracy of the residual nodes, which should be all degenerate as they do not have nodes linking to them. This could yield a problem, as those residual nodes can overshadow the actual secondary hubs. The Opposite-Phases and Alternate-Phases schemes overcome this problem, restoring the degeneration and making them residual. Thus, these two schemes have a distribution that resembles the classical one but highlighting truly secondary hubs. However, the Equal-Phases scheme yields a distribution very similar to the standard quantum one. Regarding the standard deviation of the quantum PageRanks, we have found that the Opposite-Phases and Alternate-Phases schemes can decrease it, but the Equal-Phases scheme does not. Since the effect of the different schemes is different from what was found in the small generic graph, we conclude that the effect depends on the kind of network. Moreover, in Appendix \ref{Ap_ER} we show some results for a particular instance of Erdős-Rényi graphs, observing a different behavior of some APR schemes with respect to the scale-free graphs.

Scale-free networks follow a power law distribution in the connectivity of the nodes. It was known that the classical and quantum PageRanks also have a power law behavior, being smoother in the case of the quantum algorithm since it breaks the degeneracy of the residual nodes. Comparing all of our algorithms, we have found that the standard quantum algorithm, and that with the Equal-Phases scheme have the smoothest distribution, and the power law extends to the residual nodes. The fact that the residual nodes are not degenerate and also follow a power law may indicate that these two quantum algorithms are sensitive to the outdegree distribution of the nodes, since they do not have any node linking to them, and thus they inherit the power law of the connectivity characteristic of scale-free networks. We have also seen that the Opposite-Phases and Alternate-Phases schemes have a slightly smoother distribution than the classical algorithm.

We have studied the stability of the PageRank algorithm with respect to the damping parameter $\alpha$ in the scale-free networks. In the literature it was shown that the classical algorithm was quite unstable, whereas the quantum algorithm improved in stability considerably. In the case of the APR schemes, we expected that the Equal-Phases scheme was the most stable, since its PageRank distribution is very similar to that of the standard quantum algorithm. Surprisingly, the Opposite-Phases scheme is the most stable, being comparable to the standard quantum case. The Equal-Phases and Alternate-Phases schemes have a similar intermediate stability, despite the fact that their PageRank distributions are rather different.

Taking all the results together, the fact that the algorithm with the Opposite-Phases scheme is able to highlight the secondary nodes that the classical cannot, keeps degenerate the residual nodes, reduces the standard deviation of the time-averaged PageRank, and also has a good stability similar to the original quantum algorithm, makes this new algorithm a valuable tool as an alternative to both classical and standard quantum PageRank for scale-free networks.

In the future, it would be interesting to study what happens in complex networks when we use phases other than $\pi/2$, other APR schemes, or even introduce more phases to the algorithm. It would also be interesting to apply these algorithms to other kinds of complex networks, such as hierarchical networks. Finally, although we have used Szegedy's quantum walk with APR for the PageRank algorithm, it could be of interest for other applications, such as quantum search, optimization, or machine learning.

\section*{Acknowledgments}
We acknowledge support from the CAM/FEDER Project No.S2018/TCS-4342 (QUITEMAD-CM), Spanish MINECO grants MINECO/FEDER Projects, PGC2018-099169-B-I00 FIS2018, MCIN with funding from European Union NextGenerationEU (PRTR-C17.I1) and Ministry of Economic Affairs Quantum ENIA project. M.A.M.-D. has been partially supported by the U.S.Army Research Office through Grant No. W911NF-14-1-0103. S.A.O. acknowledges support from a QUITEMAD grant, and from Universidad Complutense de Madrid - Banco Santander through Grant No. CT58/21-CT59/21.

\bibliography{MiBiblio}
\bibliographystyle{unsrt}

\onecolumngrid

\appendix
\newpage
\section{Spectral decomposition and classical simulation of Szegedy's quantum walk with APR}\label{Ap_spectral}

In \cite{Paparo1} it is shown how to decompose the time evolution operator of the quantum walk for the standard case. Here we are going to show the general decomposition for the operator, taking into account the complex phases. This decomposition will allow to simulate the final statevector in a faster manner and with less memory resources. This can be used for simulating not only the quantum PageRank, but also any Szegedy quantum walk with APR that uses either the operator $U(\theta)$ or $W(\theta_1,\theta_2)$.

We start by defining the $N \times N$ matrix $D$ whose entries are
\begin{equation}
	D_{ij} := \sqrt{G_{ij}G_{ji}},
\end{equation}
where there is no sum over repeated indexes. We also define the operator $A$ from the space of vertex $\mathbb{C}^N$ to the space of edges $\mathbb{C}^N \otimes \mathbb{C}^N$:
\begin{equation}
	A := \sum_{i=1}^N \left|\psi_i\right>\left<i\right|,
\end{equation}
which satisfies that $AA^\dagger = \Pi$, $A^\dagger A = \mathbbm{1}$, and $D = A^\dagger S A$.

The matrix $D$ is symmetric, so it can be diagonalized, yielding $N$ eigenvectors $\left|\lambda\right>$ with eigenvalues $\lambda$. With them, we define the vectors $\left|\widetilde{\lambda}\right> = A\left|\lambda\right>$, which belong to the Hilbert space where the quantum walk is performed. Now we are going to show that the vectors $\left|\widetilde{\lambda}\right>$ and $S\left|\widetilde{\lambda}\right>$ generate an invariant subspace of $U(\theta)$ for any $\theta$. We apply $U(\theta)$ to $\left|\widetilde{\lambda}\right>$ and $S\left|\widetilde{\lambda}\right>$, and using the properties of the operators defined above,
\begin{equation}\label{Ulambda}
	U(\theta)|\widetilde{\lambda}> = -e^{i\theta} S \left|\widetilde{\lambda}\right>,
\end{equation}
\begin{equation}\label{USlambda}
	U(\theta)S\left|\widetilde{\lambda}\right> = \left[1-e^{i\theta}\right] \lambda S \left|\widetilde{\lambda}\right> - \left|\widetilde{\lambda}\right>.
\end{equation}
Then, the action over these vectors yields vectors in the subspace formed by them, i.e., they generate an invariant subspace. Since these vectors are independent of $\theta$, this subspace is the same for every $U(\theta)$, and we call it $\mathcal{I}_U$. Moreover, the product of two unitaries with different phases, i.e., the operator $W(\theta_1,\theta_2) = U(\theta_1)U(\theta_2)$, also have this subspace as invariant because it does not depend on the complex phases.

The span of the vectors $\left|\widetilde{\lambda}\right>$ coincides with
the span of the vectors $\left|\psi_i\right>$, and since $\Pi$ is the projector onto the subspace formed by these vectors, the action of $\Pi$ over any vector orthogonal to them is null. The subspace orthogonal to $\mathcal{I}_U$ is orthogonal to all $\left|\widetilde{\lambda}\right>$, so the quantum walk operator in this subspace acts just like $U(\theta) = -S$ for any value of $\theta$. This means that all the eigenvalues are $\pm1$ in the orthogonal subspace. Since the subspace is independent of $\theta$, it is the same for the operator $W(\theta_1,\theta_2)$, and all the eigenvalues are equal to $1$.
Thus, all the dynamics of the walk occurs in the subspace $\mathcal{I}_U$, and the problem reduces to finding the eigenvectors and eigenvalues of $W(\theta_1,\theta_2)$ in this subspace.

We start by solving the eigenvalue problem for the simple operator $U(\theta)$. We make the following ansatz for the eigenvectors in the subspace $\mathcal{I}_U$:
\begin{equation}
	\left|\mu_\theta\right> = \left|\widetilde{\lambda}\right> - \mu_\theta S \left|\widetilde{\lambda}\right>,
\end{equation}
where $\mu_\theta$ is the eigenvalue of the eigenvector $\left|\mu_\theta\right>$, both depending on the phase $\theta$. We apply $U(\theta)$ to it, and using \eqref{Ulambda} and \eqref{USlambda},
\begin{equation}\label{Umu}
	U(\theta)\left|\mu_\theta\right> = \mu_\theta\left|\widetilde{\lambda}\right> - \left(e^{i\theta} + \left[1-e^{i\theta}\right]\mu_\theta\lambda\right)S\left|\widetilde{\lambda}\right>.
\end{equation}
By definition of eigenvector we also have
\begin{equation}
	U(\theta)\left|\mu_\theta\right> = \mu_\theta\left|\widetilde{\lambda}\right> - \mu_\theta^2 S\left|\widetilde{\lambda}\right>.
\end{equation}
Equaling both expressions, we obtain an equation for the eigenvalues,
\begin{equation}
	-\mu_\theta^2 = -e^{i\theta} - \left[1-e^{i\theta}\right]\mu_\theta\lambda,
\end{equation}
whose solution is
\begin{equation}
	\mu_\theta = \frac{\left[1-e^{i\theta}\right]\lambda \pm \sqrt{\left[1-e^{i\theta}\right]^2\lambda^2 + 4e^{i\theta}}}{2}.
\end{equation}
With this we can calculate the eigenvectors of $U(\theta)$. However, these depend on the complex phase $\theta$. This means that they are not the same for the operator $W(\theta_1,\theta_2)$ unless $\theta_1 = \theta_2$. In that case, the eigenvectors would be the same, and the eigenvalues would be $\mu_\theta^2$.

In the general case for $W(\theta_1,\theta_2)$ where $\theta_1 \neq \theta_2$, let us call $\nu \equiv \nu(\theta_1,\theta_2)$ to the eigenvalues of the eigenvectors $\left|\nu\right> \equiv \left|\nu(\theta_1,\theta_2)\right>$ in the invariant dynamical subspace $\mathcal{I}_U$. We use the following more general ansatz for the eigenvectors:
\begin{equation}
	\left|\nu\right> = \left|\widetilde{\lambda}\right> - a S \left|\widetilde{\lambda}\right>,
\end{equation}
where $a$ is a parameter to determinate that depends on the two phases as well on $\lambda$. We apply the operator to this eigenvector, and using \eqref{Ulambda}, \eqref{USlambda}, and \eqref{Umu},
\begin{equation}
	W(\theta_1,\theta_2)\left|\nu\right> = \left(e^{i\theta_1} + C_1a\lambda\right)\left|\widetilde{\lambda}\right> - \left(ae^{i\theta_2} + \left[e^{i\theta_1}+C_1a\lambda\right]C_2\lambda\right)S\left|\widetilde{\lambda}\right>,
\end{equation}
where $C_k = 1-e^{i\theta_k}$. Using the definition of eigenvector,
\begin{equation}
	W(\theta_1,\theta_2)\left|\nu\right> = \nu\left|\widetilde{\lambda}\right> - \nu a S\left|\widetilde{\lambda}\right>.
\end{equation}
We obtain a system of two equations with two variables, $a$ and $\nu$:
\begin{equation}
	\nu = e^{i\theta_1} + C_1a\lambda,
\end{equation}
\begin{equation}
	\nu a = ae^{i\theta_2} + \left[e^{i\theta_1}+C_1a\lambda\right]C_2\lambda.
\end{equation}
After substituting the first equation into the second one, we finally have a second-order equation for $a$. After solving that equation, we can calculate the (at most) $2N$ eigenvalues and eigenvectors of the operator $W(\theta_1,\theta_2)$ in the dynamical subspace $\mathcal{I}_U$.

Once the eigenvectors are obtained, we can project the initial state onto them. Raising the eigenvalues to the power of $t$, we can calculate the dynamical component of the final state without the need of the $N^2 \times N^2$ matrix. In the quantum PageRank algorithm the initial state lies in the dynamical subspace, so its component in the orthogonal subspace is null. However, for a more general initial vector, once we project it onto the dynamical subspace, we can calculate its orthogonal component. Since $W(\theta_1, \theta_2) = (-S)^2 = \mathbbm{1}$ in the orthogonal subspace, this component does not change with the time evolution and can just be added once the dynamical component is calculated.

\newpage

\section{Results for the other APR schemes of the generic graph with 7 nodes}\label{Ap_general}

\begin{figure}[H]
	\center
	\subfigure[]{\includegraphics[scale=0.375]{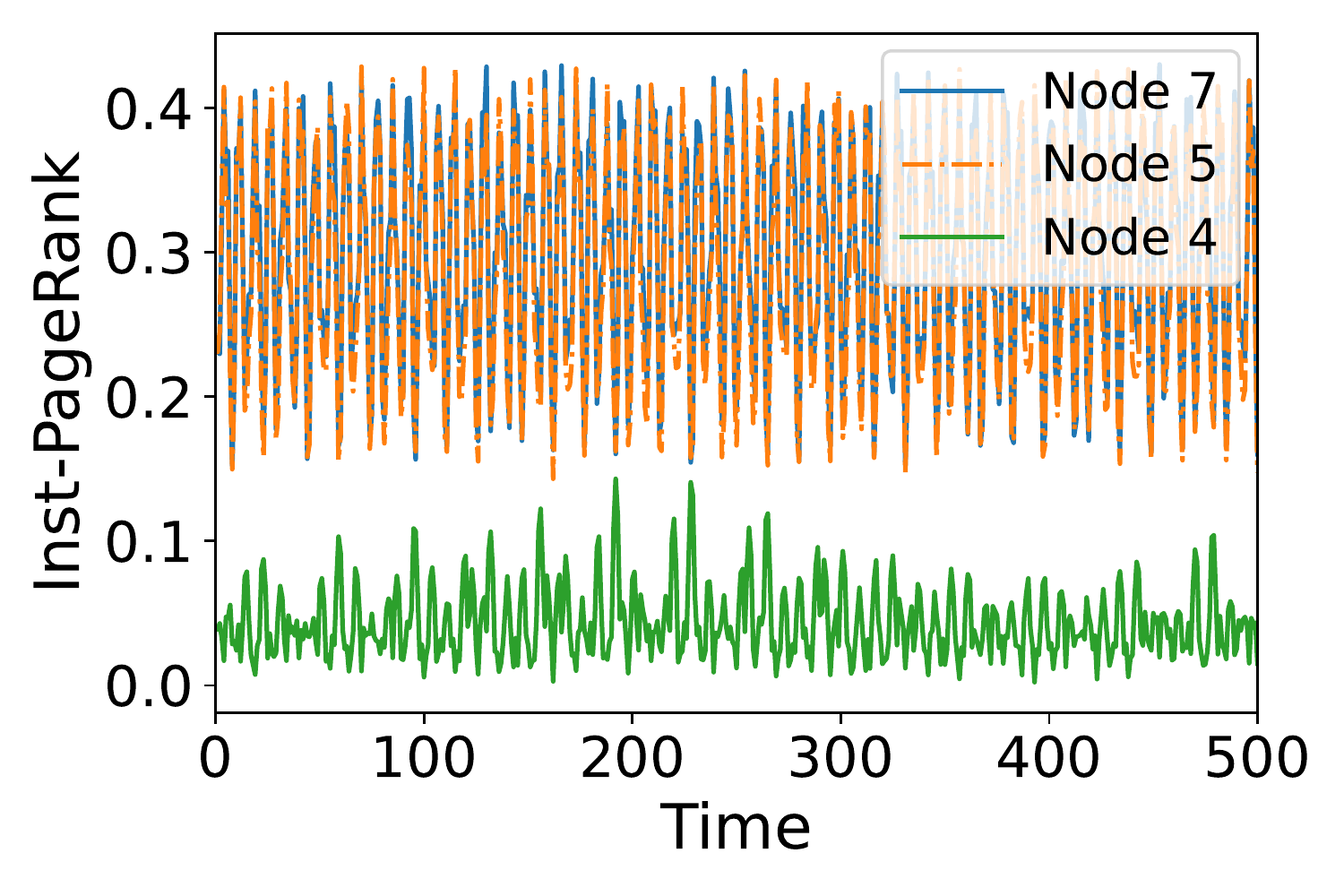}\label{F:Inst_Eq_pi2}}
	\subfigure[]{\includegraphics[scale=0.375]{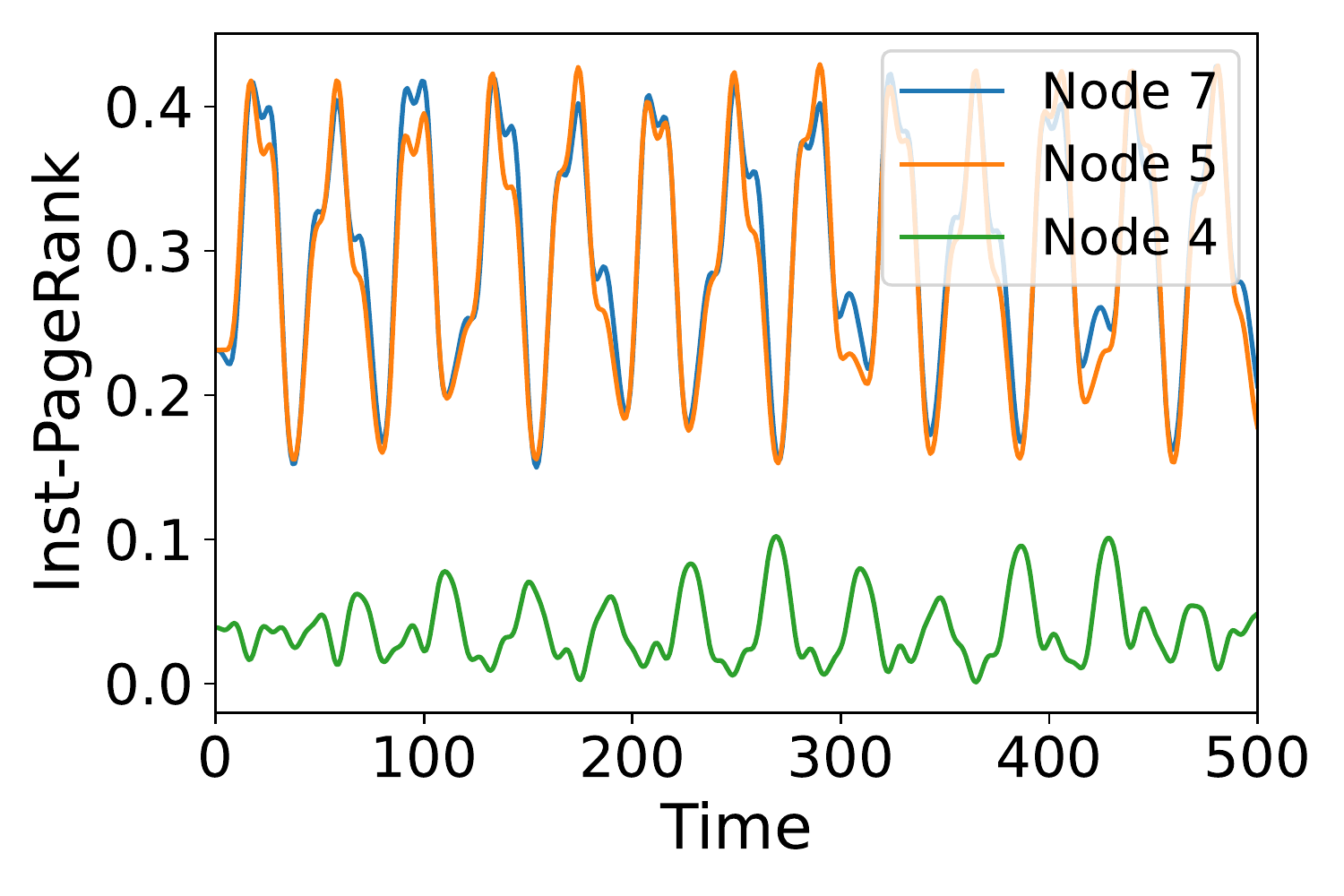}\label{F:Inst_Eq_pi10}}
	\subfigure[]{\includegraphics[scale=0.375]{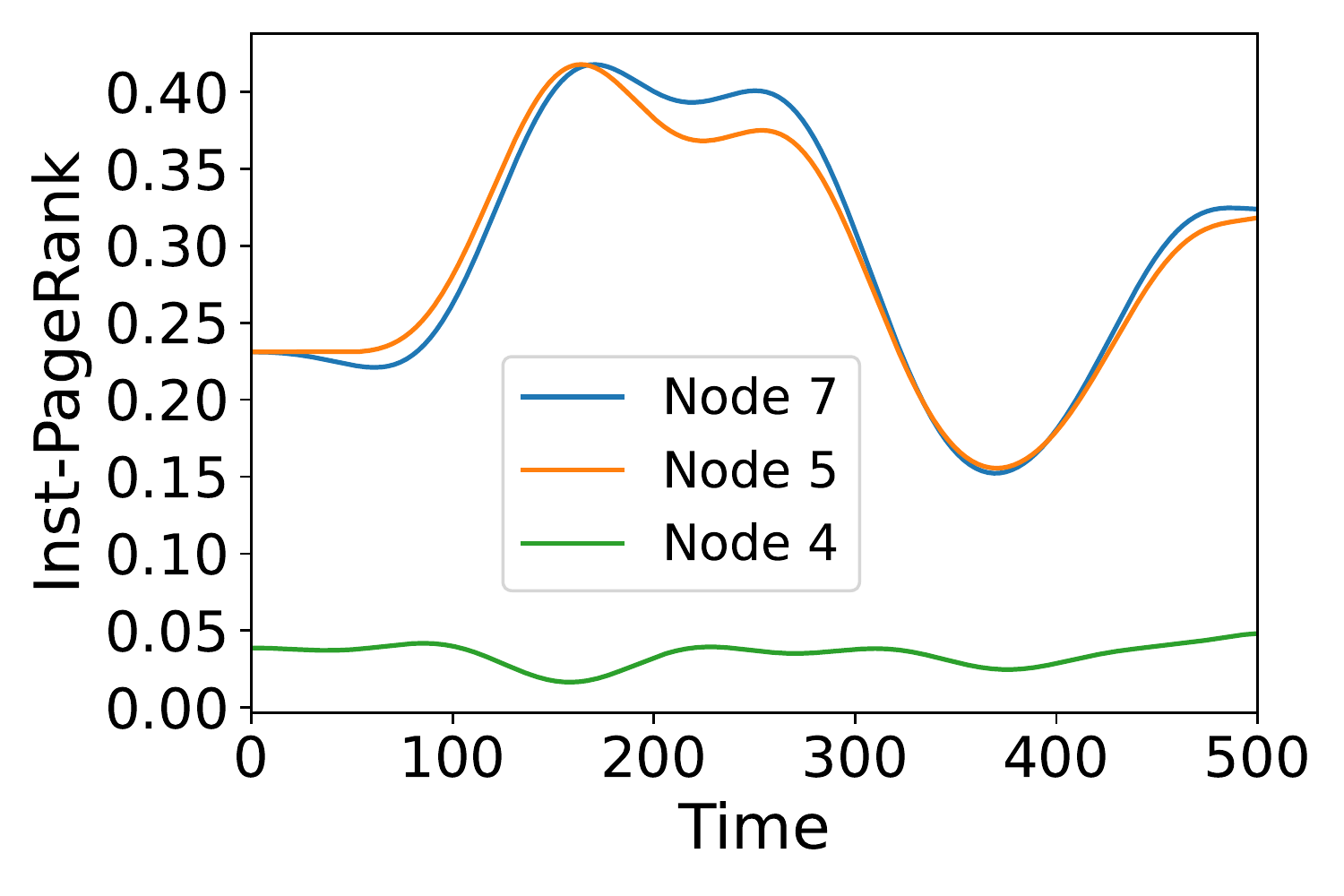}\label{F:Inst_Eq_pi100}}
	\subfigure[]{\includegraphics[scale=0.375]{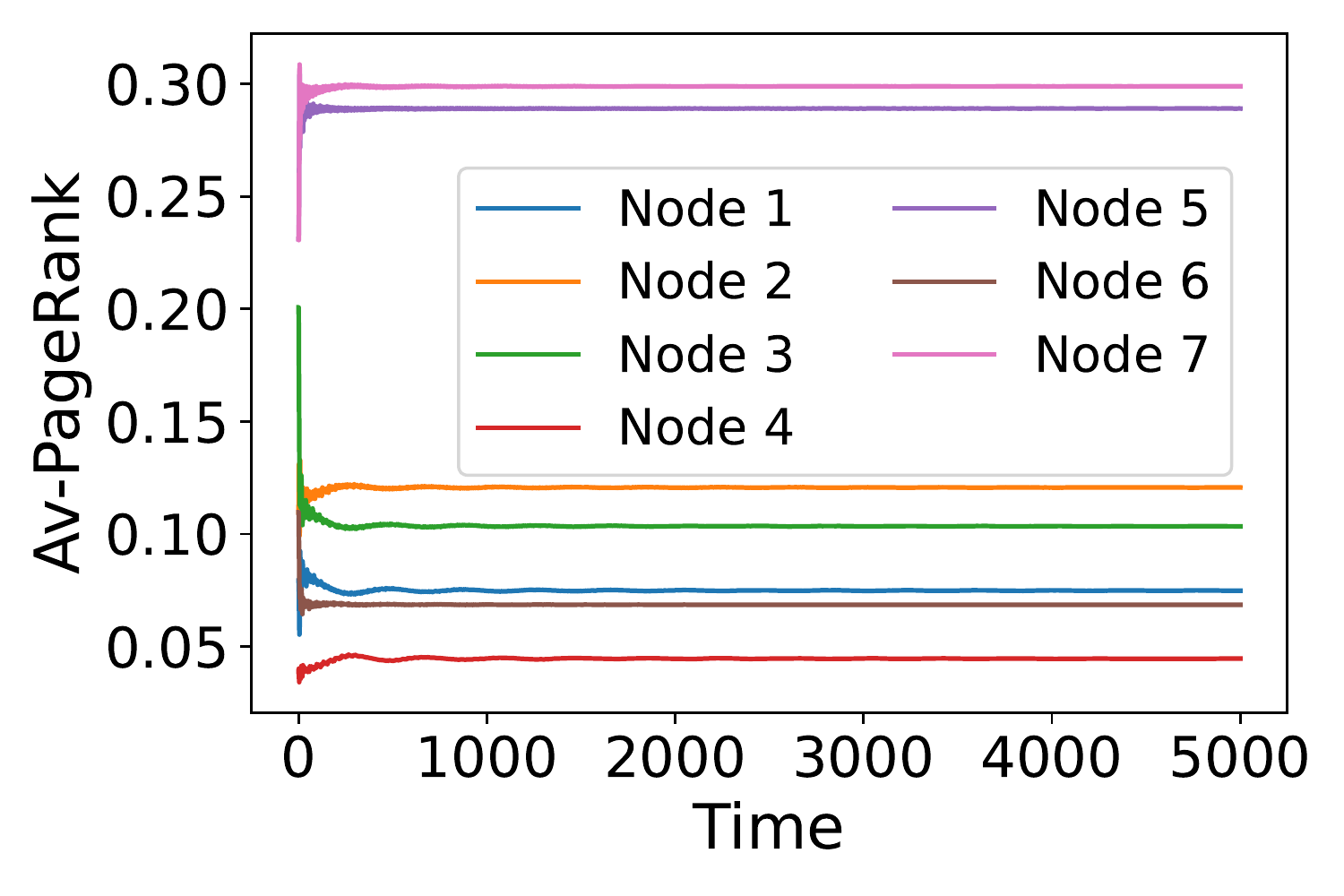}\label{F:Conv_Eq_pi2}}
	\subfigure[]{\includegraphics[scale=0.375]{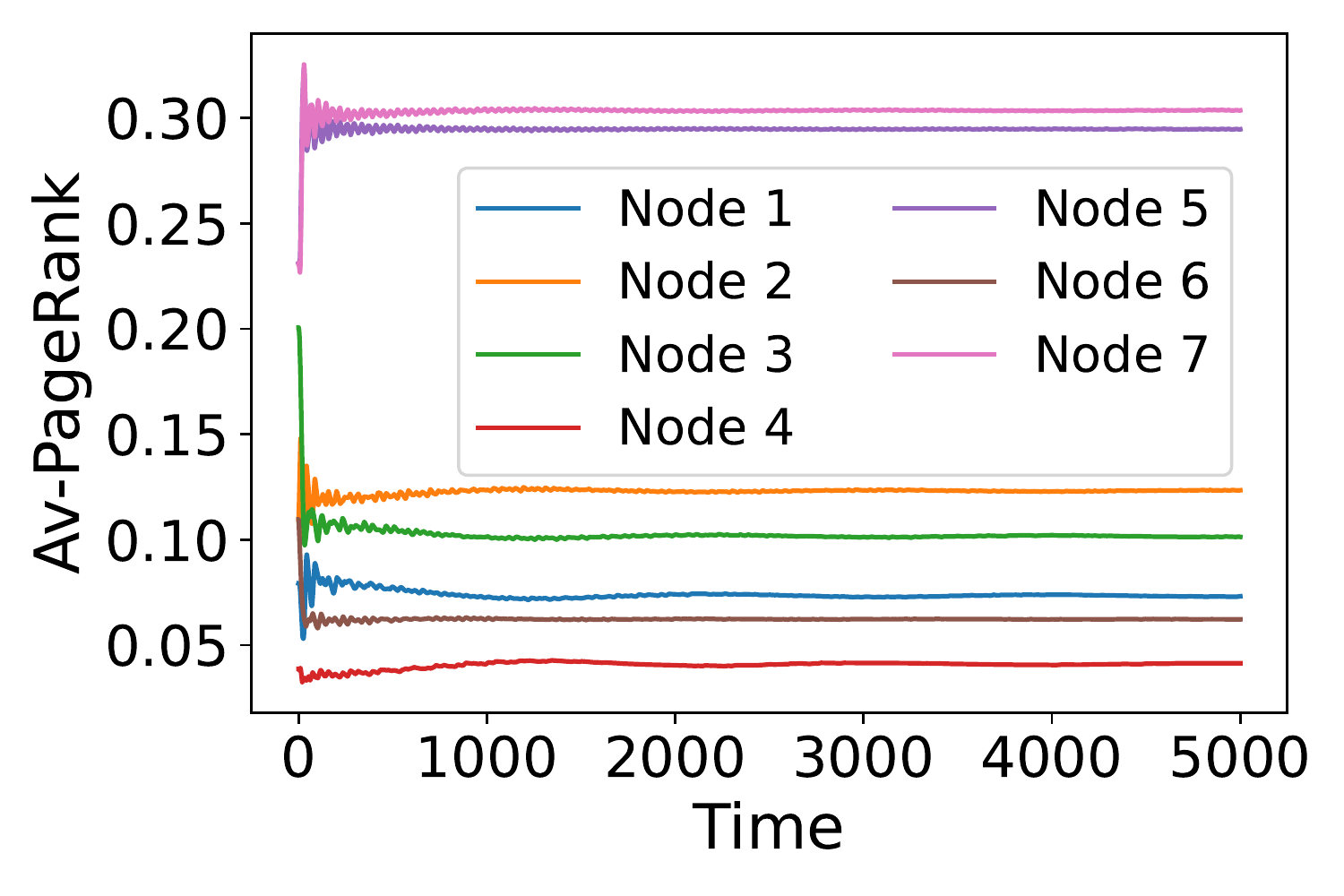}\label{F:Conv_Eq_pi10}}
	\subfigure[]{\includegraphics[scale=0.375]{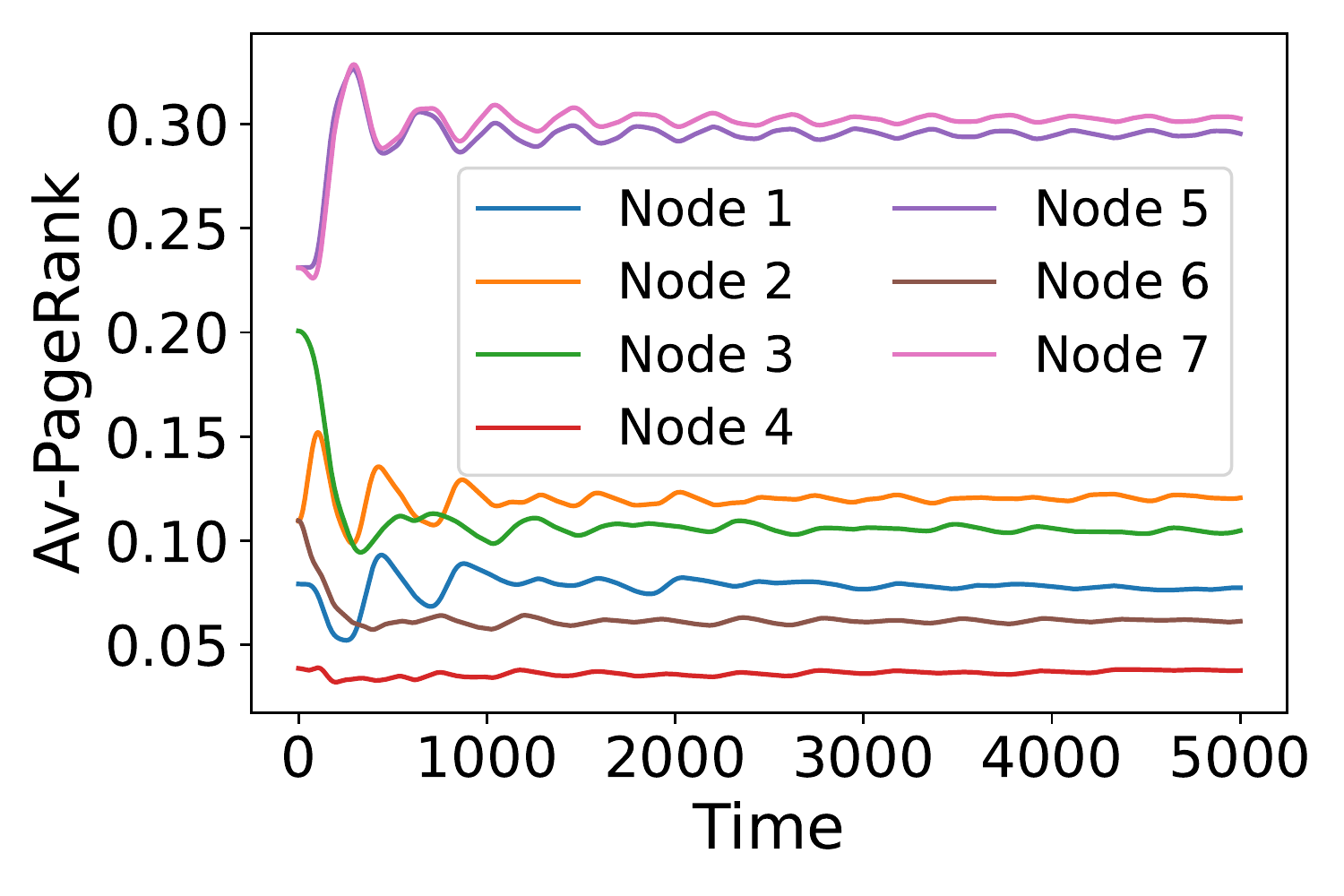}\label{F:Conv_Eq_pi100}}
	\caption{Instantaneous PageRanks of nodes $7$, $5$, and $4$ of the small generic graph for the Equal-Phases algorithm with a) $\theta = \pi/2$, b) $\theta = \pi/10$, and c) $\theta = \pi/100$. Time-averaged quantum PageRanks for all nodes vs time for the Equal-Phases algorithm with d) $\theta = \pi/2$, e) $\theta = \pi/10$, and f) $\theta = \pi/100$. It is observed that as $\theta$ decreases, the quantum fluctuations get slower and the algorithm takes more time to converge.}
	\label{F:Inst_Eq}
\end{figure}

\begin{figure}[H]
	\centering
	\includegraphics[scale=0.5]{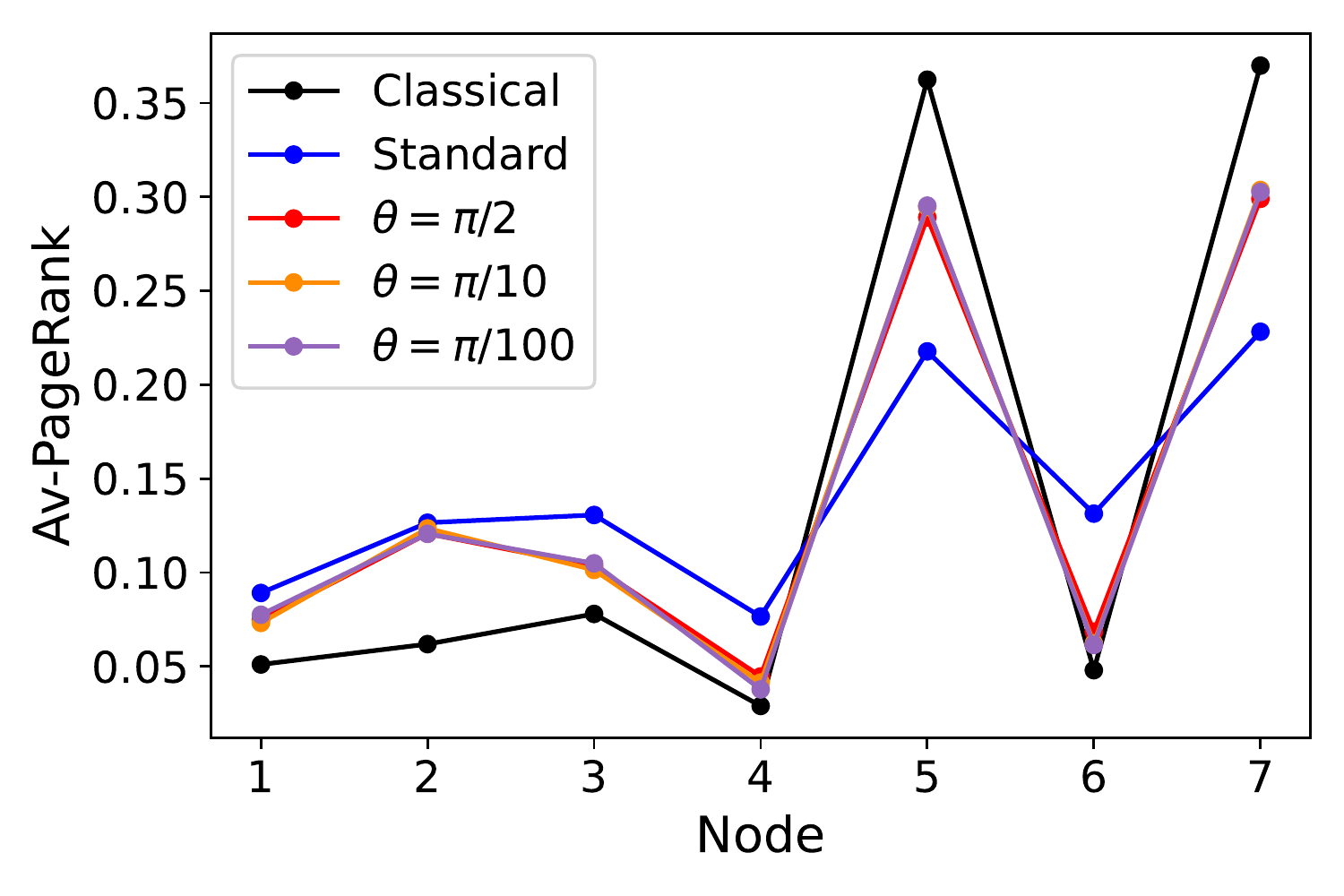}
	\caption{Time-averaged quantum PageRanks for the Equal-Phases scheme with $\theta = \pi/2$, $\pi/10$, and $\pi/100$ for the small generic graph with seven nodes. They are compared with the classical PageRanks and the standard quantum PageRanks.}
	\label{F:PR_Eq}
\end{figure}

\begin{figure}[H]
	\center
	\subfigure[]{\includegraphics[scale=0.375]{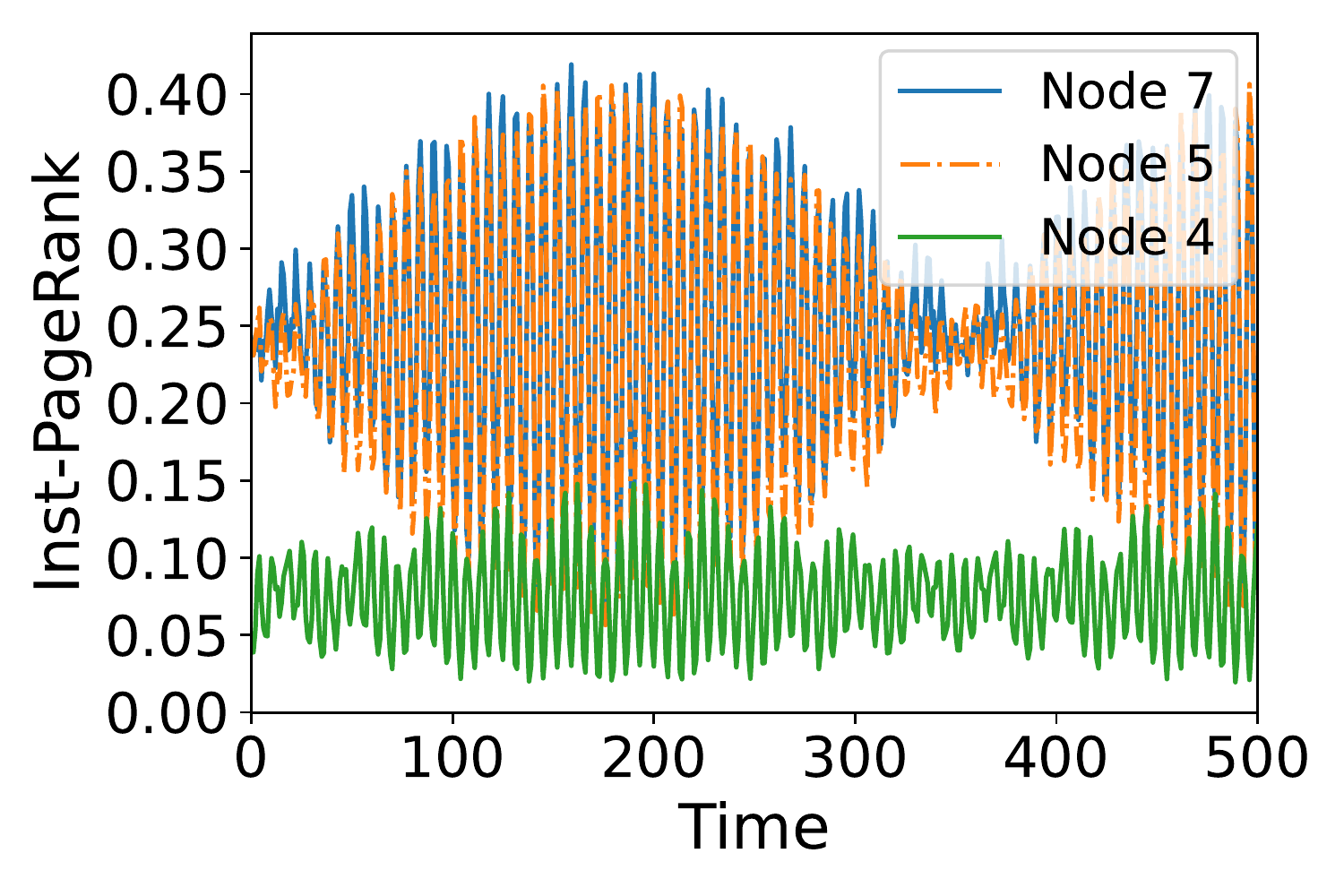}\label{F:Inst_Op_pi2}}
	\subfigure[]{\includegraphics[scale=0.375]{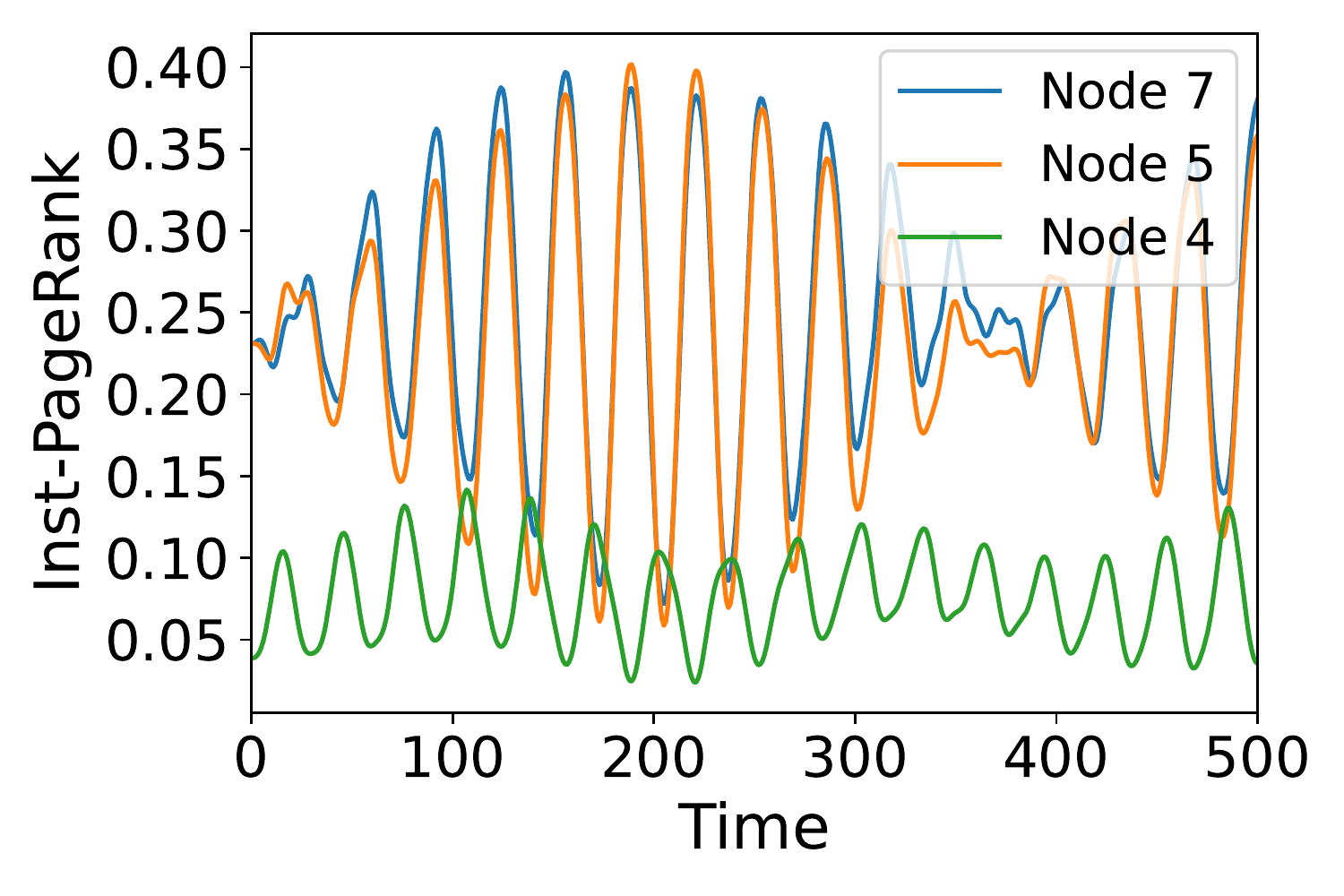}\label{F:Inst_Op_pi10}}
	\subfigure[]{\includegraphics[scale=0.375]{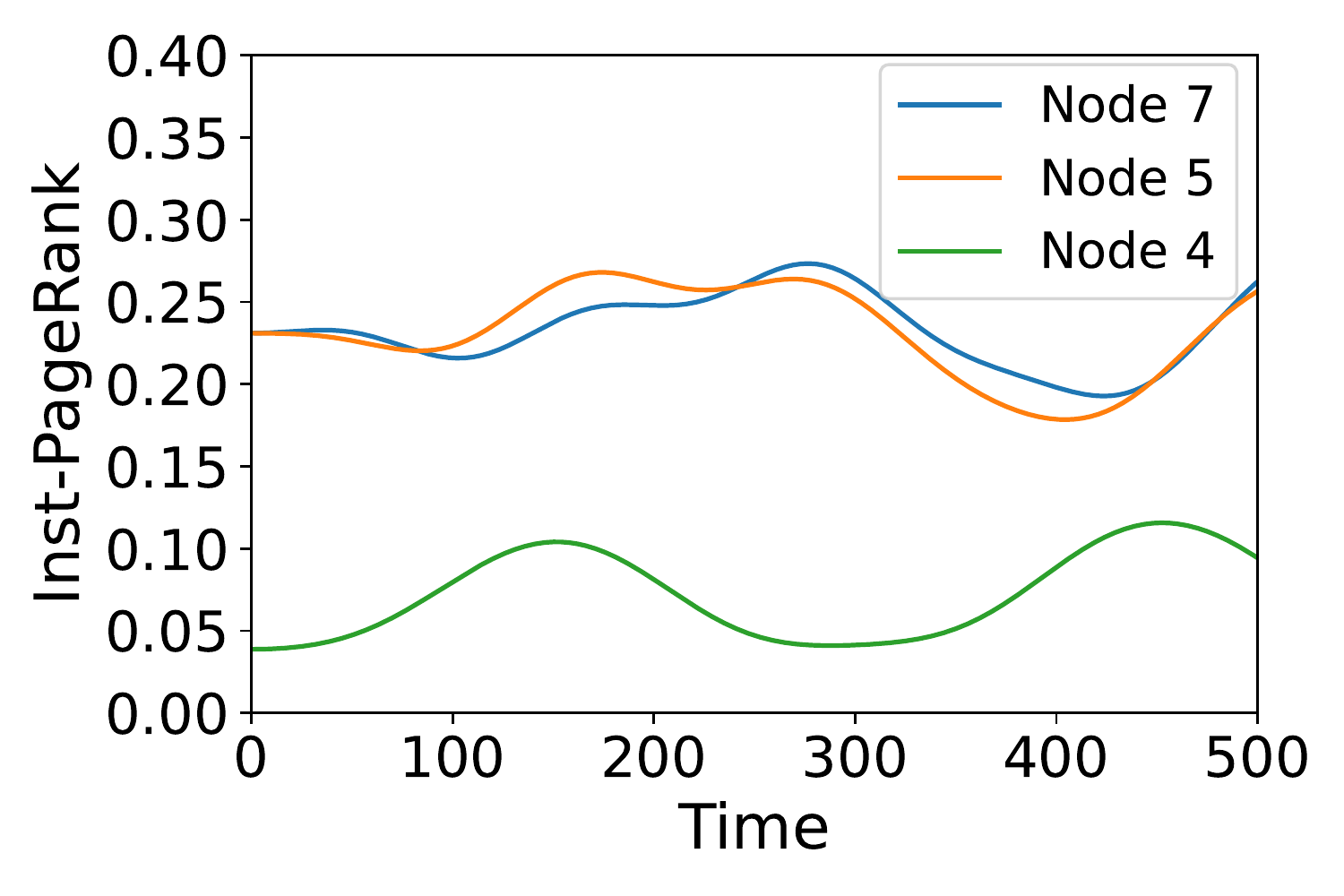}\label{F:Inst_Op_pi100}}
	\subfigure[]{\includegraphics[scale=0.375]{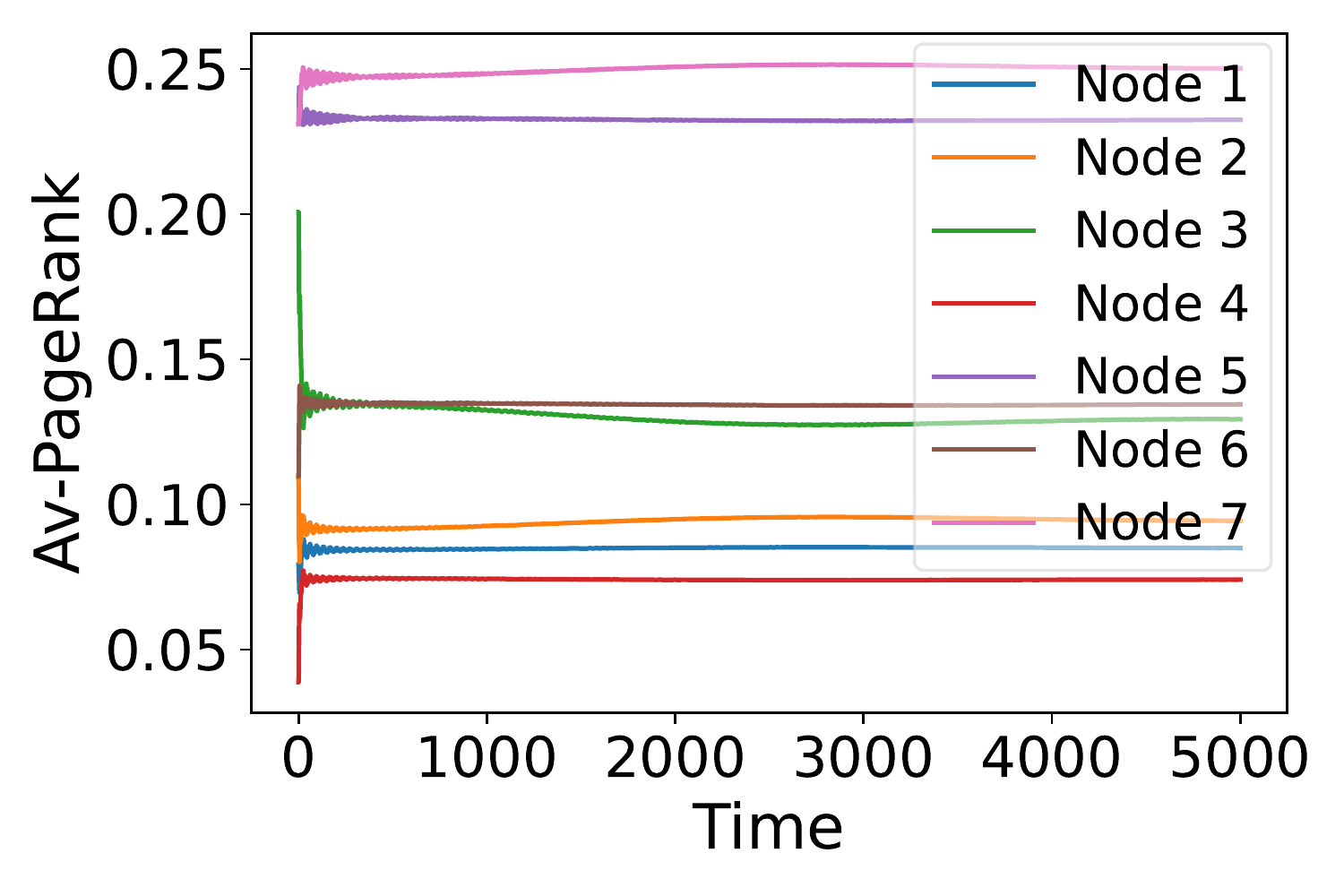}\label{F:Conv_Op_pi2}}
	\subfigure[]{\includegraphics[scale=0.375]{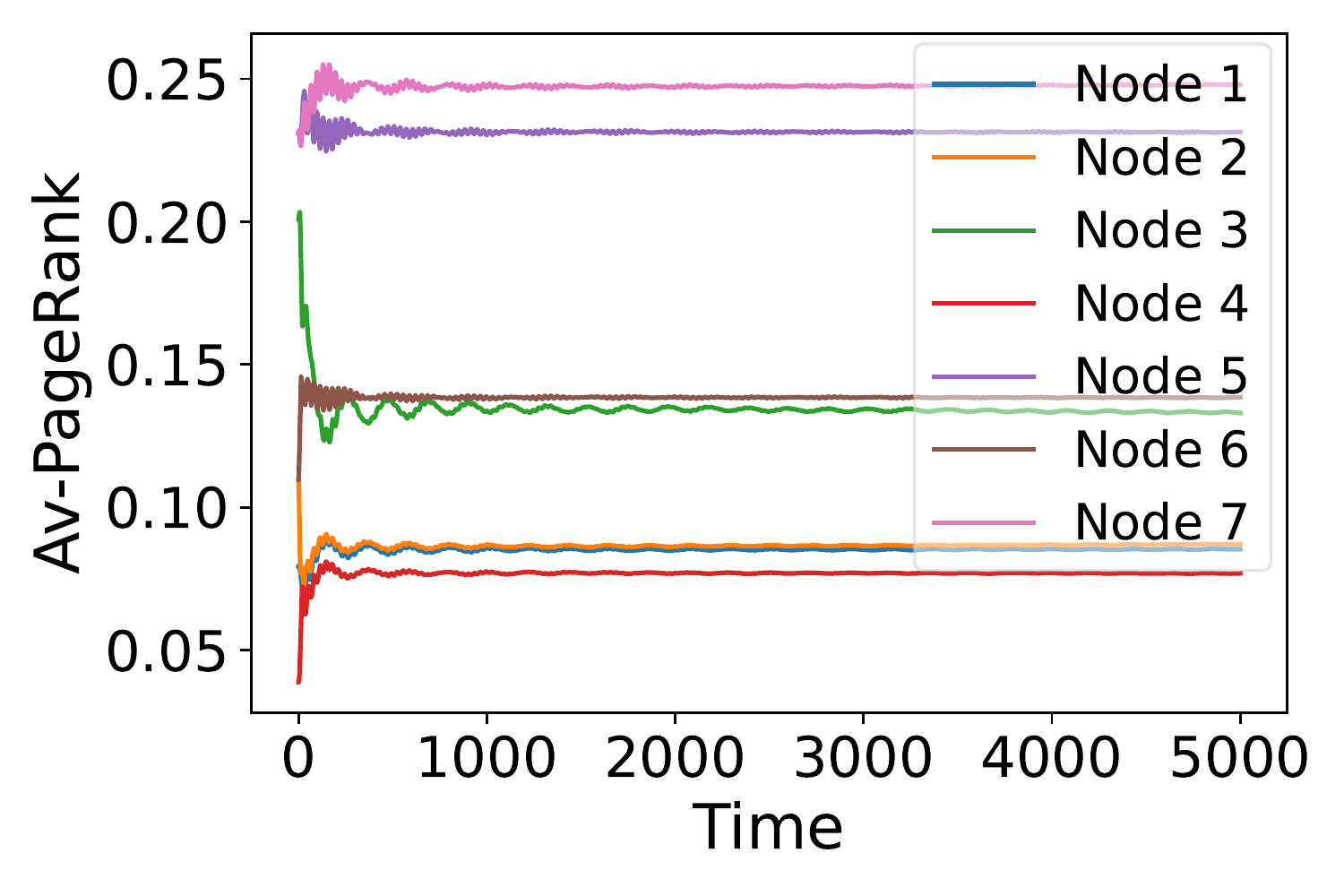}\label{F:Conv_Op_pi10}}
	\subfigure[]{\includegraphics[scale=0.375]{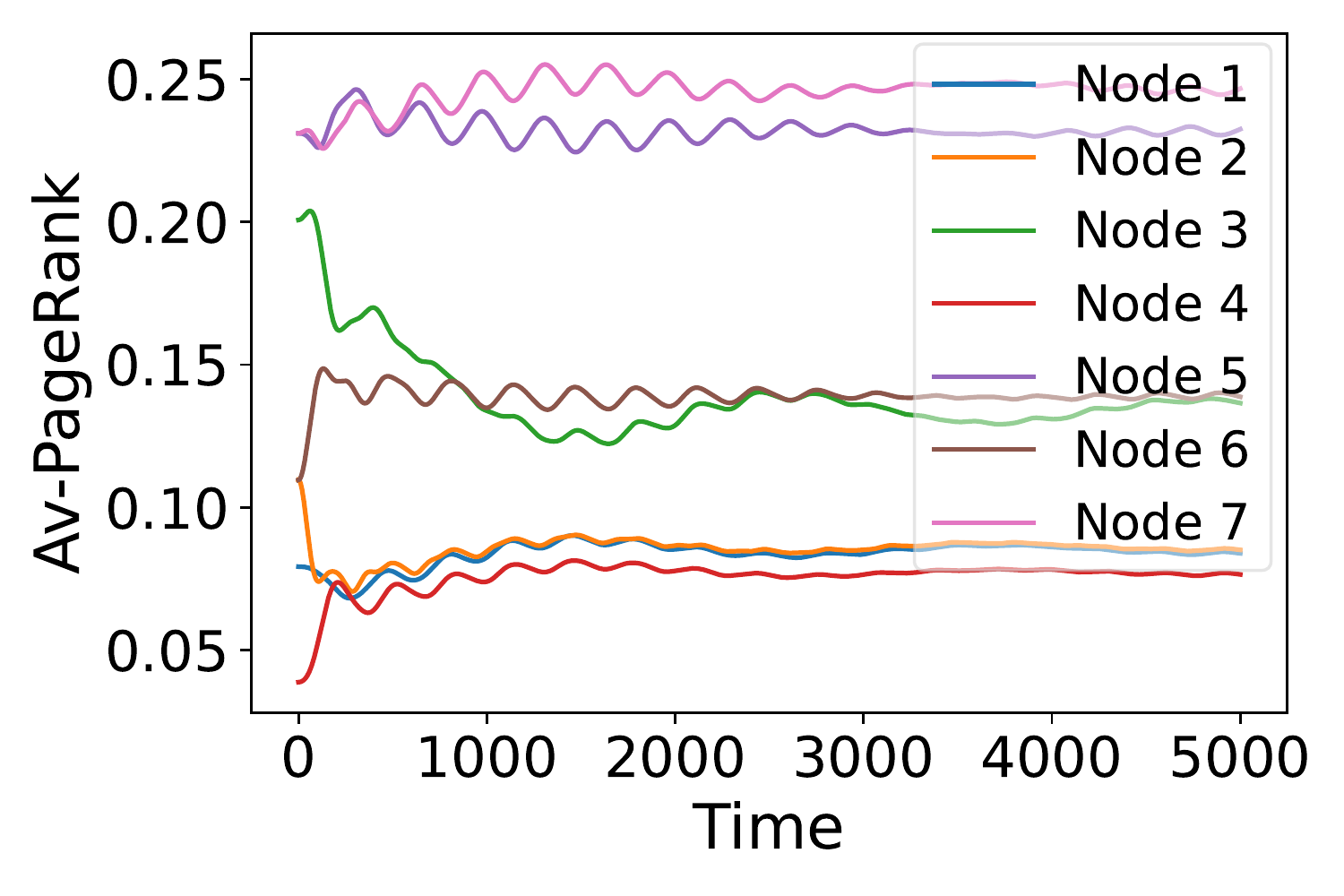}\label{F:Conv_Op_pi100}}
	\caption{Instantaneous PageRanks of nodes $7$, $5$, and $4$ of the small generic graph for the Opposite-Phases algorithm with a) $\theta = \pi/2$, b) $\theta = \pi/10$, and c) $\theta = \pi/100$. Time-averaged quantum PageRanks for all nodes vs time for the Opposite-Phases algorithm with d) $\theta = \pi/2$, e) $\theta = \pi/10$, and f) $\theta = \pi/100$. It is observed that as $\theta$ decreases, the quantum fluctuations get slower and the algorithm takes more time to converge.}
	\label{F:Inst_Op}
\end{figure}

\begin{figure}[H]
	\centering
	\includegraphics[scale=0.5]{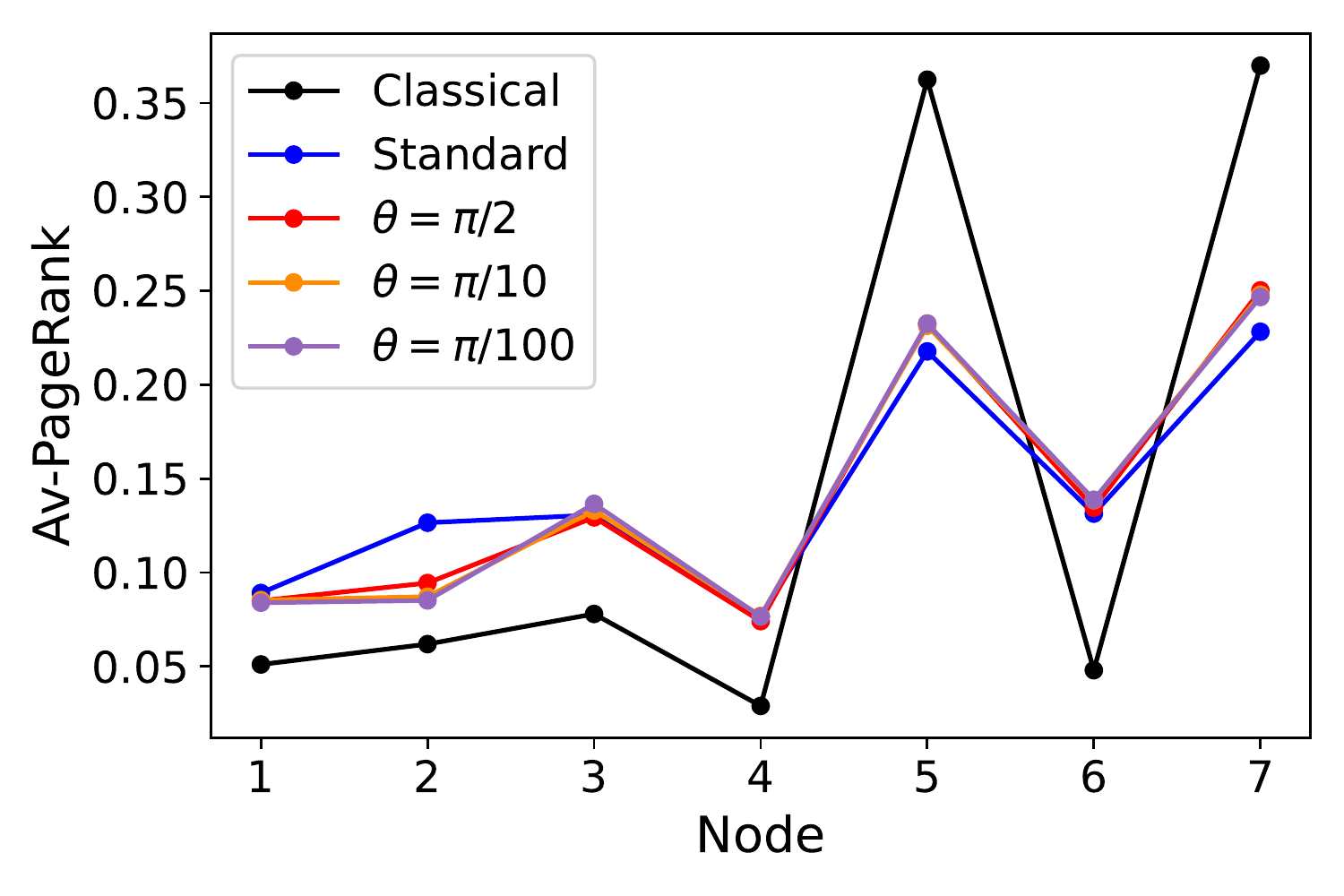}
	\caption{Time-averaged quantum PageRanks for the Opposite-Phases scheme with $\theta = \pi/2$, $\pi/10$, and $\pi/100$ for the small generic graph with seven nodes. They are compared with the classical PageRanks and the standard quantum PageRanks.}
	\label{F:PR_Op}
\end{figure}

In the main paper we have shown how the decrease of the phase $\theta$ affects the instantaneous PageRanks for the Alternate-Phases algorithm. Here we show that similar results are obtained for the other APR schemes. In Figures \ref{F:Inst_Eq} and \ref{F:Inst_Op} we show that the decrease of the phase $\theta$ increases the period of the quantum fluctuations for the Equal-phases and Opposite-phases, respectively. This results in a longer time for the time-averaged quantum PageRank to converge. Furthermore, it is interesting that in the Opposite-Phases case the quantum fluctuations get a modulated behavior. Finally, in Figures \ref{F:PR_Eq} and \ref{F:PR_Op} we show that the major effect due to the APR scheme is achieved with $\theta = \pi/2$, so we can take this value as a trade-off between having a short convergence time and a great effect due to the APR.

\newpage

\section{Erdős-Rényi graphs}\label{Ap_ER}

\begin{figure}[H]
	\centering
	\subfigure[]{\includegraphics[scale=0.5]{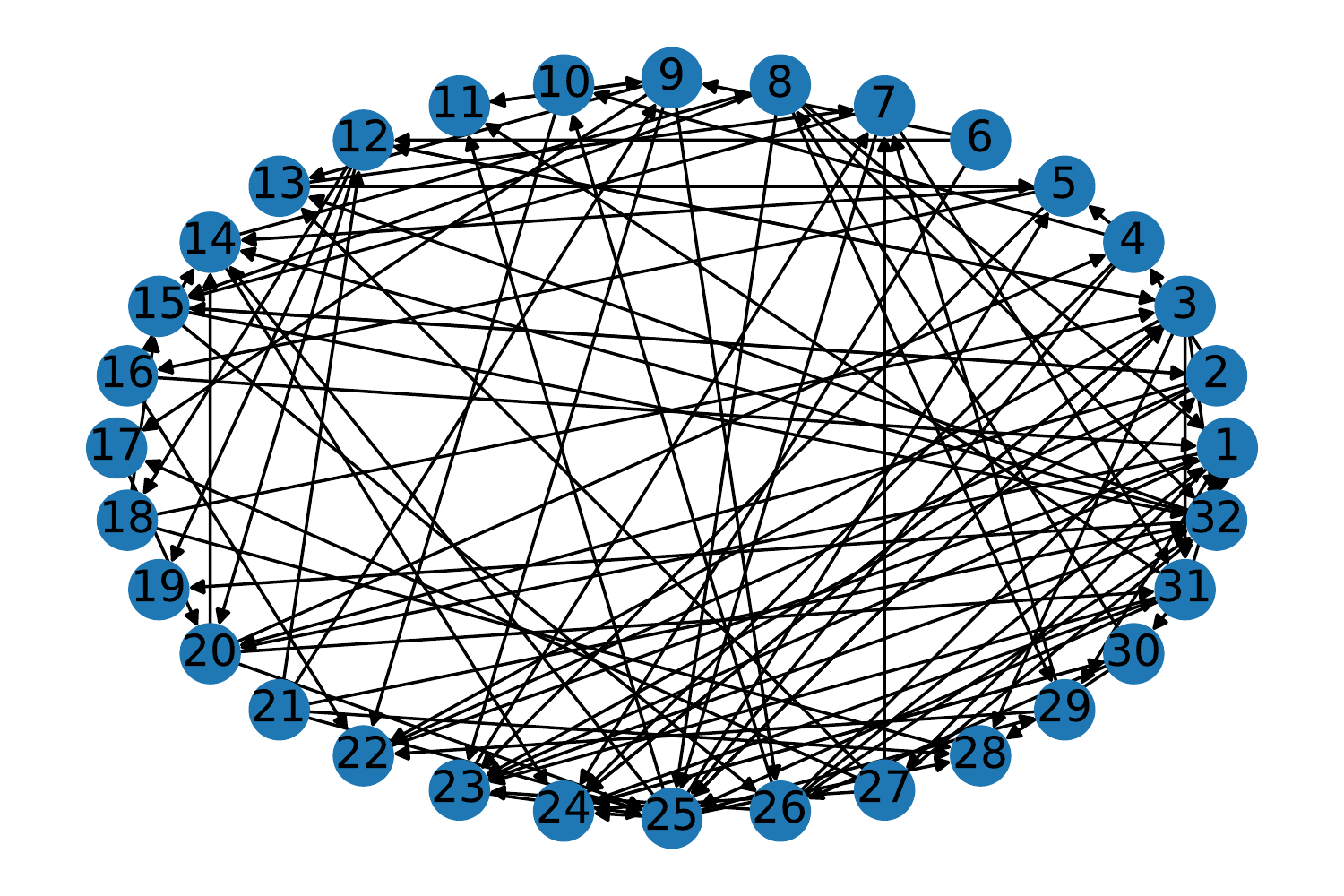}\label{F:PR_ER_32_graph}}
	\subfigure[]{\includegraphics[scale=0.25]{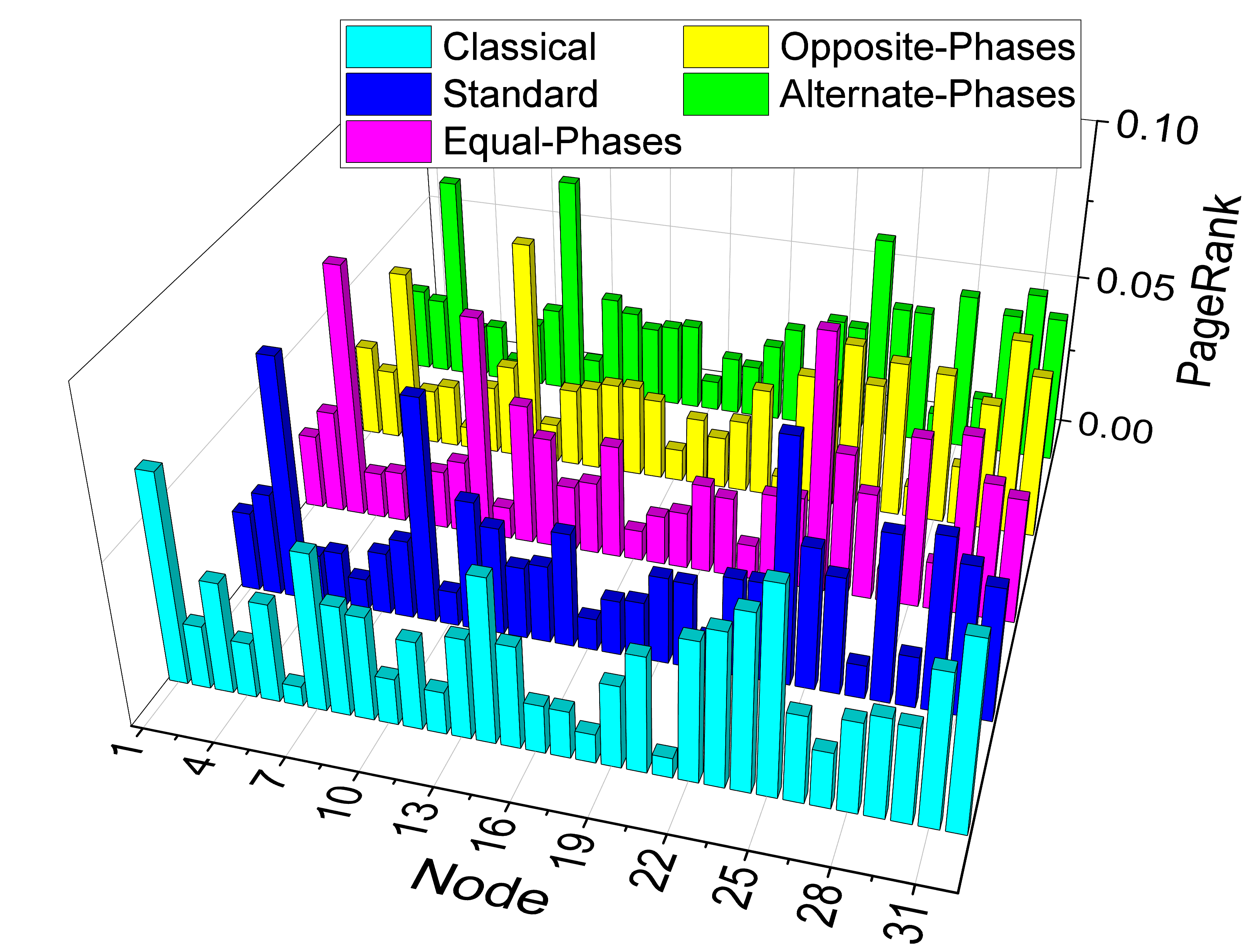}\label{F:PR_ER_32_hist}}
	\caption{a) Erdős-Rényi network with 32 nodes. b) PageRank distributions of the Erdős-Rényi network. The classical distribution is compared with all the quantum distributions, using $\theta = \pi/2$ in the three APR schemes.}
	\label{F:PR_ER_32}
\end{figure}

In order to benchmark the results found for scale-free graphs, we can use Erdős-Rényi random graphs \cite{Gilbert,ER}. Theses graphs are constructed by connecting a set of vertices randomly by adding edges with a fixed probability. In our work we are going to use a directed Erdős-Rényi network with 32 nodes created using NetworkX \cite{NetworkX}, where we have chosen the probability for adding edges as $p = 0.1$. The network is shown in Figure \ref{F:PR_ER_32_graph}. The distributions of PageRanks for the classical algorithm and all the quantum algorithms are shown in the histogram of Figure \ref{F:PR_ER_32_hist}. In \cite{Paparo2} it was found that both the classical and the standard quantum PageRank algorithms did not identify hubs in this class of networks. We observe that this is also the case for all the quantum algorithms with APR, since all the distributions are rather homogeneous. It was also observed in \cite{Paparo2} that the quantum algorithm changed the ranking of nodes with respect to the classical one, and we observe something similar. With regard to the APR schemes, the Equal-Phases algorithm has a similar distribution to the standard quantum case, in the same manner as with scale-free networks. However, the Opposite-Phases and Alternate-Phases distributions are different from both the classical and standard quantum ones. Thus, with Erdős-Rényi networks we do not obtain any quantum distribution that resembles the classical one. This highlights the fact that the effect of the APR schemes depends on the class of network that we are dealing with.

\begin{figure}[H]
	\centering
	\includegraphics[scale=0.5]{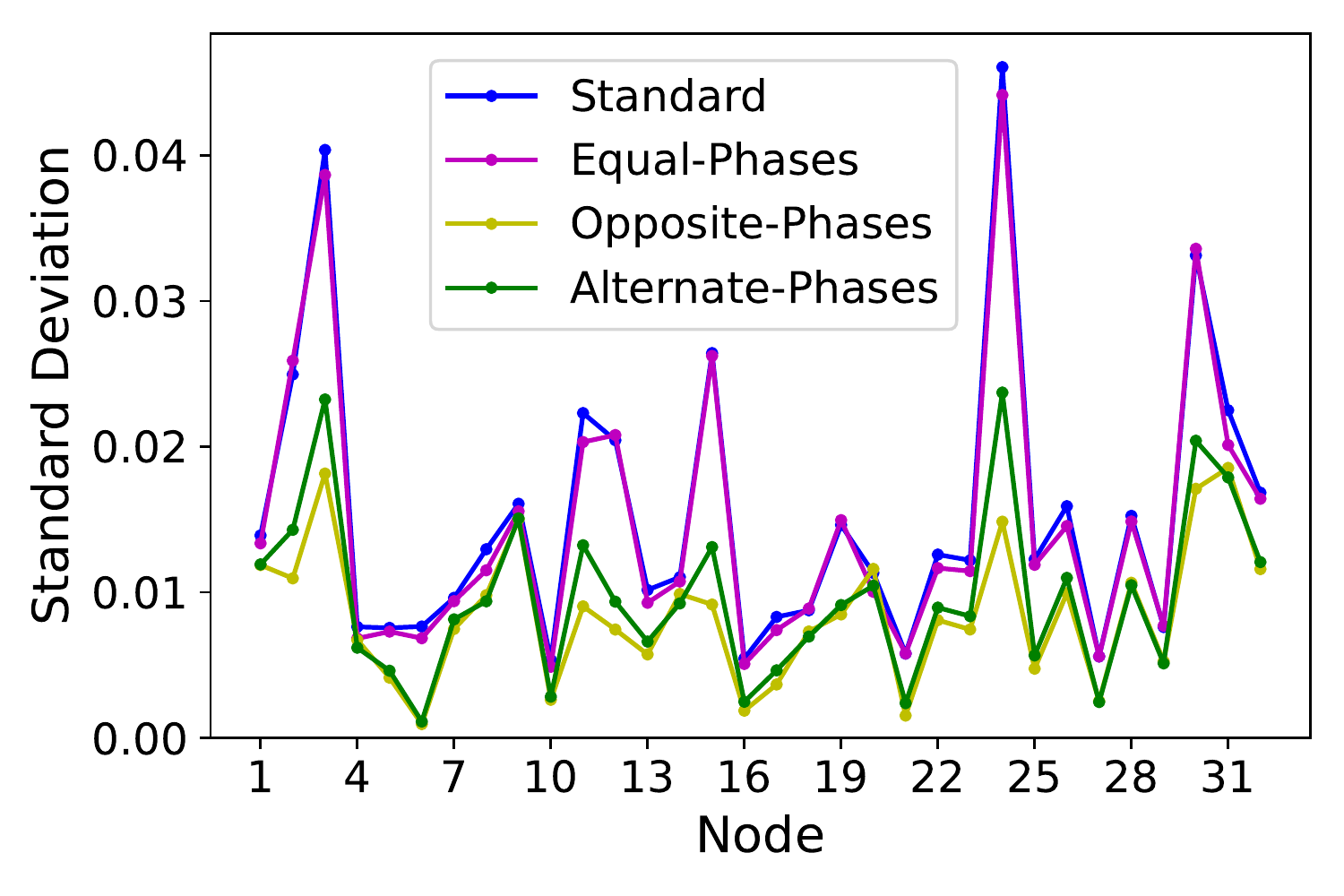}
	\caption{Standard deviations for the quantum PageRanks of a random Erdős-Rényi graph with 32 nodes. $\theta = \pi/2$ has been used for the three APR schemes. The standard deviations decrease for the Opposite-Phases and Alternate-Phases schemes.}
	\label{F:PR_ER_32_std}
\end{figure}

If we consider the standard deviations of the quantum PageRanks, however, we observe a similar effect to that with scale-free networks. In Figure \ref{F:PR_ER_32_std} we can see that for the Equal-Phases case the standard deviations are similar to those of the standard quantum algorithm, whereas for the Opposite-Phases and Alternate-Phases cases the standard deviations are significantly smaller.

\begin{figure}[H]
	\centering
	\subfigure[]{\includegraphics[scale=0.5]{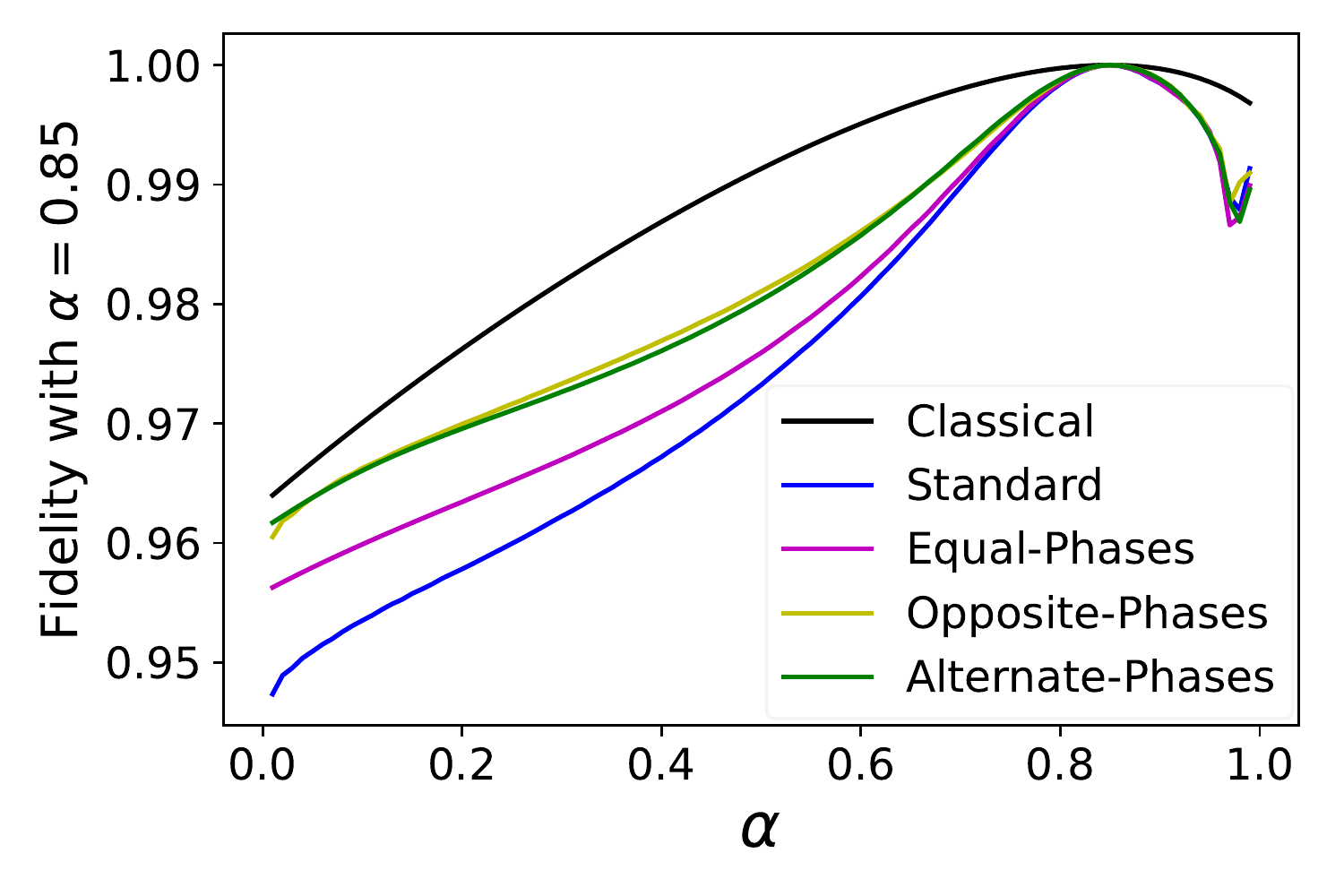}\label{F:Stability_ER_32}}
	\subfigure[]{\includegraphics[scale=0.5]{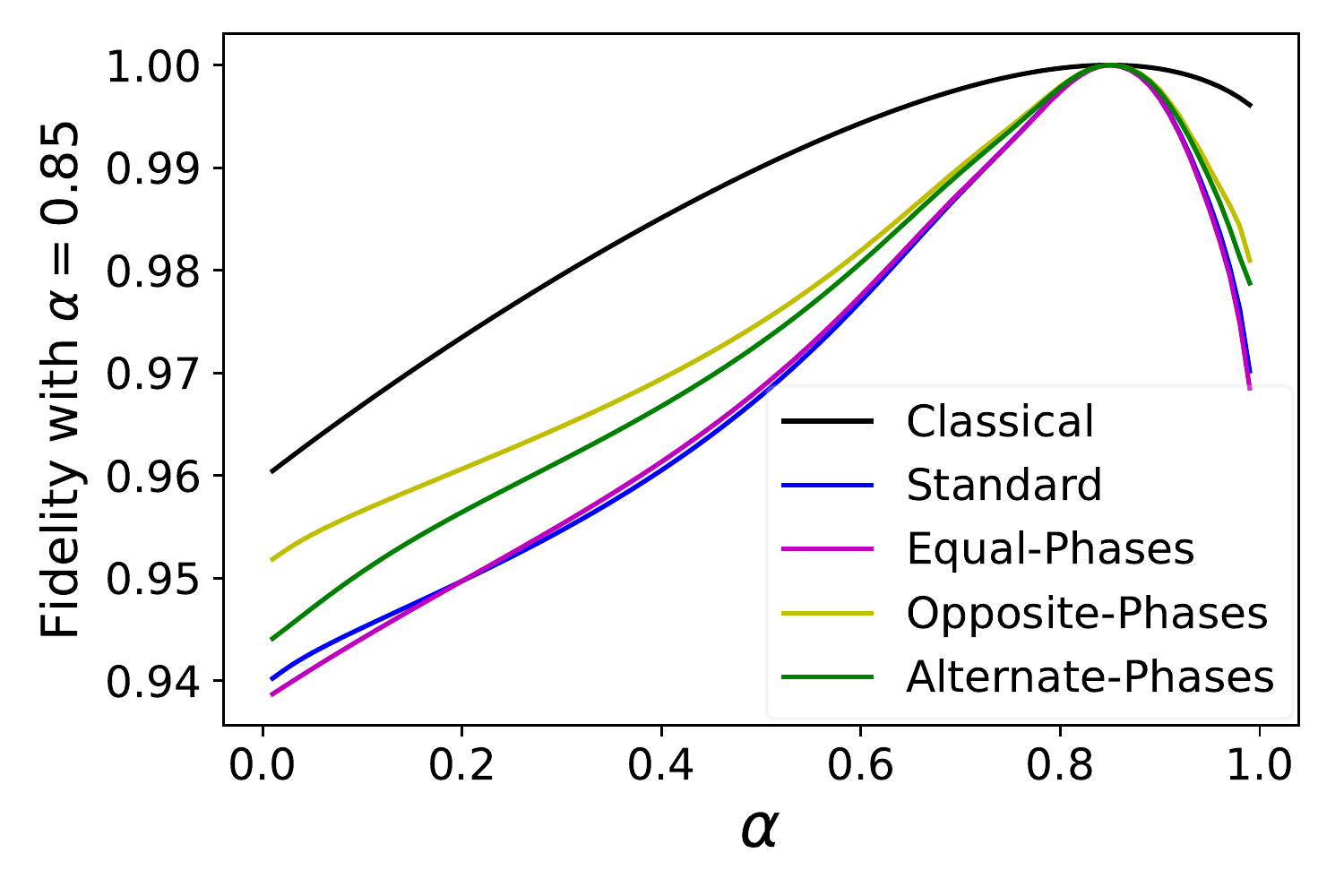}\label{F:Stability_ER_32_av}}
	\caption{a) Fidelity of the PageRank distributions vs the damping parameter $\alpha$, with respect to the distribution with $\alpha = 0.85$, for a random Erdős-Rényi graph with 32 nodes. b) Averaged fidelity of the PageRank distributions vs the damping parameter $\alpha$, with respect to the distribution with $\alpha = 0.85$, for an ensemble of 50 random Erdős-Rényi graphs with 32 nodes. The classical distribution is compared with all the quantum distributions, using $\theta = \pi/2$ in the three APR schemes. We see that all the quantum algorithms are less stable than the classical one. The Alternate-Phases and Opposite-Phases schemes improve the stability of the standard quantum algorithm.}
	\label{...}
\end{figure}

Finally, we have analyzed the stability of the algorithm with respect to the damping parameter $\alpha$ for this network. In Figure \ref{F:Stability_ER_32} we show the fidelity of the PageRank distributions for $\alpha \in [0.01,0.99]$ with respect to the distribution with $\alpha = 0.85$. We find that now the classical algorithm is more stable than all the quantum algorithms. However, all algorithms are actually quite stable, with the worst fidelity being around 0.95. This behavior can be due to the quite homogeneous pattern of the distributions. Since a decrease of alpha results in a greater importance of the random hopping, and this random-hopping tends to give a homogeneous distribution, the decrease of the parameter $\alpha$ here has little effect. However, it is interesting to note that the introduction of the APR improves the stability of the standard quantum algorithm. To ensure that these results are nor particular for this instance of Erdős-Rényi graphs, we have averaged over an ensemble of 50 random graphs with 32 nodes, obtaining Figure \ref{F:Stability_ER_32_av}. We can see that the Opposite-Phases and Alternate-Phases schemes improve the standard quantum algorithm, although the effect is small.

\end{document}